\documentclass[a4paper,11pt]{article}
\usepackage{jheppub} 
\usepackage[utf8]{inputenc}
\usepackage{physics}
\usepackage{slashed}
\usepackage{caption}
\usepackage{xcolor}
\usepackage{comment}
\usepackage{multirow}
\usepackage{graphics}
\usepackage{float}
\usepackage{cancel}
\usepackage{soul}
\usepackage{cases}
\usepackage{array}
\usepackage{mathtools}  
\usepackage{amsfonts}
\usepackage{hyperref}
\usepackage{amsmath}
\usepackage{amssymb}
\usepackage{tcolorbox}
\usepackage{tikz}     
\usetikzlibrary{positioning, arrows.meta}

\title{AdS$_4$ Boundary Wightman functions in Twistor Space: Factorization, Conformal blocks and a Double Copy
}
\author{Arhum Ansari, Sachin Jain, Dhruva K.S.}

\affiliation{Indian Institute of Science Education and Research,\\ Dr Homi Bhabha Road, Pashan, Pune, India}

\emailAdd{ansari.arhum@students.iiserpune.ac.in}
\emailAdd{sachin.jain@iiserpune.ac.in}
\emailAdd{k.s.dhruva@students.iiserpune.ac.in}

\abstract{We study real-time holographic four point Wightman functions involving scalars, photons, gluons and gravitons in the Poincare patch of AdS$_4$. We show that when the momenta of the middle two operators are spacelike, four-point exchange Wightman functions factorize into a product of three-point functions. This expression coincides with a Wightman conformal partial wave corresponding to the operator dual to the exchanged particle in the bulk. Further, we discuss and explicitly show how these results can be analytically continued to obtain Euclidean AdS correlators up to contact diagrams and avoid the need to perform any nested bulk point integrals in contrast to the traditional Witten diagram approach. We then translate our results to twistor space taking first steps towards a twistor space reformulation of four point Wightman functions. For conformally coupled scalars interacting with a cubic potential, we obtain a beautiful form for four and five point functions. For the interesting case of Yang-Mills theory and Einstein gravity, we derive a simple double copy relation. This is easiest to see when correlators are expressed in twistor space using Schwinger parameterization where the graviton correlator is simply the square of its gluon counterpart. }

\begin{document}

\maketitle

\section{Introduction}
 The modern amplitudes program has revolutionized our understanding of quantum field theory by focusing on fundamental principles and consistency conditions to directly bootstrap observables. On-shell methods using techniques such as spinor helicity have led to the development of recursion relations at both tree level and loop, unveiled double copy structures that relate gauge theory and gravity amplitudes and much more, see \cite{Elvang:2013cua,Badger:2023eqz} for a review. The related twistor \cite{Witten:2003nn, Arkani-Hamed:2009hub,Mason:2009sa} and geometric perspective has also revealed remarkable features of amplitudes such as their interpretation as volumes of positive geometries such as the amplituhedron \cite{Arkani-Hamed:2013jha}. These perspectives also help expose previously hidden or obscure symmetries like dual conformal and Yangian symmetry \cite{Drummond:2010qh,Drummond:2010uq,Korchemsky:2010ut,Arkani-Hamed:2010zjl,Arkani-Hamed:2010pyv} and shed light on interesting facts such as the Wilson-loop scattering amplitude correspondence \cite{Adamo:2011pv}.

Given these achievements, it is tempting to investigate whether  these developments of the modern on-shell ideas underlying scattering amplitudes, have natural analogues for CFT correlators, particularly in Lorentzian signature. Traditionally, these quantities have been analyzed in position space \cite{Belavin:1984vu,Rattazzi:2008pe,Poland:2022qrs,Hartman:2022zik}. Although this has the advantage of making manifest locality and the operator product expansion, it becomes cumbersome to deal with when we consider correlators of conserved currents due to the proliferation of tensor structures. Further, this language is also in contrast to the Fourier space description that is used for scattering amplitudes. In recent years, conformal field theory has been formulated in momentum space  \cite{Maldacena:2011nz,McFadden:2011kk,Coriano:2013jba,Bzowski:2013sza,Ghosh:2014kba,Bzowski:2015pba,Bzowski:2017poo,Bzowski:2018fql,Farrow:2018yni,Isono:2019ihz,Bautista:2019qxj,Gillioz:2019lgs,Baumann:2019oyu,Baumann:2020dch,Jain:2020rmw,Jain:2020puw,Jain:2021wyn,Jain:2021qcl,Jain:2021vrv,Baumann:2021fxj,Jain:2021gwa,Jain:2021whr,Gillioz:2022yze,Marotta:2022jrp,Jain:2023idr,Bzowski:2023jwt,S:2024zqp,Gupta:2024yib,Jain:2024bza,Aharony:2024nqs,Marotta:2024sce,Coriano:2024ssu,Gillioz:2025yfb,S:2025pmh,Li:2025glx} and has a variety of applications such as cosmology \cite{Maldacena:2011nz,McFadden:2011kk,Ghosh:2014kba,Arkani-Hamed:2015bza,Arkani-Hamed:2018kmz,Baumann:2022jpr}, the study of AdS amplitudes \cite{Liu:1998ty,Raju:2011mp,Raju:2012zs,Albayrak:2018tam,Albayrak:2019yve,Gadde:2022ghy,Armstrong:2022mfr,Albayrak:2023jzl,Mei:2023jkb,Chowdhury:2023khl,Albayrak:2023kfk,Chowdhury:2024wwe,Moga:2025gdy}, double copy relations \cite{Farrow:2018yni,Lipstein:2019mpu,Jain:2021qcl} and connections to one higher dimensional flat space S-matrices \cite{Maldacena:2011nz,Raju:2012zr,Marotta:2024sce}. In the case of interest to us which is 3d CFT,  spinor helicity variables lead to an advantage since they trade a spin-s symmetric traceless conserved current, (dual to a massless gauge boson in AdS$_4$), for two objects viz positive and negative helicity components \cite{Maldacena:2011nz,Jain:2021qcl,Jain:2021vrv,Jain:2021gwa,Jain:2021whr,Jain:2023idr,S:2025pmh}. Although this leads to simplifications at the level of three point functions, progress at four points and beyond has been limited. This is partly because spinor helicity variables still give rise to complicated expressions due to large  degeneracies in the allowed structures, see \cite{Armstrong:2020woi} for instance.

A recent impetus came in the form of a spacetime twistor approach which focuses on Lorentzian CFT$_3$ rather than its Euclidean counterpart \cite{Baumann:2024ttn,Bala:2025gmz,Bala:2025jbh,Bala:2025qxr,Rost:2025uyj,S:2025pmh,Mazumdar:2025egx,CarrilloGonzalez:2025qjk}. Twistor space has a long history starting with the seminal work by Penrose \cite{Penrose:1967wn}. In the context of its application to scattering amplitudes, we refer the reader to the wonderful lecture notes \cite{Wolf:2010av,Adamo:2017qyl}. In Lorentzian signature, there are many types of correlators such as time-ordered, retarded etc.. The ones that we shall be interested in particular are the real-time \textit{Wightman} functions which form the building blocks for all other correlators. As we shall make more concrete, Wightman functions (or their suitable analytic continuations) enjoy a simple description in twistor space. One reason for this fact for current correlators is that in contrast to time-ordered or retarded correlators which satisfy non trivial conservation Ward-Takahashi identities with contact terms, Wightman functions are identically conserved \cite{Bala:2025gmz}. Conformal invariance fixes two and three point functions and were found in earlier works \cite{Baumann:2024ttn,Bala:2025gmz,Bala:2025qxr}. One of the challenging problems is to understand higher point CFT correlators in twistor space.

 To obtain higher point Wightman functions, we are faced with two options. we can take an intrinsically CFT approach or use AdS/CFT \cite{Maldacena:1997re,Gubser:1998bc,Witten:1998qj} to perturbatively compute CFT correlators. The latter is also of independent interest as it extends the flat space amplitudes program to AdS. We take this holographic approach in this paper. We can either Wick rotate known Euclidean results or directly compute the real time correlators. The former option although tempting is quite difficult in practice even for three point functions \cite{Bautista:2019qxj} . Thus we will focus on the latter approach for the most part to intrinsically calculate boundary Wightman functions in AdS$_4$. There are various ways to compute real time correlators in the context of AdS/CFT, mostly using the Schwinger-Keldysh formalism or its holographic dual, see \cite{Son:2002sd,Satoh:2002bc,Skenderis:2008dh,Skenderis:2008dg,vanRees:2009rw,Barnes:2010jp,Botta-Cantcheff:2017qir,Glorioso:2018mmw,Anand:2019lkt,Loganayagam:2022teq,Loganayagam:2024mnj,Ammon:2025vod} and references therein. In this paper, we exploit the fact that Wightman functions of fundamental fields identically obey the equation of motion \cite{Steinmann:1992va}. We start with a bulk action and equation of motion which we perturbatively solve to calculate momentum space Wightman functions of interest to us order by order in the coupling constants. This method is algebraic in nature and the calculations are straightforward.
 
 We compute three and four point Wightman functions in theories with general scalars, photons, gluons, gravitons with a variety of interactions between them which reveal a striking simplicity compared to their time ordered counterparts. For example, we show that in special kinematics where the momentum of the middle two operators are space-like in a four point function, it factorizes and turns into a conformal partial wave associated to the operator dual to the exchanged particle in the bulk\footnote{Conformal partial waves are related to conformal blocks by a monodromy projection, see \cite{Simmons-Duffin:2012juh}.}. This is reminiscent of the observation previously made for time-ordered correlators when we cut the bulk to bulk propagator \cite{Meltzer:2021bmb}(see also \cite{Meltzer:2020qbr}). We also demonstrate how our factorized Wightman functions can be analytically continued to Euclidean space recovering the correlator up to contact diagram contributions. This sidesteps the nested bulk integrals that one would usually perform in Witten diagram calculations. 

Finally, returning to our main motivation for pursuing these computations, we head towards twistor space using spinor helicity variables and the half-Fourier/Witten transform  \cite{Witten:2003nn}. In the special kinematics, we find a strikingly simple form for the four and five points scalar Wightman functions with a potential generalization to the n-point case\footnote{It is tempting to directly convert the results of \cite{Meltzer:2021bmb} (for the time-ordered correlator with the bulk to bulk propagator cut) into twistor space as it is a factorized expression. However, this quantity obeys the current conservation Ward-Takahashi identities with contact terms thus preventing them from enjoying a simple representation in twistors space. One might consider performing a cut on all external legs. As one can check this would actually lead to zero! The Wightman function in the special kinematics precisely tells us what is the simplest (and non-trivial) object to consider at four points in twistor space. It is a Wightman function in the special kinematics which is obtained from the time-ordered correlator by performing three cuts: one for the exchanged momentum and two corresponding to the first and fourth operator in the Wightman function. }. Corresponding to the same kinematics for the gluon and graviton case, we establish a simple double copy relation which is easiest to see when the twistor space correlators are expressed in Schwinger parametrization: The graviton correlator is simply the square of its gluon counterpart in these variables. This double copy relation also exists in spinor helicity and momentum space but twistor space is much more efficient in establishing this result due to its effective way of dealing with degeneracies. This double copy relation allows us to systematically compute the full graviton correlator in any kinematics.

Succinctly, the main messages of this paper are,
\begin{enumerate}
\item We provide a simple and systematic method to compute Wightman functions using the bulk equations of motion.
    \item Four point Wightman functions factorize in the special kinematics when the middle operators have space-like momenta resulting in a Wightman conformal partial wave.
    \item This conformal partial wave can be used to construct the time-ordered or Euclidean correlator corresponding to the bulk process up to contact diagrams.
    \item Wightman functions in particular enjoy a simple description in twistor space and we obtain four point functions corresponding to these special kinematics.
    \item We discovered a simple double copy relation between gluon and graviton four point functions in twistor space and spinor helicity variables.
\end{enumerate}

Overall, our results indicate that twistor space is a promising avenue to pursue for the study of higher point functions.

A detailed outline of the paper is as follows:
\subsection*{Outline of the paper}
In section \ref{sec:WightmanBasics}, we review the basics of Wightman functions in field theory followed by setting up their holographic calculation in AdS$_4$. In section \ref{sec:propagatorsAction}, we set up the quadratic actions and compute the Wightman and Feynman propagators in the theories of interest to us. We then illustrate our formalism with many examples to compute momentum space Wightman functions in section \ref{sec:WightmanMomentumSpace}, first with two point functions in subsection \ref{subsec:Twopoint}, three point functions in subsection \ref{subsec:Threepoint} and four point functions in subsection \ref{subsec:fourpoint}, subsection \ref{subsec:fourpointspecial}. Having observed that four point functions factorize into a product of three point functions, we organize and re-interpret these results in terms of conformal partial waves in subsection \ref{sec:FactorizationandCPW}. Finally, we discuss one five point example in subsection \ref{subsec:fivepoint}.  In section \ref{sec:WightmanToEuclid}, we show how these factorized Wightman functions can be analytically continued to Euclidean AdS/CFT correlators. With all this formalism and results at hand, we convert our results into the twistor space framework in section \ref{sec:Twistors}. We begin by reviewing the construction of twistor space via a half-Fourier transform from momentum space and spinor helicity variables in subsection \ref{subsec:SHtoTwistor}. We also make clear in this section, the precise relationship between Wightman functions and their twistor counterparts. In subsection \ref{subsec:twistor2point} and subsection \ref{subsec:twistor3point}, we respectively discuss the construction of two and three point twistor space correlators with many examples. We then initiate the construction of four point functions in twistor space in subsection \ref{subsec:twistor4point} with examples of conformally coupled scalars, gluons and gravitons. We also present an example of a five point scalar correlator in subsection \ref{subsec:twistorfivepoint}. As an application of our results, we show a beautiful double copy relation between the Yang-Mills and Einstein gravity Wightman functions in section \ref{sec:DoubleCopy}. Finally, we summarize our findings and discuss potential future directions in section \ref{sec:Discussion}.
\subsection*{Outline of the appendices}
We have a few appendices to clarify and support the material in the main-text. In appendix \ref{app:Notation}, we set our notation straight. In appendix \ref{app:CCscalarsReview}, we review the basics of conformally coupled scalars with an emphasis on $i\epsilon$ prescription and boundary conditions. In appendix \ref{app:LongitudinalContributions}, we show that the conserved current Wightman functions we construct identically obey the Ward-Takahashi identity without contact terms, serving as an additional consistency check of our formalism. In appendix \ref{app:CPWexamples}, we present the expressions for our factorized four point functions in terms of conformal partial waves. In appendix \ref{app:HalfFourierDetails}, we derive the spinor helicity counterpart of the general solution to the twistor space conformal Ward identities for current three point functions. Finally, we tabulate in appendix \ref{app:YmandGRallhelicities}, the results in all helicities in both spinor helicity and twistor space for three point Yang-Mills and Einstein gravity Wightman functions.

\section{The Anatomy of Wightman Functions in AdS/CFT}\label{sec:WightmanBasics}
The aim of this section is to set the stage for the calculation of Wightman functions in the context of the AdS/CFT correspondence. There are a few possible approaches to compute these real-time correlators such as, (i) Solving the bulk equation of motion generalizing the methods of \cite{Steinmann:1992va} to AdS, (ii) Through analytic continuation from Euclidean AdS such as in \cite{Bautista:2019qxj} and (iii) Via the Skenderis-van Rees construction \cite{Skenderis:2008dg,Skenderis:2008dh}. For the vacuum Wightman functions we are interested in, we choose the first option as it is straightforward, algebraic and simple to implement even when dealing with spinning particles. The second option of Wick rotation is tempting but in momentum space is actually quite involved for generic momentum configurations. We shall however, use it as a check of our results whenever possible.  

Our focus is on AdS$_4$/CFT$_3$ for the most part but the methods outlined below extend to arbitrary dimensions. To set the stage, we begin our discussion with the general properties of Wightman functions in quantum field theories (conformal field theories in particular) in subsection \ref{subsec:generalities}. In subsection \ref{subsec:AdSCFTextrapolate}, we discuss the AdS/CFT extrapolate dictionary which we shall use to compute tree level Wightman functions in the Poincare patch. We then present the three possible approaches discussed above to perform these calculations in subsection \ref{subsec:WaysToComputeWightman}.

\subsection{Generalities}\label{subsec:generalities}
A Wightman function is an ordered expectation value in a particular state \cite{Streater:1989vi}, which for us is the vacuum. Let $\mathcal{O}_i$ be generic operators with $i$ shorthand for possible spinor or vector indices. A Wightman function with a particular ordering is denoted as,
\begin{align}\label{WightmanPosSpace1}
    \langle 0|\mathcal{O}_1(x_1)\cdots \mathcal{O}_n(x_n)|0\rangle.
\end{align}
$|0\rangle$ is the Poincar\'e invariant vacuum and $x_i=(t_i,\Vec{x}_i)$ labels the spacetime location of the operator. With $n$ operators, there are $n!$ such Wightman functions which in general are distinct. However, they are all equal when every operator is space-like separated by virtue of micro-causality. Wightman functions also come equipped with a particular $i\epsilon$ prescription. To see this, let us write \eqref{WightmanPosSpace1} in the Heisenberg picture, placing all operators at the origin in space for simplicity. Writing the $i\epsilon$ dependence explicitly we have,
\begin{align}\label{WightmanDefPosSpace}
    \langle 0|\mathcal{O}_1(0)e^{-i H(t_1-t_2)-H(\epsilon_1-\epsilon_2)}\mathcal{O}_2(0)\cdots e^{-iH(t_{n-1}-t_n)-H(\epsilon_{n-1}-\epsilon_n)}\mathcal{O}_n(0)|0\rangle,
\end{align}
with $\epsilon_1>\epsilon_2>\cdots>\epsilon_n>0$ that ensures that the contribution of high energy states to the correlator are damped. The limit $\epsilon_i\to 0$ keeping fixed their ordering is to be taken only after smearing the Wightman functions with Schwartz functions. As such, Wightman functions are actually distributions (tempered distributions to be precise) but we shall refer to them as functions as is common practice. This $i\epsilon$ prescription also has the consequence that Wightman functions are not quite real and obey the following reality property:
\begin{align}\label{WightmanRealityPos}
    \langle 0|\mathcal{O}_1(x_1)\cdots\mathcal{O}_n(x_n)|0\rangle^*=\langle 0|\mathcal{O}_n(x_n)\cdots \mathcal{O}_1(x_1)|0\rangle,
\end{align}
that is, complex conjugating the correlator reverses the operator ordering.

Another important property of Wightman functions is that they satisfy the spectral condition. To see this, we need to go to momentum space. The fact that the Fourier transform of a tempered distribution is also a tempered distribution makes momentum space Wightman functions well defined. Poincar\'e invariance in Fourier space implies momentum conservation so we can strip off this universal factor to obtain,
\begin{align}\label{WightmanDefMomSpace1}
    \langle 0|\mathcal{O}_1(p_1)\cdots \mathcal{O}_n(p_n)|0\rangle=(2\pi)^d\delta^d(p_1+\cdots+p_n)\langle\langle 0|\mathcal{O}_1(p_1)\cdots \mathcal{O}_n(p_n)|0\rangle\rangle,
\end{align}
that is, single bra-ket quantities in momentum space $\langle 0|\cdots |0\rangle$ including the momentum conserving delta function whereas double bra-ket quantities $\langle\langle 0|\cdots|0\rangle\rangle$ indicated that this factor is stripped off.
The spectral condition is the statement that Wightman function are non-zero iff,
\begin{align}\label{spectralcondition}
    \bigg(\sum_{i=j}^{n}p_i\bigg)^2<0,~~ \sum_{i=j}^{n}p_i^{0}>0,~~j\in\{2,3,\cdots,n-1,n\},
\end{align}
that is, the sum of all the momenta of the operators acting on $|0\rangle$ should be time-like and have positive energy. For example, take $j=n$. This implies that $p_n^2<0,p_n^0>0$. Taking $j=2$ yields $(p_{2}^{\mu}+\cdots p_{n}^{\mu})^2<0,(p_2^0+\cdots p_n^0)>0$ which by momentum conservation implies that $p_1^2<0, p_1^0<0$. Similarly, taking $j=n-1$ implies $(p_n^\mu+p_{n-1}^\mu)^2<0,p_n^0+p_{n-1}^0>0$ and so on. Thus, for every Wightman function, the rightmost and leftmost operators have to have time-like momenta with positive and negative energy respectively. The operators in the middle can in general have time-like or space-like momenta as long as \eqref{spectralcondition} is satisfied\footnote{For example, the middle operator in a three point function can in general have time-like or space-like momenta.\label{footnote:threepointkinematics}}. Finally, the reality condition \eqref{WightmanRealityPos} in Fourier space takes the form,
\begin{align}\label{WightmanRealityMom}
    \langle 0|\mathcal{O}_1(p_1)\cdots \mathcal{O}_n(p_n)|0\rangle^*=\langle 0|\mathcal{O}_n(-p_n)\cdots \mathcal{O}_1(-p_1)|0\rangle.
\end{align}
Wightman functions satisfy other properties such as cluster decomposition and positivity for which we refer the reader to \cite{Kravchuk:2021kwe,Gillioz:2022yze} for a detailed discussion of all the Wightman axioms in the context of CFT.
\subsection{The AdS/CFT extrapolate dictionary}\label{subsec:AdSCFTextrapolate}
With the above Wightman axioms in mind, we proceed to discuss our objects of interest which are tree level boundary $\mathbb{R}^{2,1}$ conformal Wightman functions of scalars, photons, gluons and gravitons in the Poincar\'e patch of AdS$_4$. The metric is given by,
\begin{align}
    ds^2=\frac{dz^2+\eta_{\mu\nu}dx^\mu dx^\nu}{z^2},
\end{align}
with $\eta_{\mu\nu}=\text{diag}(-1,1,1)$, is the three dimensional Minkowski metric on $\mathbb{R}^{2,1}$. The $z$ coordinate lies in the range $0<z<\infty$ with the conformal boundary located at $z=0$. 

We obtain boundary Wightman functions via the AdS/CFT extrapolate dictionary \cite{Harlow:2011ke}\footnote{This equation contains an overall factor that we suppress. This is easy to compute at tree level. At loop level, this factor renormalizes non-trivially and thus \eqref{AdSCFTextrapolate} would have to be modified \cite{Banados:2022nhj}. We thank Kostas Skenderis for this comment.},
\begin{align}\label{AdSCFTextrapolate}
    \langle 0|\mathcal{O}_1(p_1)\cdots \mathcal{O}_n(p_n)|0\rangle\sim \lim_{z_i\to 0} z_1^{-\Delta_1}\cdots z_n^{-\Delta_n}\langle 0| \Phi_1(z_1,p_1)\cdots \Phi_n(z_n,p_n)|0\rangle.
\end{align} 
$\Phi_i(z,p)$ is the bulk dual of the CFT operator $\mathcal{O}_i(z,p)$ which has scaling dimension $\Delta_i$ and spin $s_i$\footnote{The  dictionary \eqref{AdSCFTextrapolate} also generalizes naturally for spinning cases. In our normalizations, we scale by $z^{-1}$ for photons and gluons and $z^{-3}$ for gravitons, when taking the boundary limit of bulk correlators involving the fields corresponding to these operators.}. These boundary correlators obey the properties discussed in subsection \ref{subsec:generalities} as will also be clear from explicit calculations in the later sections. Before we proceed to the computations, let us discuss the different types of propagators that we shall require. First of all, since we are interested in boundary correlators we require both bulk to bulk propagators representing particle exchanges in the bulk as well as bulk to boundary propagators.  In Lorentzian signature AdS, there are many possibilities for propagators such as Feynman, anti-Feynman, retarded, advanced and the two Wightman propagators. The Wightman two point functions in terms of the bulk field two point function are given by,
\begin{align}
    &W_{+}(z,z',x-x')=\langle 0|\Phi(z,x)\Phi(z',x')|0\rangle,\notag\\
    &W_{-}(z,z',x-x')=\langle 0|\Phi(z',x')\Phi(z,x)|0\rangle.
\end{align}
These quantities as a consequence of the $i\epsilon$ prescription (which is the same as in \eqref{WightmanDefPosSpace} since it has to do only with the time coordinates which the presence of the extra bulk spatial coordinate does not affect) are also complex conjugates of each other,
\begin{align}\label{Wplusminusconjugation}
    W_{+}(z,z',x-x')^*=W_{-}(z,z',x-x').
\end{align}
The remaining propagators such as Feynman ($G_F$), anti-Feynman ($G_{\Bar{F}}$), retarded ($G_R$) and advanced ($G_A$) are determined in terms of these fundamental quantities. 
\small
\begin{align}\label{propagatordefs}
    &G_F(z,z',x-x')=\langle 0|T\{\Phi(z,x)\Phi(z',x')\}|0\rangle=\theta(t-t')W_{+}(z,z',x-x')+\theta(t'-t)W_{-}(z,z',x-x'),\notag\\
    &G_{\Bar{F}}(z,z',x-x')=\langle 0|\Bar{T}\{\Phi(z,x)\Phi(z',x')\}|0\rangle=\theta(t'-t)W_{+}(z,z',x-x')+\theta(t-t')W_{-}(z,z',x-x'),\notag\\
    &G_R(z,z',x-x')=-i\theta(t-t')\langle 0|[\Phi(z,x),\Phi(z',x')]|0\rangle=-i\theta(t-t')\big(W_{+}(z,z',x-x')-W_{-}(z,z',x-x')\big),\notag\\&G_A(z,z',x-x')=i\theta(t'-t)\langle 0|[\Phi(z,x),\Phi(z',x')]|0\rangle=i\theta(t'-t)\big(W_{+}(z,z',x-x')-W_{-}(z,z',x-x')\big).
\end{align}
\normalsize
The bulk to boundary propagators are obtained by taking either $z$ or $z'$ to the boundary and rescaling to obtain a finite non-zero result. Not all these propagators are linearly independent. They satisfy what is known as the largest time equation viz,
\begin{align}
    &~~~G_F(z,z',x-x')+G_{\Bar{F}}(z,z',x-x')=W_{+}(z,z',x-x')+W_{-}(z,z',x-x'),\notag\\&i(G_R(z,z',x-x')-G_A(z,z',x-x'))=W_{+}(z,z',x-x')-W_{-}(z,z',x-x').
\end{align}
Finally, let us note that once we calculate all the $n!$ different $n-$point Wightman functions, all other correlators of interest can be obtained. For instance the time-ordered correlator is given in terms of Wightman functions as follows,
\begin{align}
    \langle 0|T\{\mathcal{O}_1(x_1)\cdots \mathcal{O}_n(x_n)\}|0\rangle=\sum_{\pi\in S_n}\theta(t_{\pi(1)}>t_{\pi(2)}>\cdots t_{\pi(n)})\langle 0|\mathcal{O}_{\pi(1)}(x_{\pi(1)})\cdots \mathcal{O}_{\pi(n)}(x_{\pi(n)})|0\rangle,
\end{align}
where $S_n$ is the symmetric group on $n$ elements which is the group whose elements are all possible permutations of $n$ objects, which are $n!$ in total. Each object on the RHS of this equation are Wightman functions with different orderings. Similarly, one can derive formulae that relate other types of correlators to Wightman functions. 

\subsection{Wightman functions and How to Compute Them}\label{subsec:WaysToComputeWightman}
 With all the above facts in mind, let us discuss three different ways to calculate holographic Wightman functions.
\subsubsection{Boundary Wightman functions via Bulk EOM}
The first and most straightforward way to calculate tree level holographic correlators is to perturbatively solve the non-linear bulk equation of motion. To illustrate this method, let us consider a conformally coupled scalar field in AdS$_4$ with a cubic self-interaction.
The action is given by\footnote{Note that all integrals over $z$ run from $0$ to $\infty$ even when not explicitly mentioned.},
\begin{align}\label{ccscalaraction}
    S=-\frac{1}{2}\int\frac{dz}{z^4} d^3 x\bigg((\partial_z\Phi)^2+(\partial_\mu\Phi)^2-2 \Phi^2\bigg)-\frac{g}{3!}\int\frac{dz}{z^4}d^3 x \Phi^3.
\end{align}
Let us perform a Weyl rescaling,
\begin{align}\label{ccscalarWeyl}
    \Phi(z,x)=z \phi(z,x).
\end{align}
This converts \eqref{ccscalaraction} into an action in half of flat space with a $z$ dependent interaction:
\begin{align}
    S=-\frac{1}{2}\int dz~d^3 x~\big((\partial_z\phi)^2+(\partial_\mu\phi)^2\big)-\frac{g}{3!}\int\frac{dz}{z}d^3 x ~\phi^3.
\end{align}
The equation of motion reads,
\begin{align}\label{ccphicubeEOMfull}
    (\partial_z^2+\Box)\phi(z,x)=\frac{g}{2z}\phi^2(z,x),
\end{align}
 We perturbatively expand the field in $g$ as,
\begin{align}\label{ccphicubeperturbation}
    \phi(z,x)=\sum_{n=0}^{\infty}g^n \phi^{(n)}(z,x).
\end{align}
Substituting this expansion into the equation of motion \eqref{ccphicubeEOMfull} results in a set of equations organized by the powers of $g$.
\small
\begin{align}\label{ccphicubeEOM}
&\sum_{n=0}^{\infty}g^{n}(\partial_z^2+\Box)\phi^{(n)}(z,x)=\sum_{n=0}^{\infty}\sum_{m=0}^{\infty}\frac{g^{n+m+1}}{2z}\phi^{(n)}(z,x)\phi^{(m)}(z,x)\notag\\
&\implies (\partial_z^2+\Box)\phi^{(0)}(z,x)=0,~(\partial_z^2+\Box)\phi^{(1)}(z,x)=\frac{1}{2z}(\phi^{(0)}(z,x))^2,(\partial_z^2+\Box)\phi^{(2)}(z,x)=\frac{1}{z}\phi^{(0)}(z,x)\phi^{(1)}(z,x),\cdots.
\end{align}
\normalsize
The first equation is the free equation of motion whereas the higher order equations determine the field corrections in terms of the free field $\phi_0$. We can solve \eqref{ccphicubeEOM} using the technique of Green's functions as follows:
\begin{align}\label{phicubefieldcorrections}
    &\phi^{(1)}(z,x)=\frac{i}{2}\int \frac{dz' d^3 x'}{z'}\mathcal{G}(z,z',x-x')(\phi^{(0)}(z',x'))^2,\notag\\
    &\phi^{(2)}(z,x)=i\int \frac{dz' d^3 x'}{z'}\mathcal{G}(z,z',x-x')\phi^{(0)}(z',x')\phi^{(1)}(z'',x''),\cdots,
\end{align}
where $\mathcal{G}(z,z',x-x')$ solves,
\begin{align}\label{mathcalGeqn}
    (\partial_z^2+\Box)\mathcal{G}(z,z',x-x')=-i \delta(z-z')\delta^3(x-x').
\end{align}
 At this point a natural choice could be to take $\mathcal{G}(z,z',x-x')$ as the Feynman propagator. However, note that we can also freely add  Wightman functions to these solutions since they are homogeneous quantities\footnote{One can easily show that the definition \eqref{propagatordefs} implies that any of the propagators there obey \eqref{mathcalGeqn} using the fact that Wightman functions identically solve the equation of motion.},
 \begin{align}
     (\partial_z^2+\Box)W_{\pm}(z,z',x-x')=0.
 \end{align}
 These ambiguities can be fixed by demanding appropriate boundary conditions as well as reality conditions. We impose Dirichlet boundary conditions for the field so $\mathcal{G}(z,z',x-x')$ must go to zero as we take either $z$ or $z'$ to zero. Further, since our scalar field is real, $\phi^{(1)},\phi^{(2)}$ etc.. all have to be real. Due to the overall factors of $i$ in \eqref{phicubefieldcorrections}, this implies that the propagator has to be purely imaginary to ensure that the field corrections are real. Thus, the propagator relevant for our computation of Wightman functions is the following combination of the Feynman and Wightman propagators:
\begin{align}\label{newpropagator}
    \mathcal{G}(z,z',x-x')=&G_F(z,z',x-x')-\frac{1}{2}\big(W_{+}(z,z',x-x')+W_{-}(z,z',x-x')\big)\notag\\
    &=i \Im{G_F(z,z',x-x')}.
\end{align}
Note that this propagator is purely imaginary, satisfying,
\begin{align}\label{newpropagatorreality}
    \mathcal{G}(z,z',x-x')^*=-\mathcal{G}(z,z',x-x').
\end{align}
We call this propagator the EOM (equation of motion) inverter propagator. It will also ensure that the Wightman functions we construct respect the conjugation property \eqref{WightmanRealityPos} ensuring that our construction is correct. 
In the next section, we will be more explicit about the form of these propagators.
Thus, \eqref{phicubefieldcorrections} with \eqref{newpropagator} can be used to calculate Wightman functions perturbatively. For example, consider the $\order{g}$ correction to the three point Wightman function. This receives contributions from three terms,
\begin{align}\label{phicubethreepoint}
    &\langle 0|\phi(z_1,x_1)\phi(z_2,x_2)\phi(z_3,x_3)|0\rangle_{\order{g}}=g\bigg(\langle 0|\phi^{(1)}(z_1,x_1)\phi^{(0)}(z_2,x_2)\phi^{(0)}(z_3,x_3)|0\rangle\notag\\&+\langle 0|\phi^{(0)}(z_1,x_1)\phi^{(1)}(z_2,x_2)\phi^{(0)}(z_3,x_3)|0\rangle+\langle 0|\phi^{(0)}(z_1,x_1)\phi^{(0)}(z_2,x_2)\phi^{(1)}(z_3,x_3)|0\rangle\bigg).
\end{align}
After solving for $\phi^{(1)}$ in terms of $(\phi^{(0)})^2$ as in \eqref{phicubefieldcorrections}, we obtain,
\begin{align}
    &\langle 0|\phi(z_1,x_1)\phi(z_2,x_2)\phi(z_3,x_3)|0\rangle_{\order{g}}\notag\\=&\frac{ig}{2}\bigg(\int \frac{dz d^3 x}{z}\mathcal{G}(z,z_1,x-x_1)\langle 0|(\phi^{(0)}(z,x))^2\phi^{(0)}(z_2,x_2)\phi^{(0)}(z_3,x_3)|0\rangle\notag\\
    &~~+\int \frac{dz d^3 x}{z}\mathcal{G}(z,z_2,x-x_2)\langle 0|\phi^{(0)}(z_1,x_1)(\phi^{(0)}(z,x))^2\phi^{(0)}(z_3,x_3)|0\rangle\notag\\
    &~~+\int \frac{dz d^3 x}{z}\mathcal{G}(z,z_3,x-x_3)\langle 0|\phi^{(0)}(z_1,x_1)\phi^{(0)}(z_2,x_2)(\phi^{(0)}(z,x))^2|0\rangle\bigg).
\end{align}
As the correlators in the RHS of the above expressions are free theory Wightman functions, we can perform Wick contractions using Wightman propagators\footnote{Note that this is in contrast to what we usually do in standard QFT for time-ordered correlators, which involve Feynman, rather than Wightman propagators in Wick's theorem.}. It is also important to note that when we encounter composite operators, we follow the standard practice of normal ordering them to avoid self contractions. Thus, we obtain,
\begin{align}
     &\langle 0|\phi(z_1,x_1)\phi(z_2,x_2)\phi(z_3,x_3)|0\rangle_{\order{g}}\notag\\=&i g\bigg(\int \frac{dz d^3 x}{z}\mathcal{G}(z,z_1,x-x_1)W_{+}(z,z_2,x-x_2)W_{+}(z,z_3,x-x_3)\notag\\
    &~~+\int \frac{dz d^3 x}{z}\mathcal{G}(z,z_2,x-x_2)W_{-}(z,z_1,x-x_1)W_{+}(z,z_3,x-x_3)\notag\\
    &~~+\int \frac{dz d^3 x}{z}\mathcal{G}(z,z_3,x-x_3)W_{-}(z,z_1,x-x_1)W_{-}(z,z_2,x-x_2)\bigg).
\end{align}
Note that the choice of using the propagator \eqref{newpropagator} and its reality condition \eqref{newpropagatorreality}, the Wightman propagator conjugation property \eqref{Wplusminusconjugation}, ensures the Wightman conjugation property \eqref{WightmanRealityPos}. Given this expression, one can then obtain the dual boundary conformal correlator via the extrapolate dictionary,
\begin{align}
    \langle 0|O_{\Delta}(x_1)O_{\Delta}(x_2)O_{\Delta}(x_3)|0\rangle_{\order{g}}=\lim_{z_i\to 0}(z_1 z_2 z_3)^{1-\Delta}\langle 0|\phi(z_1,x_1)\phi(z_2,x_2)\phi(z_3,x_3)|0\rangle_{\order{g}},
\end{align}
where $\Delta=1,2$ depending on whether we impose Neumann or Dirichlet boundary conditions for $\phi_0(z,x)$ at $z=0$ which will also determine the exact forms of the propagators in the above expression. The extra $1$ in the rescaling exponents is due to the fact that we are working with Weyl rescaled fields \eqref{ccscalarWeyl}. The analysis for higher point functions and those involving operators with arbitrary scaling dimension and spin is analogous. Also, as mentioned earlier, we will compute these objects in momentum space. This is obtained via a Fourier transform of the above equations but it is more convenient to directly calculate momentum space Wightman functions as we shall illustrate in the subsequent sections. Now, we proceed to another method to obtain Wightman functions: by Wick rotating their Euclidean AdS counterparts.

\subsubsection{From Euclidean correlators to Wightman functions via Wick rotation}
The conformal boundary of Euclidean AdS$_4$ is $\mathbb{R}^3$. Conformal correlators in Euclidean space are single-valued and thus there exists only one kind of correlator which is unsurprisingly called the Euclidean correlator. In this section, we discuss how one can Wick rotate these quantities appropriately to obtain Wightman functions. We start in position space where the procedure is clear\footnote{We closely follow David Duffins' TASI lecture notes on Conformal Field Theory in Lorentzian Signature available at his \href{http://theory.caltech.edu/~dsd/.}{Caltech home page}.}. Consider the following Euclidean conformal correlator,
\begin{align}
    \langle \mathcal{O}_1(\tau_1,\vec{x}_1)\cdots\mathcal{O}_n(\tau_n,\vec{x}_n)\rangle,
\end{align}
where $\tau_i$ denotes the Euclidean time and $\vec{x}_i$ are the spatial locations of the operators. To obtain the Wightman function in \eqref{WightmanPosSpace1} we set $\tau_i=it_i+\epsilon_i$ fixing $\epsilon_1>\cdots>\epsilon_n$ which are infinitesimal quantities. This results in the Wightman function \eqref{WightmanPosSpace1} while also explaining the origin of the $i\epsilon$ prescription in \eqref{WightmanDefPosSpace}. The ordering of the $\epsilon_i$ is required for the Euclidean correlator to be finite since the Euclidean time evolution operator is $e^{-H\tau}$ in contrast to the oscillating $e^{iH t}$ in Lorentzian signature. The $n!$ different Wightman functions can thus be obtained by taking different Euclidean orderings and performing an analytic continuation as above. The corresponding process in momentum space is more involved. Let us consider a Euclidean momentum space correlator,
\begin{align}
     \langle\langle \mathcal{O}_1(p_{1E},\vec{p}_1)\cdots\mathcal{O}_n(p_{nE},\vec{p}_n)\rangle\rangle=\sum_{\mathcal{I}}\mathcal{T}_{\mathcal{I}}\mathcal{F}_{\mathcal{I}},
\end{align}
where $\mathcal{T}_{\mathcal{I}}$ denotes possible tensor structures and $\mathcal{F}_{\mathcal{I}}$ are form factors that depend on the independent invariants formed out of the momenta. For example $p_i=\sqrt{p_{iE}^2+(\vec{p}_i)^2}$ is one possible class of invariants. $s_{ij}=\sqrt{(p_{iE}+p_{jE})^2+(\vec{p}_i+\vec{p}_j)^2}$ is another.
The Wick rotation we need to perform depends on whether the momentum we want to reach to is space-like or time-like. For the space-like case $p_i^2>0$, we do not require an $i\epsilon$ prescription and one can simply set $p_{iE}=-ip_{i}^{0}$ resulting in,
\begin{align}
    p_i=\sqrt{p_{iE}^2+(\vec{p}_i)^2}\to \sqrt{-(p_{i}^0)^2+(\vec{p}_i)^2}=\sqrt{|p_i|^2}=|p_i|~~~\text{if}~~~p_i^2>0.
\end{align}
On the other hand, we need to specify an $i\epsilon$ prescription for time-like momenta $p_i^2<0$. Let us see this for the Feynman and Wightman propagators where we have two distinct $i\epsilon$ prescriptions to consider. The Feynman $i\epsilon$ prescription is,
\begin{align}\label{Feynmaniepsilon}
    p_i=\sqrt{p_{iE}^2+(\vec{p}_i)^2}\to \sqrt{-(p_{i}^0)^2+(\vec{p}_i)^2-i\epsilon}=-i \sqrt{(p_{i}^0)^2-(\vec{p}_i)^2}=-i|p_i|~~~\text{if}~~~p_i^2<0.
\end{align}
The Wightman $i\epsilon$ prescription on the other hand is,
\begin{equation}\label{Wightmaniepsilon}
     p_i=\sqrt{p_{iE}^2+(\vec{p}_i)^2}\to \sqrt{-(p_{i}^0)^2+(\vec{p}_i)^2+i\epsilon p_{i}^{0}}=\left\{\begin{split}
       e^{-\frac{i\pi}{2}}\sqrt{|p_i|^2}=-i|p_i|,~~p_i^2<0,p_{i}^{0}<0\\e^{+\frac{i\pi}{2}}\sqrt{|p_i|^2}=i|p_i|,~~p_i^2<0,p_{i}^{0}>0
    \end{split}\right\}.
\end{equation}
Further, to obtain a Wightman function we need to take particular discontinuities of the Euclidean form factors.

Let us illustrate this with a three point example. The spectral conditions require that the first and third operators have time-like momenta which we take to be $p_1^2<0,p_3^2<0$ with $p_1^0<0,p_3^0>0$. For concreteness, we consider the middle operator to have space-like momenta $p_2^2>0$ although as mentioned in footnote \ref{footnote:threepointkinematics}, the middle operator can in general also have time-like momenta. Consider the Euclidean correlator,
\begin{align}
    \langle \langle\mathcal{O}_1(p_{1E},\vec{p}_1)\mathcal{O}_2(p_{2E},\vec{p}_2)\mathcal{O}_3(p_{3E},\vec{p}_3)\rangle\rangle=\sum_{\mathcal{I}}\mathcal{T}_{\mathcal{I}}\mathcal{F}_{\mathcal{I}}(p_1,p_2,p_3).
\end{align}
To obtain the Wightman function with the middle operator having space-like momenta we need to do the following. First, we take a discontinuity with respect to $p_1^2$ and $p_3^2$. Then, we Wick rotate the momenta following the Wightman $i\epsilon$ prescription \eqref{Wightmaniepsilon}. The result is,
\footnotesize
\begin{align}\label{threepointdiscexample}
    &\langle\langle 0|\mathcal{O}_1(p_1)\mathcal{O}_2(p_2)\mathcal{O}_3(p_3)|0\rangle\rangle\theta(p_2^2)\notag\\&=\theta(-p_1^2)\theta(-p_3^2)\theta(-p_1^0)\theta(p_3^0)\theta(p_2^2)\sum_{\mathcal{I}}\mathcal{T}_{\mathcal{I}}\text{Disc}_{p_1^2}\text{Disc}_{p_3^2}\mathcal{F}_{\mathcal{I}}(p_1,p_2,p_3)\big|_{p_1\to -i |p_1|,p_3\to i |p_3|,p_2\to |p_2|}.
\end{align}
\normalsize
One can also first Wick rotate and then take the discontinuities, the order of these operations does not matter. Note that the tensor structures $\mathcal{T}_{\mathcal{I}}$ are at most analytic in $p_i^2$ so one does not need to take their discontinuity. The above procedure in these kinematics can be ascertained by seeing the structure of the three point scalar Wightman function in \cite{Bautista:2019qxj}, and generalizing it to arbitrary spin.

For the middle operator also having time-like momenta, the analytic continuation is more complicated. Even at the level of higher point functions, the analogous procedure increases in complexity if we have more and more time-like momenta and the general procedure is not quite known. However, the direct computation using the equation of motion discussed earlier allows us to still obtain these results in a straightforward manner. In the special (but as we shall see, important!) case when all the middle operators all have space-like momenta, we will show that the procedure is identical to \eqref{threepointdiscexample}. We simply take discontinuities with respect to the squares of all the momenta we want to take to be time-like (including the exchanged momenta such as $s^\mu=p_1^\mu+p_2^\mu$) and then Wick rotate following the $i\epsilon$ prescription \eqref{Wightmaniepsilon}. We will discuss this in much more detail in section \ref{subsec:fourpoint} as well as other cases where we have more operators in the Wightman function with time-like momenta. 

\subsubsection{The Skenderis-Van Rees prescription}
 The standard way to calculate non time-ordered correlators such as Wightman functions in QFT is using the Schwinger-Keldysh formalism. In the context of AdS/CFT, Skenderis and Van Rees developed a formalism to compute these real time quantities holographically \cite{Skenderis:2008dg,Skenderis:2008dh}. Essentially, their idea is to map every segment of the Schwinger Keldysh contour to a bulk geometry. For instance, let us consider the process required to calculate a two point Wightman function in the vacuum state. In the Schwinger Keldysh formalism, we need a contour with two segments, one going forward in time and the other backwards, with one operator insertion in each. First we prepare the vacuum state by a Euclidean path integral. At some reference time, we join it to the first Lorentzian segment and insert one operator at time $t_1$. We then move till $t=\infty$ and then fold and evolve back to $t=-\infty$ inserting the second operator at some time $t_2$ on the way. We then connect this segment to another Euclidean one to get back the vacuum state. The holographic prescription is to identify the Euclidean segments with (half of) Euclidean AdS where we prepare the vacuum state and the Lorentzian segments with the Lorentzian AdS. First, we solve the field equations in each of these segments. We then apply matching conditions wherever these geometries join and obtain a solution with sources inserted at the conformal boundary of both Lorentzian AdS spacetimes. Plugging these solutions into the on-shell action and taking one functional derivative with respect to each source results in the Wightman function. For three and four point Wightman functions, we need to use multi-fold contours which entails joining four Lorentzian AdS spacetimes with two Euclidean caps to prepare the vacuum state. 
Although we shall not take this approach in this work, it is essentially equivalent to our approach of using the equation motion. We choose the latter as it is completely algebraic and sufficient for our purposes.

Armed with all these tools and techniques, let us set the stage for the theories of interest to us in this work.

\section{Setting the Stage: Propagators in the Theories of Interest}\label{sec:propagatorsAction}
Having outlined a formalism for the computation of AdS boundary Wightman functions and their relation via analytic continuation to EAdS correlators, we turn our attention towards specific theories. These include theories with scalars, photons, gluons and gravitons with a variety of interactions between them as we shall discuss case by case in the sections to follow. Here, our aim is to calculate the Wightman, Feynman, and the propagators we use to invert the free equations of motion \eqref{newpropagator}. For readers wanting a refresher on the different $i\epsilon
$ prescriptions for the propagators, we have discussed the case of conformally coupled scalars in detail in appendix \ref{app:CCscalarsReview}.

\subsection{Scalars}\label{subsec:Scalars}
The free action for a massive scalar field in the Poincare patch of AdS$_4$ is given by,
\begin{align}\label{ActionForScalarnonminimal}
    S_{KG}=\int \frac{dz d^3 x}{z^4}\bigg(-\frac{1}{2}g^{AB}\partial_{A}\Phi \partial_{B}\Phi-\frac{m^2}{2}\Phi^2\bigg),
\end{align}
in units where the $AdS_4$ radius is set to unity. Let us perform the following Weyl transformation:
\begin{align}\label{scalarWeyltransform}
    \Phi(z,x)=z \phi(z,x).
\end{align}
The resulting action for $\phi$ is simply,
\begin{align}
    S_{KG,\text{massive}}=\int dz d^3 x~ \bigg(-\frac{1}{2}\eta^{AB}\partial_A \phi \partial_B \phi-\frac{(m^2+2)}{2z^2}\phi^2\bigg).
\end{align}
The free equation of motion is,
\begin{align}\label{scalarfreeEOM1}
    (\partial_z^2+\Box-\frac{(m^2+2)}{z^2})\phi(z,x)=0.
\end{align}
The above equation is solved by $\phi(z,x)=\frac{z^{\Delta-1}}{(z^2+x^2)^{\Delta}}$ with $m^2=\Delta(\Delta-3)$ which is the usual AdS/CFT relation. We impose Dirichlet boundary conditions for the general scalar field at the $z=0$ conformal boundary.

The Feynman propagator is a Green's function of the operator $(\partial_z^2+\Box-\frac{(m^2+2)}{z^2})$ which appears above. In particular, it satisfies,
\begin{align}\label{scalarFeynmanbulktobulk1}
    (\partial_z^2+\Box-\frac{(m^2+2)}{z^2})G_{F,\Delta}(z,z',x-x')=-i\delta(z-z)\delta^3(x-x').
\end{align}
The Wightman propagators on the other hand are homogeneous solution to \eqref{scalarFeynmanbulktobulk1}. 
\begin{align}
    (\partial_z^2+\Box-\frac{(m^2+2)}{z^2})W_{\Delta,\pm}(z,z',x-x')=0,
\end{align}
Converting these equations to momentum space, imposing Dirichlet boundary conditions, the spectral conditions for the Wightman functions and matching the normalization via Fourier transforming the position space results yields the expressions in table \ref{tab:deltagenmomspacescalarpropagators}. Apart from the bulk to bulk propagators, we have also listed the bulk to boundary propagators obtained by taking one of the bulk points to the boundary and rescaling to obtain a finite result. Further, we have presented the expressions for the propagator used to invert the equation of motion obtained via its definition viz \eqref{newpropagator}. For conformally coupled scalars which have $m^2=-2$, we provide the explicit details of computation in appendix \ref{app:CCscalarsReview}.
\begin{table}[h]
\centering
\renewcommand{\arraystretch}{2.5} 
\begin{tabular}{|>{\bfseries}l|m{0.80\linewidth}|}
\hline
Feynman BtB & $G_{F,\Delta}(z,z',p)=\frac{\sqrt{\pi}(z z')^{\frac{1}{2}}\Gamma(\Delta-\frac{1}{2})}{\Gamma(\Delta)}\bigg[\theta(-p^2)\bigg(H^{(1)}_\nu(|p|z)J_{\nu}(|p|z')\theta(z-z')+(z\leftrightarrow z')\bigg)-\frac{2i}{\pi}\theta(p^2)\bigg(K_{\nu}(|p|z)I_{\nu}(|p|z')\theta(z-z')+(z\leftrightarrow z')\bigg)\bigg]$ \\ \hline
Feynman Btb & $G_{F,\Delta}(z,p)=\frac{2^{\frac{3}{2}-\Delta}\sqrt{z}|p|^{\nu}}{\sqrt{\pi}\Gamma(\Delta)}\bigg(\theta(-p^2)\pi H^{(1)}_{\nu}(|p|z)-2i \theta(p^2)K_{\nu}(|p|z)\bigg)$ \\ \hline
Wightman BtB & $W_{\Delta,\pm}(z,z',p)=\frac{2\sqrt{\pi}(zz')^{\frac{1}{2}}\Gamma(\Delta-\frac{1}{2})}{\Gamma(\Delta)}J_{\nu}(|p|z)J_{\nu}(|p|z')\theta(-p^2)\theta(\mp p^0)$ \\ \hline
Wightman Btb & $W_{\Delta,\pm}(z,p)=\frac{2^{\frac{5}{2}-\Delta}\sqrt{\pi}\sqrt{z}|p|^{\nu}}{\Gamma(\Delta)}J_{\nu}(|p|z)\theta(-p^2)\theta(\mp p^0)$ \\ \hline
EOM Inverter BtB & $\mathcal{G}_{\Delta}(z,z',p)=\frac{i\sqrt{\pi}(z z')^{\frac{1}{2}}\Gamma(\Delta-\frac{1}{2})}{\Gamma(\Delta)}\bigg[\theta(-p^2)\bigg(Y_\nu(|p|z)J_{\nu}(|p|z')\theta(z-z')+(z\leftrightarrow z')\bigg)-\frac{2}{\pi}\theta(p^2)\bigg(K_{\nu}(|p|z)I_{\nu}(|p|z')\theta(z-z')+(z\leftrightarrow z')\bigg)\bigg]$\\ \hline

EOM Inverter Btb & $\mathcal{G}_{\Delta}(z,p)=\frac{i2^{\frac{3}{2}-\Delta}\sqrt{z}|p|^\nu}{\sqrt{\pi}\Gamma(\Delta)}\bigg(\theta(-p^2)\pi Y_{\nu}(|p|z)-2 \theta(p^2)K_{\nu}(|p|z)\bigg)$\\ \hline
\end{tabular}
\caption{Momentum space Scalar BtB (Bulk to Bulk) and Btb (Bulk to boundary) propagators for generic $\Delta=\nu+\frac{3}{2}$ scalars. We use the notation $|p|=\sqrt{|p^2|}$. $H_{\nu}^{(1)}$ is the Hankel function of the first kind, $K_\nu, J_{\nu}$ and $Y_{\nu}$ are the BesselK, BesselJ and BesselY functions. Gauge and gravity Wightman propagators \eqref{spin1Wightmanprops},\eqref{spin2Wightmanprops} can be obtained by multiplying $\Delta=2$ and $\Delta=3$ propagators by appropriate projectors respectively (with an extra $z z'$ rescaling for the latter case) and adding the longitudinal parts for the Feynman and EOM Inverter propagators \eqref{photonEOMinverterprops}, \eqref{gravitonEOMinverterprops}.}
\label{tab:deltagenmomspacescalarpropagators}
\end{table}
\subsection{Photons and Gluons}
The next two theories of interest to use are those involving Photons and gluons. The free Maxwell action is given by,
\begin{align}
    S_{EM}=-\frac{1}{4}\int \frac{dz}{z^4}~d^3 x z^4\bigg((F_{\mu\nu}F^{\mu\nu}+2F_{\mu z}F^{\mu z}\bigg).
\end{align}
where the field strength is given in terms of the gauge field $A_\mu$ as follows:
\begin{align}
    F_{\mu\nu}=\partial_\mu A_\nu-\partial_\mu A_\mu, F_{\mu z}=\partial_\mu A_z-\partial_z A_\mu.
\end{align}
The Maxwell action has the gauge redundancy,
\begin{align}\label{MaxwellGaugeSym}
   (A_z,A_\mu)\to (A_z-\partial_z \alpha,A_\mu-\partial_\mu \alpha).
\end{align}
Thanks to the fact that Maxwell theory in four dimensions enjoys conformal invariance, the action is the same as in half of flat space:
\begin{align}
    S_{EM}=-\frac{1}{2}\int dz d^3 x~\bigg((\partial_\mu A_\nu)^2-(\partial_\mu A_\nu)(\partial^\nu A^\mu)+(\partial_\mu A_z-\partial_z A_\mu)^2)\bigg).
\end{align}
To fix the gauge redundancy \eqref{MaxwellGaugeSym} we work in the gauge $A_z=0$. In the action, this corresponds to adding a gauge fixing term $\zeta A_z^2$ and then taking $\zeta\to -\infty$ thus effectively freezing the value $A_z=0$. This still leaves behind residual gauge transformations with $z-$independent gauge parameters $\alpha$. To fix this, we then set $\partial_\mu A^\mu=0$, thus fully fixing the gauge. Thus, the action in axial gauge up to boundary terms is given by,
\begin{align}
    S_{EM,axial}=-\frac{1}{2}\int dz d^3 x~\bigg((\partial_z A_\mu)^2+(\partial_\nu A_\mu)^2-\partial_\nu A_\mu \partial^\mu A^\nu\bigg).
\end{align}
The equation of motion with the constraint equation are respectively given by,
\begin{align}\label{gaugefieldfreeEOM}
    (\partial_z^2+\Box)A_\mu(z,x)=0,
\end{align}
and,
\begin{align}\label{gaugefieldconstraint}
    \partial_z(\partial_\mu A^\mu(z,x))=0.
\end{align}
First, note that the constraint equation \eqref{gaugefieldconstraint} which arises due to the equation of motion of $A_z$ is automatically satisfied in our choice $\partial_\mu A^\mu=0$.
The equation of motion \eqref{gaugefieldfreeEOM} is simply that of a free conformally coupled scalar obtained by setting $m^2=-2$ in \eqref{scalarfreeEOM1}.  The Wightman propagators and the EOM inverter propagators are thus obtained by dressing $\Delta=2$ scalar propagators obtained from table \ref{tab:deltagenmomspacescalarpropagators} by projectors that take into account the tensor structures and gauge constraint.
\begin{align}\label{spin1Wightmanprops}
    &W_{\mu\nu,\pm}(z,z',p)=\pi_{\mu\nu}(p)W_{\Delta=2}(z,z',p),W_{\mu\nu,\pm}(z,z',p)=\pi_{\mu\nu}(p)W_{\Delta=2}(z,p),
\end{align}
with the transverse projector in the given by,
\begin{align}\label{pimunu}
    \pi_{\mu\nu}(p)=\eta_{\mu\nu}-\frac{p_\mu p_\nu}{p^2}.
\end{align}
The EOM inverter propagator found using its definition \eqref{newpropagator} on the other hand is given by,
\begin{align}\label{photonEOMinverterprops}
     &\mathcal{G}_{\mu\nu}(z,z',p)=\pi_{\mu\nu}(p)\mathcal{G}_{\Delta=2}(z,z',p)-i\frac{p_\mu p_\nu}{p^2}\bigg(\theta(z-z') z'+\theta(z'-z)z\bigg),\notag\\&\mathcal{G}_{\mu\nu}(z,p)=\pi_{\mu\nu}(p)\mathcal{G}_{\Delta=2}(z,p)-i\frac{p_\mu p_\nu}{p^2}.
\end{align}
The expression for the Feynman propagator can be found in \cite{Moga:2025gdy}. Note in particular that although the Wightman propagators are identically transverse whereas the EOM inverter propagators have a longitudinal contribution\footnote{These are terms that arise due to the Feynman propagator in $\mathcal{G}$ via \eqref{newpropagator}. This fact can be traced back to the Heaviside theta functions that enforce time-ordering in a Feynman propagator \eqref{propagatordefs}.}.

Finally, for the gluon case, we just need to dress the gauge fields with colour indices which we take as the adjoint indices of $SU(N)$. This is because the free $SU(N)$ Yang-Mills theory is simply $N^2-1$ copies of the Maxwell action. Thus, the gauge fixing remains identical to the above discussion. The propagators are also the same as in Maxwell's theory with a simple additional contribution taking account of the colour indices.
\begin{align}\label{gluonprops}
    &W_{\mu\nu,\pm}^{AB}(z,z',p)=\delta^{AB}W_{\mu\nu,\pm}(z,z',p),W_{\mu\nu,\pm}^{AB}(z,p)=\delta^{AB}W_{\mu\nu,\pm}(z,p)\notag\\
    &\mathcal{G}_{\mu\nu}^{AB}(z,z',p)=\delta^{AB}\mathcal{G}_{\mu\nu}(z,z',p),\mathcal{G}_{\mu\nu}^{AB}(z,p)=\delta^{AB}\mathcal{G}_{\mu\nu}(z,p).
\end{align}
\subsection{Gravity}
Finally, we turn towards gravity. The Einstein Hilbert action is given by,
\begin{align}
    S=\frac{1}{16\pi G}\int d^4 x \sqrt{-g}(R-2\Lambda),
\end{align}
with $\kappa=8 \pi G$ defined for future use and with the metric expanded about the AdS$_4$ Poincare patch metric as follows,
\begin{align}
    g_{\mu\nu}=\frac{(\eta_{\mu\nu}+h_{\mu\nu})}{z^2},g_{\mu z}=\frac{(\eta_{\mu z}+h_{\mu z})}{z^2},g_{z z}=\frac{(\eta_{zz}+h_{zz})}{z^2}.
\end{align}
We follow exactly the conventions of \cite{Liu:1998ty}. We work in the axial gauge,
\begin{align}
    h_{\mu z}=0.
\end{align}
The equations of motion are,
\begin{align}
 \bigg(\partial_z^2(h_{\mu\nu}-\eta_{\mu\nu}h_\rho^\rho)-\frac{2}{z}\partial_z(h_{\mu\nu}-\eta_{\mu\nu}h_\rho^\rho)+\Box \tilde{h}_{\mu\nu}-\partial_\rho\partial_\mu \tilde{h}_{\nu}^{\rho}-\partial_\rho\partial_\nu\tilde{h}_{\mu}^{\rho}+\eta_{\mu\nu}\partial_\rho\partial_\alpha\tilde{h}^{\alpha\rho}\bigg)=0,
\end{align}
with $\tilde{h}_{\mu\nu}=h_{\mu\nu}-\frac{1}{2}\eta_{\mu\nu}h^\rho_\rho$. The constraint equations due to the $h_{zz}$ and $h_{z\mu}$ equations of motion are,
\begin{align}
    \partial_z(\partial_\nu h_{\mu}^{\nu}-\partial_\mu h^\rho_\rho)=0,-\partial_\mu\partial_\nu h^{\mu\nu}+\Box h^\rho_\rho-\frac{2}{z}\partial_z h^\rho_\rho=0.
\end{align}
If we use the residual gauge transformations to set $\partial_\mu h^{\mu\nu}=0$ and $h^\mu_\mu=0$, the constraint equations are automatically satisfied and the equation of motion becomes extremely simple viz,
\begin{align}
    &\partial_z^2 h_{\mu\nu}-\frac{2}{z}\partial_z h_{\mu\nu}+\Box h_{\mu\nu}=0\implies \partial_z^2 h_{\mu\nu}-\frac{2}{z}\partial_z h_{\mu\nu}-p^2 h_{\mu\nu}=0.
\end{align}
The Wightman propagators solve the equation of motion and satisfy all the gauge constraints identically and are given in terms of $\Delta=3$ scalar propagators found in table \ref{tab:deltagenmomspacescalarpropagators} as follows:
\begin{align}\label{spin2Wightmanprops}
    W_{\mu\nu\rho\sigma,\pm}(z,z',p)=\Pi_{\mu\nu\rho\sigma}(p)z  z' W_{\Delta=3,\pm}(z,z',p), W_{\mu\nu\rho\sigma}(z,p)=\Pi_{\mu\nu\rho\sigma}(p)z W_{\Delta=3,\pm}(z,p),
\end{align}
The transverse traceless projector is given by,
\begin{align}
    \Pi_{\mu\nu\rho\sigma}(p)=\pi_{\mu\nu}(p)\pi_{\rho\sigma}(p)-\pi_{\mu\rho}(p)\pi_{\nu\sigma}(p)-\pi_{\mu\sigma}(p)\pi_{\nu\rho}(p),
\end{align}
with the transverse projector itself given in \eqref{pimunu}.
The Feynman propagator on the other hand is given by \cite{Raju:2011mp}\footnote{Note that our normalization differs from \cite{Raju:2011mp} by a factor of $\frac{1}{z^2 (z')^2}$.},
\begin{align}
    G_F^{\mu\nu\rho\sigma}(z,z',p)=\int \frac{-i dk^2}{2}\frac{(z z')^{3/2} J_{3/2}(k z)J_{3/2}(k z')}{p^2+k^2-i\epsilon}\mathcal{T}^{\mu\nu\rho\sigma},
\end{align}
 with,
\begin{align}
    \mathcal{T}^{\mu\nu\rho\sigma}=&(\tilde{\pi}^{\mu\rho}\tilde{\pi}^{\nu\sigma}+\tilde{\pi}^{\mu\sigma}\tilde{\pi}^{\nu\rho}-\tilde{\pi}^{\mu\nu}\tilde{\pi}^{\rho\sigma}),\tilde{\pi}^{\mu\nu}=\eta^{\mu\nu}+\frac{p^\mu p^\nu}{k^2}.
\end{align}
The Feynman propagator contains both transverse and longitudinal contributions which are picked up by enclosing the different poles in the above integral. This fact can again be traced back to the relation that relates Feynman propagators to Wightman functions \eqref{propagatordefs}. Finally, the EOM inverter propagator can be found using its definition \eqref{newpropagator},
\begin{align}\label{gravitonEOMinverterprops}
    \mathcal{G}_{\mu\nu\rho\sigma}(z,z',p)=G_{F\mu\nu\rho\sigma}(z,z',p)-\frac{1}{2}\bigg(W_{\mu\nu\rho\sigma,+}(z,z',p)+W_{\mu\nu\rho\sigma,-}(z,z',p)\bigg).
\end{align}
The bulk to boundary propagators can be found by rescaling by $\frac{1}{(z')^3}$ and taking the limit $z'\to 0$. 
\section{Wightman functions in momentum space}\label{sec:WightmanMomentumSpace}
Having discussed the theories of interest to us, we calculate several examples of two, three and four point Wightman functions in this section to illustrate our formalism.
\subsection{Two point functions}\label{subsec:Twopoint}
Let us begin with the simplest case of two point Wightman functions. These quantities can be obtained by extrapolating the bulk point in the bulk to boundary propagators to the boundary. First for scalars, consider the Wightman plus propagator in table \ref{tab:deltagenmomspacescalarpropagators}. Taking $z\to 0$ along with rescaling results in,
\begin{align}\label{ODeltaODelta2point}
    \langle\langle 0|O_{\Delta}(p_1)O_{\Delta}(p_2)|0\rangle\rangle&=\lim_{z\to 0}\frac{1}{z^{\Delta-1}}W_{\Delta,+}(z,p_1)\notag\\&=\lim_{z\to 0}\frac{1}{z^{\Delta-1}}\frac{2^{\frac{5}{2}-\Delta}\sqrt{\pi~z}}{\Gamma(\Delta)}|p_1|^{\Delta-\frac{3}{2}}J_{\Delta-\frac{3}{2}}(|p_1|z)\theta(-p_1^2)\theta(-p_1^0)\notag\\&=\frac{4}{\Gamma(2\Delta-1)}|p_1|^{2\Delta-3}\theta(-p_1^2)\theta(-p_1^0),
\end{align}
which is indeed the correct scalar two point Wightman plus function, spectral theta functions and all. Similarly, we can do the same for the Wightman minus propagator which can also be found in table \ref{tab:deltagenmomspacescalarpropagators}. 
\begin{align}
    \langle\langle 0|O_{\Delta}(p_2)O_{\Delta}(p_1)|0\rangle\rangle&=\lim_{z\to 0}\frac{1}{z^{\Delta-1}}W_{\Delta,-}(z,p_1)\notag\\
    &=\lim_{z\to 0}\frac{1}{z^{\Delta-1}}\frac{2^{\frac{5}{2}-\Delta}\sqrt{\pi~z}}{\Gamma(\Delta)}|p_1|^{\Delta-\frac{3}{2}}J_{\Delta-\frac{3}{2}}(|p_1|z)\theta(-p_1^2)\theta(p_1^0)\notag\\&=\frac{4}{\Gamma(2\Delta-1)}|p_1|^{2\Delta-3}\theta(-p_1^2)\theta(p_1^0).
\end{align}

Similarly, for the spin-1 case we find using \eqref{spin1Wightmanprops},
\begin{align}\label{JJtwopoint}
    &\langle\langle 0|J^{\mu}(p_1)J^{\nu}(p_2)|0\rangle\rangle=\lim_{z\to 0}\frac{1}{z}\pi^{\mu\nu}(p_1)2\sin(|p_1|z)\theta(-p_1^2)\theta(-p_1^0)=2|p_1|\pi^{\mu\nu}(p_1)\theta(-p_1^2)\theta(-p_1^0),\notag\\
    &\langle\langle 0|J^{\mu}(p_2)J^{\nu}(p_1)|0\rangle\rangle=\lim_{z\to 0}\frac{1}{z}\pi^{\mu\nu}(p_1)2\sin(|p_1|z)\theta(-p_1^2)\theta(p_1^0)=2|p_1|\pi^{\mu\nu}(p_1)\theta(-p_1^2)\theta(p_1^0),
\end{align}
which are the correct results. For gluons the result is the above with an extra factor of $\delta^{AB}$, demanding orthogonality in colour space.

Finally, let us consider gravitons. We find using the bulk to boundary propagator \eqref{spin2Wightmanprops},
\begin{align}\label{TTtwopoint}
    &\langle\langle 0| T_{\mu\nu}(p_1)T_{\rho\sigma}(p_2)|0\rangle\rangle=\lim_{z\to 0}\frac{1}{z^3}\Pi_{\mu\nu\rho\sigma}(p_1)\big(\sin(|p_1|z)-|p_1|z\cos(|p_1|z)\big)\theta(-p_1^2)\theta(-p_1^0)\notag\\
    &~~~~~~~~~~~~~~~~~~~~~~~~~~~~~=\frac{\Pi_{\mu\nu\rho\sigma}(p_1)}{6}|p_1|^3\theta(-p_1^2)\theta(-p_1^0),\notag\\
    & \langle\langle 0| T_{\mu\nu}(p_2)T_{\rho\sigma}(p_1)|0\rangle\rangle=\frac{\Pi_{\mu\nu\rho\sigma}(p_1)}{6}|p_1|^3\theta(-p_1^2)\theta(p_1^0),
\end{align}
as desired. 

\subsection{Three point functions}\label{subsec:Threepoint}
Moving on to three points, we begin with three illustrative examples in detail. Generic scalars, scalar QED and Yang-Mills. We then provide the results for more examples including Wightman functions involving gravitons. To evaluate some of the integrals, the following nice integral is useful (see  6.578.2 in \cite{gradshteyn2014table}),
\small
\begin{align}\label{AppellF4Integral}
    &\int_{0}^{\infty}dz~z^{\rho-1}J_{\lambda}(a z)J_{\mu}(b z)K_{\nu}(c z)\notag\\&=\frac{2^{\rho-2}a^{\lambda}b^\mu c^{-\rho-\lambda-\mu}}{\Gamma(\lambda+1)\Gamma(\mu+1)}\Gamma(\frac{\rho+\lambda+\mu-\nu}{2})\Gamma(\frac{\rho+\lambda+\mu+\nu}{2})F_4(\frac{\rho+\lambda+\mu-\nu}{2},\frac{\rho+\lambda+\mu+\nu}{2};\lambda+1,\mu+1;-\frac{a^2}{c^2},-\frac{b^2}{c^2}),
\end{align}
\normalsize
which also requires $\rho+\lambda+\mu-\nu>0,c>0$. $F_4(\alpha,\beta;\gamma,\gamma';x,y)$ is the generalized hypergeometric Appell $F_4$ function. For $\sqrt{|x|}+\sqrt{|y|}<1$, it admits the series expansion,
\begin{align}\label{AppellF4series}
    F_4(\alpha,\beta;\gamma,\gamma';x,y)=\sum_{m,n=0}^{\infty}\frac{(\alpha)_{n+m}(\beta)_{n+m}}{(\gamma)_m(\gamma')_{n}m!n!}x^m y^n,
\end{align}
where $(\alpha)_m$ denotes the rising Pochhammer symbol. 

We perform all computations using the equation of motion as well as analytic continuation when the middle operator has spacelike momenta which also serves to verify our results.
\subsubsection{Scalars}
Let us calculate the three point function of arbitrary scalar operators dual to bulk fields with any mass. We do so in the two ways we discussed previously. First, via the bulk equation of motion and second, via analytic continuation.
\subsubsection*{Via the Equation of Motion}
The relevant action reads,
\begin{align}
    &S=\sum_{i=1}^{3}\int dz d^3 x\bigg(-\frac{1}{2}(\partial_z\phi_i)^2-\frac{1}{2}\eta^{\mu\nu}\partial_\mu\phi_i\partial_\nu\phi_i-\frac{(m_i^2+2)}{2 z^2}\phi_i^2\bigg)-\lambda \int \frac{dz}{z} d^3 x~\phi_1 \phi_2 \phi_3,
\end{align}
with $m_i^2=\Delta_i(\Delta_i-3)$. The equations of motion are given by,
\begin{align}\label{ScalarthreepointEOM}
    &(\partial_z^2+\Box-\frac{(m_1^2+2)}{z^2})\phi_1(z,x)=\frac{\lambda}{z}\phi_2(z,x)\phi_3(z,x),\notag\\
    &(\partial_z^2+\Box-\frac{(m_2^2+2)}{z^2})\phi_2(z,x)=\frac{\lambda}{z}\phi_1(z,x)\phi_3(z,x),\notag\\
    &(\partial_z^2+\Box-\frac{(m_3^2+2)}{z^2})\phi_3(z,x)=\frac{\lambda}{z}\phi_1(z,x)\phi_2(z,x).
\end{align}
We solve these equations perturbatively in the coupling $\lambda$ via the expansions,
\begin{align}
    \phi_i(z,x)=\sum_{n=0}^{\infty}\lambda^n \phi_{i}^{(n)}(z,x).
\end{align}
For the tree level three point function only the zeroth order and $\order{\lambda}$ corrections contribute. Solving the equations of motion \eqref{ScalarthreepointEOM} at this order we get,
\begin{align}\label{scalar3pteomsolfirstorder}
    &\phi_i^{(1)}(z_i,x_i)=i\int_{0}^{\infty}\frac{dz}{z}\int d^3 x \mathcal{G}(z_i,z,x_i-x)\phi_j^{(0)}(z,x)\phi^{(0)}_k(z,x)\notag\\
    &\implies \phi_i(z_i,p_i)=i\int_0^\infty \frac{dz}{z}\mathcal{G}(z_i,z,p_i)(\phi^{(0)}_j \phi^{(0)}_k)(z,p_i),
\end{align}
where $i\ne j\ne k\in\{1,2,3\}$ and $\mathcal{G}$ is the EOM inverter propagator defined in \eqref{newpropagator} with its explicit expression given in table \ref{tab:deltagenmomspacescalarpropagators}. Further, the free theory composite operators are defined by (with an implicit normal ordering),
    \begin{align}
        (\phi_{i}^{(0)}\phi_{j}^{(0)})(z,p_k)=\int \frac{d^3 l}{(2\pi)^3}~\phi_{i}^{(0)}(z,l)\phi_{j}^{(0)}(z,p_k-l).
    \end{align}
Thus, we obtain for the boundary three point function the following expression:
\small
\begin{align}
    &\langle 0|O_{\Delta_1}(p_1)O_{\Delta_2}(p_2)O_{\Delta_3}(p_3)|0\rangle_{\order{\lambda}}=\lim_{z_1,z_2,z_3\to 0}z_1^{1-\Delta_1}z_2^{1-\Delta_2}z_3^{1-\Delta_3} \langle 0|\phi_1(z_1,p_1)\phi_2(z_2,p_2)\phi_3(z_3,p_3)|0\rangle_{\order{\lambda}}\notag\\&=\lambda \lim_{z_1,z_2,z_3\to 0}z_1^{1-\Delta_1}z_2^{1-\Delta_2}z_3^{1-\Delta_3} \bigg(\langle 0|\phi_{1}^{(1)}(z_1,p_1)\phi_{2}^{(0)}(z_2,p_2)\phi_{3}^{(0)}(z_3,p_3)|0\rangle+\langle 0|\phi_{1}^{(0)}(z_1,p_1)\phi_{2}^{(1)}(z_2,p_2)\phi_{3}^{(0)}(z_3,p_3)|0\rangle\notag\\&\qquad\qquad\qquad\qquad\qquad\qquad~~~~~+\langle 0|\phi_{1}^{(0)}(z_1,p_1)\phi_{2}^{(0)}(z_2,p_2)\phi_{3}^{(1)}(z_3,p_3)|0\rangle\bigg)\notag\\&=i\lambda\lim_{z_1,z_2,z_3\to 0}z_1^{1-\Delta_1}z_2^{1-\Delta_2}z_3^{1-\Delta_3} \int_{0}^{\infty}\frac{dz}{z}\bigg(\mathcal{G}_{\Delta_1}(z,z_1,p_1)\langle 0|(\phi_{2}^{(0)}\phi_{3}^{(0)})(z,p_1)\phi_{2}^{(0)}(z_2,p_2)\phi_{3}^{(0)}(z_3,p_3)|0\rangle\notag\\&~~~~~~~~~~~~~~~~~~~~~~~~~~~~~~~~~~~~~~~~~~~~~~~~+\mathcal{G}_{\Delta_2}(z,z_2,p_2)\langle 0|\phi_{1}^{(0)}(z_1,p_1)(\phi_{1}^{(0)}\phi_{3}^{(0)})(z,p_2)\phi_{3}^{(0)}(z_3,p_3)|0\rangle\notag\\&~~~~~~~~~~~~~~~~~~~~~~~~~~~~~~~~~~~~~~~~~~~~~~~~+\mathcal{G}_{\Delta_3}(z,z_3,p_3)\langle 0|\phi_{1}^{(0)}(z_1,p_1)\phi_{2}^{(0)}(z_2,p_2)(\phi_{1}^{(0)}\phi_{2}^{(0)})(z,p_3)|0\rangle\bigg),
    \end{align}
    \normalsize
   
    Performing the Wick contractions we obtain (stripping off the momentum conserving delta function),
    \small
    \begin{align}\label{scalar3pointimpstep}
    &i\lim_{z_1,z_2,z_3\to 0}z_1^{1-\Delta_1}z_2^{1-\Delta_2}z_3^{1-\Delta_3} \int\frac{dz}{z}\bigg(\mathcal{G}_{\Delta_1}(z,z_1,p_1)W_{\Delta_2,+}(z,z_2,-p_2)W_{\Delta_3,+}(z,z_3,-p_3)\notag\\&+W_{\Delta_1,-}(z,z_1,-p_1)\mathcal{G}_{\Delta_2}(z,z_2,p_2)W_{\Delta_3,+}(z,z_3,-p_3)+W_{\Delta_1,-}(z,z_1,-p_1)W_{\Delta_2,-}(z,z_2,-p_2)\mathcal{G}_{\Delta_3}(z,z_3,p_3)\bigg)\notag\\&=i\lambda\int_{0}^{\infty}\frac{dz}{z}\bigg(\mathcal{G}_{\Delta_1}(z,p_1)W_{\Delta_2,+}(z,-p_2)W_{\Delta_3,+}(z,-p_3)+W_{\Delta_1,-}(z,-p_1)\mathcal{G}_{\Delta_2}(z,p_2)W_{\Delta_3,+}(z,-p_3)\notag\\
    &\qquad~~~~~~~~+W_{\Delta_1,-}(z,-p_1)W_{\Delta_2,-}(z,-p_2)\mathcal{G}_{\Delta_3}(z,p_3)\bigg).
\end{align}
\normalsize
Before plugging in the explicit forms of the propagators, let us note a few important points. In the first term of the above equation, we have Wightman plus propagators with momentum $-p_2^\mu$ and $-p_3^\mu$ which imply that $p_2^2<0,p_3^2<0,p_{2}^{0}>0,p_{3}^{0}>0$. By momentum conservation this implies that $p_{1}^{0}<0$ as well as $p_1^2<0$. This implies that only the time-like part of $\mathcal{G}_{\Delta_1}(z,p_1)$ with negative energy contributes to this correlator. Similarly, in the third term, we find that $p_1^2<0,p_2^2<0,p_{1}^{0}<0,p_{2}^{0}<0$ implies that $p_3^2<0,p_{3}^{0}>0$ which tells us that only the time-like part of $\mathcal{G}_{\Delta_3}(z,p_3)$ with positive energy contributes. For the second term, which involves $\mathcal{G}_{\Delta_2}(z,p_2)$, we can make no such kinematic statement and thus both its time-like (positive and negative energy components) and space-like contributions are present. Using these facts we obtain,
\begin{align}\label{genericscalarthreepoint}
    &\langle\langle 0|O_{\Delta_1}(p_1)O_{\Delta_2}(p_2)O_{\Delta_3}(p_3)|0\rangle\rangle\notag\\&=\frac{\lambda\sqrt{\pi}}{\Gamma(\Delta_1)\Gamma(\Delta_2)\Gamma(\Delta_3)}2^{\frac{13}{2}-\Delta_1-\Delta_2-\Delta_3}\theta(-p_1^2)\theta(-p_3^2)\theta(-p_{1}^{0})\theta(p_{3}^{0})\mathcal{A}_{\Delta_1,\Delta_2,\Delta_3},
\end{align}
with,
\begin{align}\label{threepointgenscalar}
&\mathcal{A}_{\Delta_1,\Delta_2,\Delta_3}=|p_1|^{\nu_1}|p_2|^{\nu_2}|p_3|^{\nu_3}\int_{0}^{\infty}dz \sqrt{z}\bigg(2~\theta(p_2^2)J_{\nu_1}(z|p_1|)K_{\nu_2}(z|p_2|)J_{\nu_3}(z |p_3|)\notag\\
&-\pi \theta(-p_2^2)\theta(-p_{2}^{0})J_{\nu_1}(z|p_1|)\bigg(J_{\nu_2}(z|p_2|)Y_{\nu_3}(z|p_3|)+Y_{\nu_2}(z|p_2|)J_{\nu_3}(z|p_3|)\bigg)\notag\\
&-\pi \theta(-p_2^2)\theta(p_{2}^{0})\bigg(J_{\nu_1}(z|p_1|)Y_{\nu_2}(z|p_2|)+Y_{\nu_1}(z|p_1|)J_{\nu_2}(z|p_2|)\bigg)J_{\nu_3}(z|p_3|),
\end{align}
and $\nu_i=\Delta_i-\frac{3}{2}$. These integrals can be evaluated in terms of Appell functions using the formulae found in GR such as \eqref{AppellF4Integral} if desired. 
\subsubsection*{Via analytic continuation}
Let us now compare our answer \eqref{threepointgenscalar} with the results of \cite{Bautista:2019qxj} where the authors obtained the CFT three point function of generic scalar operators via analytic continuation from Euclidean space. The Euclidean correlator is given by,
\begin{align}
    \langle\langle \mathcal{O}_{\Delta_1}(p_1)\mathcal{O}_{\Delta_2}(p_2)\mathcal{O}_{\Delta_3}(p_3)\rangle\rangle&=c_{123}\frac{2^{\frac{17}{2}-(\Delta_1+\Delta_2+\Delta_3)}\pi^3}{\Gamma(\frac{\Delta_1+\Delta_2-\Delta_3}{2})\Gamma(\frac{\Delta_1-\Delta_2+\Delta_3}{2})\Gamma(\frac{-\Delta_1+\Delta_2+\Delta_3}{2})\Gamma(\frac{\Delta_1+\Delta_2+\Delta_3-3}{2})}\notag\\&\times p_1^{\nu_1}p_2^{\nu_2}p_3^{\nu_3}\int dz \sqrt{z}K_{\nu_1}(zp_1)K_{\nu_2}(zp_2)K_{\nu_3}(zp_3).
\end{align}
One then translates the Wightman $i\epsilon$ prescription in position space \eqref{WightmanPosSpace1} to momentum space using the Fourier transform and carefully analytically continue the result taking care of the branch cuts in the momentum space correlator. Their result matches exactly with ours \eqref{genericscalarthreepoint} with our coefficient $g$ being determined in terms of theirs through the relation,
\begin{align}\label{gtobautistaanswer}
    g=2c_{123}\pi^{\frac{9}{2}}\frac{\Gamma(\Delta_1)\Gamma(\Delta_2)\Gamma(\Delta_3)}{\Gamma(\frac{\Delta_1+\Delta_2-\Delta_3}{2})\Gamma(\frac{\Delta_1-\Delta_2+\Delta_3}{2})\Gamma(\frac{-\Delta_1+\Delta_2+\Delta_3}{2})\Gamma(\frac{\Delta_1+\Delta_2+\Delta_3-3}{2})}.
\end{align}
This comparision serves as a rigorous check on our method as well as highlight its utility. In contrast to performing involved analytic continuations from the Euclidean CFT correlator to obtain the Wightman function, we are able to directly calculate the real-time correlator and arrive at the correct answer quite easily.

\subsubsection{Photon-Scalar-Scalar}
We move on to an example involving a photon in the scalar QED theory. Again, we perform the computation in the two distinct ways.
\subsubsection*{Via the Equation of Motion}
We consider the scalar QED action in the axial gauge $A_z=0$.
\small
\begin{align}\label{scalarqedaction1}
     S_{\text{scalar-QED}}&=-\int dz d^3 x\bigg((\partial_z\phi)(\partial_z\phi^*)+\eta^{\mu\nu}\partial_\mu\phi\partial_\nu\phi^*+\frac{(m^2+2)}{z^2}\phi^*\phi+\frac{1}{2}(\partial_z A_\mu)^2+\frac{1}{2}\partial_\mu A_\nu \partial^\mu A^\nu-\frac{1}{2}\partial_\mu A_\nu \partial^\nu A^\mu\bigg)\notag\\
    &~~+\int dz d^3 x \bigg(-i e A_\mu\big(\phi^*\partial^\mu \phi-\phi \partial^\mu \phi^*\big)+e^2 A_\mu A^\mu \phi \phi^*\bigg).
\end{align}
\normalsize
The first line of the above equation represents the quadratic portion of the action while the second line encodes the interactions. Note that there is no $z$ dependence in the interaction terms which is a consequence of massless scalar QED being classically conformally invariant. The equations of motion are,
\begin{align}\label{scalarqedEOM}
    &(\partial_z^2+\Box-\frac{(m^2+2)}{z^2})\phi=i e A^\mu \partial_\mu \phi-e^2 A_\mu A^\mu \phi,\notag\\
    &(\partial_z^2+\Box-\frac{(m^2+2)}{z^2})\phi^*=-i e A^\mu\partial_\mu \phi^*-e^2 A_\mu A^\mu \phi^*,\notag\\
    &(\partial_z^2+\Box)A_\mu=i e(\phi^*\partial_\mu \phi-\phi \partial_\mu \phi^*)-2 e^2 A_\mu \phi^* \phi.
\end{align}
We expand all the fields perturbatively in $e$ as follows:
\begin{align}
    \phi=\sum_{n=0}^{\infty}e^n\phi^{(n)},\phi^*=\sum_{n=0}^{\infty}e^n\phi^{*(n)},A_\mu=\sum_{n=0}^{\infty}e^n A_{\mu}^{(n)}.
\end{align}
We are interested in the $\order{e}$ tree level contribution to the boundary three point function,
\begin{align}
    &\langle 0|J^\mu(p_1)O_\Delta(p_2)O_\Delta^*(p_3)|0\rangle_{\order{e}}=\lim_{z_1,z_2,z_3\to 0}\frac{(z_2 z_3)^{1-\Delta}}{z_1}\Bigg(\langle 0|A^{\mu(1)}(z_1,p_1)\phi^{(0)}(z_2,p_2)\phi^{*(0)}(z_3,p_3)|0\rangle\notag\\
    &\langle 0|A^{\mu(0)}(z_1,p_1)\phi^{(1)}(z_2,p_2)\phi^{*(0)}(z_3,p_3)|0\rangle+\langle 0|A^{\mu(0)}(z_1,p_1)\phi^{(0)}(z_2,p_2)\phi^{*(1)}(z_3,p_3)|0\rangle\Bigg).
\end{align}
For concreteness, we focus on the case where $p_2^2>0$ although one can do the analysis for timelike $p_{2\mu}$ similarly. Following the steps we did for the scalar three point function, we solve the equations of motion at first order to obtain the analog of \eqref{scalar3pteomsolfirstorder} ultimately leading to \eqref{scalar3pointimpstep}. From there we see the only contributions with $p_{2\mu}$ space-like, occur due to the EOM inverter propagator $\mathcal{G}$. Any term without it will vanish since it contains only Wightman propagators which have support only for time-like momenta. Both these facts can be seen from table \ref{tab:deltagenmomspacescalarpropagators}. The same is true for this spinning three point function or any other one for that matter. Thus, only one term contributes to this correlator which is where the middle operator $\phi$ receives the $\order{e}$ correction and thus results in $\mathcal{G}_{\Delta}(z,p_2)$ in the expression. Thus we have,
\begin{align}\label{JOOexp1}
    &\langle 0|J^\mu(p_1)O_{\Delta}(p_2)O^*_{\Delta}(p_3|0\rangle\theta(p_2^2)\notag\\&=ie\theta(p_2^2)\lim_{z_1,z_2,z_3\to 0}\frac{(z_2 z_3)^{1-\Delta}}{z_1}\int_{0}^{\infty}dz\int \frac{d^3 l}{(2\pi)^3} \mathcal{G}_{\Delta}(z,z_2,p_2)\langle 0|A^{\mu}(z_1,p_1)(A^{\nu}(z,p_2-l)l_{\nu}\phi(z,l))\phi^*(z_3,p_3)|0\rangle\notag\\
    &=i e\theta(p_2^2)~\pi^{\mu\nu}(p_1)(p_1+p_2)_{\nu}\int_{0}^{\infty}dz~\mathcal{G}_{\Delta}(z,p_2)W_{\Delta=2,-}(z,-p_1)W_{\Delta,+}(z,-p_3)\notag\\
    &=\frac{e 4^{3-\Delta}}{\Gamma(\Delta)^2} \theta(-p_1^2)\theta(p_2^2)\theta(-p_3^2)\theta(-p_{1}^{0})\theta(p_{3}^{0})\mathcal{A}_{\gamma\phi_{\Delta}\phi^*_{\Delta}},
\end{align}
where,
\begin{align}\label{JOOgendelta}
    &\mathcal{A}_{\gamma\phi_{\Delta}\phi^*_{\Delta}}=\pi^{\mu\nu}(p_1)p_{2\nu}|p_2|^{\Delta-\frac{3}{2}}|p_3|^{\Delta-\frac{3}{2}}\int_{0}^{\infty}dz~z \sin(|p_1|z)K_{\Delta-\frac{3}{2}}(|p_2|z)J_{\Delta-\frac{3}{2}}(|p_3|z)\notag\\
    &=\pi^{\mu\nu}(p_1)p_{2\nu}\frac{\sqrt{\pi}\Gamma(\Delta)}{\Gamma(\Delta-\frac{1}{2})}\frac{|p_1||p_3|^{2\Delta-3}}{|p_2|^3}F_4(\frac{3}{2},\Delta;\frac{3}{2},\Delta-\frac{1}{2};-\frac{|p_1|^2}{|p_2|^2},-\frac{|p_3|^2}{|p_2|^2}).
\end{align}
For example , let us take $\Delta=1$ and $\Delta=2$. Explicitly evaluating the Appell $F_4$ for these arguments using \eqref{AppellF4series} yields,
\begin{align}
    &\mathcal{A}_{\gamma\phi_{\Delta=1}\phi^*_{\Delta=1}}=\frac{|p_1|(|p_1|^2+|p_2|^2-|p_3|^2)}{|p_2||p_3|(|p_1|^4+2|p_1|^2(|p_2|^2-|p_3|^2)+(|p_2|^2+|p_3|^2)^2)},\notag\\
    &\mathcal{A}_{\gamma\phi_{\Delta=2}\phi^*_{\Delta=2}}=\frac{2|p_1||p_2||p_3|}{|p_1|^4+2|p_1|^2(|p_2|^2-|p_3|^2)+(|p_2|^2+|p_3|^2)^2}.
\end{align}
Let us write the results in terms of $p=\sqrt{p^2}$ rather than $|p|=\sqrt{|p|^2}$. First of all, since $p_2^2>0$, $|p_2|=p_2$. However, since $p_1^2<0$ and $p_3^2<0$ we need to be a bit more careful taking into account the Wightman $i\epsilon$ prescription \eqref{Wightmaniepsilon}. Using the facts that $p_{1}^{0}<0,p_{3}^{0}>0$, we have $|p_1|=i p_1, |p_3|=-i p_3$. This results in,
\begin{align}\label{JO1O1andJO2O2}
    &\mathcal{A}_{\gamma\phi_{\Delta=1}\phi^*_{\Delta=1}}=\frac{1}{4p_2p_3}\bigg(\frac{1}{E}-\frac{1}{E-2p_1}+\frac{1}{E-2p_2}+\frac{1}{E-2p_3}\bigg),\notag\\
    &\mathcal{A}_{\gamma\phi_{\Delta=2}\phi^*_{\Delta=2}}=\frac{1}{4}\bigg(\frac{1}{E}-\frac{1}{E-2p_1}-\frac{1}{E-2p_2}-\frac{1}{E-2p_3}\bigg),
\end{align}
where $E=p_1+p_2+p_3$.
\subsubsection*{Via analytic continuation}
Let us start with the Euclidean correlator. Computing the Witten diagram for this interaction results in,
\begin{align}
    \langle\langle J^\mu(p_1)O_{\Delta}(p_2)O_{\Delta}(p_3)\rangle\rangle\propto \pi^{\mu\nu}(p_1)p_{2\nu}p_2^{\Delta-\frac{3}{2}}p_3^{\Delta-\frac{3}{2}}\int_{0}^{\infty}dz~z~ e^{-p_1z}K_{\Delta-\frac{3}{2}}(p_2 z)K_{\Delta-\frac{3}{2}}(p_3 z).
\end{align}
We want to Wick rotate this result to a Wightman function with $p_1^2<0,p_1^0<0,p_3^2<0,p_3^0>0$ and $p_2^2>0$. For such a configuration, we need to take first take discontinuities with respect to the squares of the momenta we want to be time-like.
\footnotesize
\begin{align}
    \text{Disc}_{p_1^2}\text{Disc}_{p_3}^2\langle\langle J^\mu(p_1)O_{\Delta}(p_2)O_{\Delta}(p_3)\rangle\rangle\propto \pi^{\mu\nu}(p_1)p_{2\nu}p_2^{\Delta-\frac{3}{2}}p_3^{\Delta-\frac{3}{2}}\int_{0}^{\infty}dz~z~ (-2\sinh(p_1 z))K_{\Delta-\frac{3}{2}}(p_2 z)((-1)^{\Delta-1}\pi I_{\Delta-\frac{3}{2}}(p_3 z))
\end{align}
\normalsize
Wick rotating following the Wightman $i\epsilon$ prescription \eqref{Wightmaniepsilon} results in,
\begin{align}
    2\pi \pi^{\mu\nu}(p_1)p_{2\nu}|p_2|^{\Delta-\frac{3}{2}}|p_3|^{\Delta-\frac{3}{2}}\theta(-p_1^2)\theta(-p_1^0)\theta(p_2^2)\theta(-p_3^2)\theta(p_3^0)\int_0^{\infty}z \sin(|p_1|z)K_{\Delta-\frac{3}{2}}(|p_2|z)J_{\Delta-\frac{3}{2}}(|p_3|z),
\end{align}
which is the same result as we obtained earlier via the EOM viz \eqref{JOOexp1}, \eqref{JOOgendelta}.
\subsubsection{Yang-Mills theory}
We now move on to the gluon three point function in Yang-Mills theory with the gauge group $SU(N)$. 
\subsubsection*{Via the Equation of Motion}
The Yang-Mills action in the axial gauge is given by,
\begin{align}\label{YMaction}
    S=-\frac{1}{2}\int dz d^3 x\bigg(&(\partial_z A_\mu^A)^2+(\partial_\nu A_\mu^A)^2-\partial_\nu A_{\mu}^A \partial^\mu A^{\nu A}+gf^{ABC}(\partial_\mu A_\nu^A-\partial_\nu A_\mu^{A})A^{\mu B}A^{\nu C}\notag\\&+g^2 f^{ABC}f^{ADE}A_{\mu}^B A_{\nu}^C A^{\mu D}A^{\nu E}\bigg).
\end{align}
The equation of motion for $A^{\alpha A}$ is,
\begin{align}\label{YMEOM}
    (\partial_z^2+\Box)A^{\alpha A}=-g f^{ABC}(2 A^{\nu B}\partial_\nu A^{\alpha C}+A^{\nu C}\partial^\alpha A_{\nu}^B)+g^2 f^{ABC}f^{CDE}A_{\mu}^{E}A^{\mu B}A^{\alpha D}.
\end{align}
we expand the gauge field in $g$ as follows:
\begin{align}\label{YMperturbation1}
A^{\mu A}=\sum_{n=0}^{\infty}g^n A^{\mu A(n)}.
\end{align}
Let us compute the three point function of the non-abelian conserved currents dual to this gauge field. Again, we focus on the case where the middle operator has space-like momenta. The result of the calculation after contracting with transverse polarization vectors is,
\begin{align}\label{JJJWightman}
    &\epsilon_{1\mu}\epsilon_{2\nu}\epsilon_{3\rho}\langle\langle 0|J^{\mu A}(p_1)J^{\mu_2 B}(p_2)J^{\rho C}(p_3)|0\rangle\rangle\theta(p_2^2)\notag\\&=\frac{ig}{2}\theta(p_2^2)\theta(-p_1^2)\theta(-p_3^2)\theta(-p_{1}^{0})\theta(p_{3}^{0})f^{ABC}V_{3,YM}\int_{0}^{\infty}dz~\sin(|p_1|z)e^{-|p_2|z}\sin(|p_3|z)\notag\\&=\frac{ig}{2}\theta(p_2^2)\theta(-p_1^2)\theta(-p_3^2)\theta(-p_{1}^{0})\theta(p_{3}^{0})f^{ABC}V_{3,YM}\bigg(\frac{1}{E}-\frac{1}{E-2p_1}-\frac{1}{E-2p_2}-\frac{1}{E-2p_3}\bigg)\notag\\
    &=4ig\theta(p_2^2)\theta(-p_1^2)\theta(-p_3^2)\theta(-p_{1}^{0})\theta(p_{3}^{0})f^{ABC}V_{3,YM}\frac{p_1p_2p_3}{E(E-2p_1)(E-2p_2)(E-2p_3)},
\end{align}
where the Yang-Mills three point factor is,
\begin{align}\label{YMvertex3pt}
    V_{3,YM}=\bigg((\epsilon_1\cdot \epsilon_2)(\epsilon_3\cdot p_1)+(\epsilon_2\cdot \epsilon_3)(\epsilon_1\cdot p_2)+(\epsilon_3\cdot \epsilon_1)(\epsilon_2\cdot p_3)\bigg),
\end{align}
which is also the flat space scattering amplitude for three gluons in YM theory.
\subsubsection*{Via Analytic Continuation}
Consider the Euclidean AdS Yang-Mills three point correlator. It is given by,
\begin{align}
    \epsilon_{1\mu}\epsilon_{2\nu}\epsilon_{3\rho}\langle\langle J^{A\mu}(p_1)J^{B\nu}(p_2)J^{C\rho}(p_3)\rangle\rangle=\frac{i g}{2} V_{3}^{YM}f^{ABC}\frac{1}{E}.
\end{align}
To obtain the Wightman function with $p1$ and $p3$ timelike and past and future pointing and $p2$ spacelike, we take discontinuities with respect to $p_3^2$ and $p_1^2$. This yields,
\begin{align}
    4ig V_{3}^{YM}f^{ABC}\frac{p_1p_2p_3}{E(E-2p_1)(E-2p_3)(E-2p_4)}.
\end{align}
Inserting the $\theta$ functions and taking care of the Wightman $i\epsilon$ prescription, we obtain the same answer as above \eqref{JJJWightman}.
\subsubsection{Examples involving gravitons}
To conclude our section on three point functions, we present some results for Wightman functions involving gravitons. We also focus on the case $p_2^2>0$ for simplicity.
\subsubsection*{Minimal coupling of scalars to gravitons: $\langle TO_{\Delta}O_{\Delta}\rangle$}
Consider the minimal coupling of scalar fields to gravity via their kinetic term. Calculating the resulting Wightman function with $p_2^2>0$ yields,
\small
\begin{align}
    &\langle\langle 0|T(p_1,\epsilon_1)O_{\Delta}(p_2)O_{\Delta}(p_3)|0\rangle\rangle\theta(p_2^2)=\kappa(\epsilon_1\cdot p_2)^2\theta(p_2^2)\int_{0}^{\infty}dz W_{\Delta=3,-}(z,-p_1)\mathcal{G}_{\Delta}(z,p_2)W_{\Delta,+}(z,-p_3)\notag\\
    &=\kappa\theta(-p_1^2)\theta(-p_1^0)\theta(p_2^2)\theta(-p_3^2)\theta(p_3^0)(\epsilon_1\cdot p_2)^2\frac{4^{2-\Delta}}{\Gamma(\Delta)^2}|p_2|^{\nu}|p_3|^{\nu}\int_{0}^{\infty}(\sin(z |p_1|)-z |p_1|\cos(z|p_1|))K_{\nu}(z|p_2|)J_{\nu}(|p_3|z)dz,\notag\\
    &= \kappa\theta(-p_1^2)\theta(-p_1^0)\theta(p_2^2)\theta(-p_3^2)\theta(p_3^0)(\epsilon_1\cdot p_2)^2\frac{2\sqrt{2}|p_1|^3|p_3|^{2\nu}}{|p_2|^5\Gamma(\nu+1)}F_4(\frac{5}{2},\frac{5}{2}+\nu;1+\nu,\frac{5}{2};-\frac{|p_3|^2}{|p_2|^2},-\frac{|p_1|^2}{|p_2|^2}).
\end{align}
\normalsize
where $\nu=\Delta-\frac{3}{2}$. For example, the $\Delta=1$ result is given by,
\begin{align}\label{TO1O1Wightman}
    \kappa\theta(-p_1^2)\theta(-p_1^0)\theta(p_2^2)\theta(-p_3^2)\theta(p_3^0)(\epsilon_1\cdot p_2)^2\frac{8p_1^3(p_1^4+p_2^4+6p_2^2p_3^2+p_3^4-2p_1^2(p_2^2+p_3^2))}{p_2 p_3 E^2(E-2p_1)^2(E-2p_2)^2(E-2p_3)^2}.
\end{align}
One can also easily confirm that this expression matches with the Wick rotation of the Euclidean correlator following the discontinuity procedure \eqref{threepointdiscexample}.
\subsubsection*{Minimal coupling of photons to gravity: $\langle TJJ\rangle$}
We now consider the minimal coupling of photons to gravity.
Following similar steps to the previous examples we obtain,
\begin{align}
    \langle 0|T(p_1,\epsilon_1)J(p_2,\epsilon_2)J(p_3,\epsilon_3)|0\rangle\theta(p_2^2)=\theta(-p_1^2)\theta(-p_1^0)\theta(p_2^2)\theta(-p_3^2)\theta(p_3^0)\mathcal{A}_{3,h\gamma\gamma},
\end{align}
where,
\begin{align}
    &\mathcal{A}_{3,h\gamma\gamma}=\frac{16 p_1^3 p_2 p_3}{(p_1^4+(p_2^2-p_3^2)^2-2p_1^2(p_2^2+p_3^2))^2}\bigg(-(\epsilon_1\cdot \epsilon_2)(\epsilon_1\cdot\epsilon_3)(p_1^4+(p_2^2-p_3^2)^2-2p_1^2(p_2^2+p_3^2))\notag\\&+4(p_1^2-p_2^2-p_3^2)\big((\epsilon_1\cdot p_2)^2(\epsilon_2\cdot \epsilon_3)+(\epsilon_1\cdot p_2)(\epsilon_1\cdot \epsilon_2)(\epsilon_3\cdot p_1)-(\epsilon_1\cdot p_2)(\epsilon_1\cdot \epsilon_3)(\epsilon_2\cdot p_1)\big)\bigg).
\end{align}
\subsubsection*{Einstein Gravity three point function: $\langle TTT\rangle$}
We consider the $\order{\kappa}$ contribution to the three point function where the middle operator is spacelike. Following the same steps as in the previous examples and performing some algebra results in,
\begin{align}\label{TTTWightman}
    &\langle\langle 0|T(p_1,\epsilon_1)T(p_2,\epsilon_2)T(p_3,\epsilon_3)|0\rangle\rangle\theta(p_2^2)\notag=\kappa \theta(-p_1^2)\theta(p_2^2)\theta(-p_3^2)\theta(-p_1^0)\theta(p_3^0)\mathcal{A}_{3,\text{GR}},\\
    &\mathcal{A}_{3,\text{GR}}=\sqrt{\frac{\pi}{2}}V_{3,GR}(\epsilon_1,\epsilon_2,\epsilon_3)|p_1|^{\frac{3}{2}}|p_2|^{\frac{3}{2}}|p_3|^{\frac{3}{2}}\int_{0}^{\infty}dz~z^{\frac{5}{2}}~J_{\frac{3}{2}}(z|p_1|)K_{\frac{3}{2}}(z|p_2|)J_{\frac{3}{2}}(z|p_3|)\notag\\
    &=\frac{16p_1^3 p_2^3 p_3^3}{E^2(E-2p_1)^2(E-2p_2)^2(E-2p_3)^2}\bigg((\epsilon_1\cdot p_2)(\epsilon_2\cdot \epsilon_3)+(\epsilon_2\cdot p_3)(\epsilon_3\cdot \epsilon_1)+(\epsilon_3\cdot p_1)(\epsilon_1\cdot \epsilon_2)\bigg)^2,
\end{align}
where we identified the Einstein gravity three point vertex as the square of its Yang-Mills counterpart (Double copy!). Wick rotation from the Euclidean result yields the same answer confirming the result.
\subsection{Four point functions}\label{subsec:fourpoint}
In this subsection and the next, we proceed to the more complicated case of four point Wightman functions. We calculate explicitly examples of scalar, photon, gluon and graviton Wightman functions involving a variety of contact and exchange interactions. The general expression for the Wightman functions are quite complicated since the spectral condition \eqref{spectralcondition} allows for a lot more possibilities of momenta configurations. However, we shall find that when we take the middle two operators to have space-like momenta, it leads to dramatic simplifications. For scalars, we will do the general kinematics computation to illustrate this in this subsection and then for the spinning case in the next subsection, we present only the results in these special (but as we shall see, interesting!) kinematics. 
\subsubsection{Quartic contact interactions}
We begin with the simple example of four distinct scalar fields interacting with a quartic contact interaction. As usual, we shall obtain this result both by the EOM as well as Wick rotation from Euclidean space for various kinematic regimes when possible.
\subsubsection*{Via Equation of Motion}
The action we work with is,
\begin{align}
    S=-\frac{1}{2}\sum_{i=1}^{4}\int dz d^3 x\big( (\partial_z \phi_i)^2+(\partial_\mu \phi_i)^2+\frac{(m_i^2+2)}{z^2}\phi_i^2\big)-\lambda \int dz d^3 x \phi_1\phi_2\phi_3\phi_4,
\end{align}
with $m_i^2=\Delta_i(\Delta_i-3)$. The equations of motion read,
\begin{align}
    (\partial_z^2+\Box-\frac{(m_i^2+2)}{z^2})\phi_i=\lambda \phi_j \phi_k \phi_l~,i\ne j\ne k\ne l\in\{1,2,3,4\}.
\end{align}
Let us expand every field in $\lambda$ as follows:
\begin{align}
    \phi_i(z,x)=\sum_{n=0}^{\infty}\lambda^n \phi_i^{(n)}(z,x).
\end{align}
For the four point contact Wightman function we need the $\order{\lambda}$ corrections to the fields. Inverting the above equation of motion at $\order{\lambda}$ yields,
\begin{align}
    &\phi_i^{(1)}(z_i,x_i)=i\int_0^\infty dz \int d^3 x \mathcal{G}_{\Delta_i}(z_i,z,x_i-x)\phi^{(0)}_j(z,x)\phi^{(0)}_k(z,x)\phi^{(0)}_l(z,x)\notag\\
    &\implies \phi^{(1)}_i(z_i,p_i)=i\int_0^\infty dz~\mathcal{G}_{\Delta_i}(z_i,z,p_i)\int \frac{d^3 q_1 d^3 q_2}{(2\pi)^6}\phi^{(0)}_j(z,q_1)\phi^{(0)}_k(z,q_2)\phi^{(0)}_l(z,p_i-q_1-q_2),
\end{align}
with $i\ne j\ne k\ne l\in\{1,2,3,4\}$ and $\mathcal{G}$ is the EOM inverter propagator that can be found in table \ref{tab:deltagenmomspacescalarpropagators}. We use this to compute,
\footnotesize
\begin{align}
    &\langle 0|O_{\Delta_1}(p_1)O_{\Delta_2}(p_2)O_{\Delta_3}(p_3)O_{\Delta_4}(p_4)|0\rangle_{\order{\lambda}}=\lambda \bigg(\prod_{i=1}^{4}\lim_{z_i\to 0}z_i^{1-\Delta_i}\big)\bigg(\langle \phi_1^{(1)}(z_1,p_1)\phi_2^{(0)}(z_2,p_2)\phi_3^{(0)}(z_3,p_3)\phi_4^{(0)}(z_4,p_4)|0\rangle\notag\\
    &+\langle \phi_1^{(0)}(z_1,p_1)\phi_2^{(1)}(z_2,p_2)\phi_3^{(0)}(z_3,p_3)\phi_4^{(0)}(z_4,p_4)|0\rangle+\langle \phi_1^{(0)}(z_1,p_1)\phi_2^{(0)}(z_2,p_2)\phi_3^{(1)}(z_3,p_3)\phi_4^{(0)}(z_4,p_4)|0\rangle\notag\\
    &+\langle \phi_1^{(0)}(z_1,p_1)\phi_2^{(0)}(z_2,p_2)\phi_3^{(0)}(z_3,p_3)\phi_4^{(1)}(z_4,p_4)|0\rangle\bigg)\notag\\&=\lambda\bigg(\int_{0}^{\infty}dz~\mathcal{G}_{\Delta_1}(z,p_1)W_{\Delta_2,+}(z,-p_2)W_{\Delta_3,+}(z,-p_3)W_{\Delta_4,+}(z,-p_4)\notag\\&~~~~+\int_{0}^{\infty}dz~ W_{\Delta_1,-}(z,-p_1)\mathcal{G}_{\Delta_2}(z,p_2)W_{\Delta_3,+}(z,-p_3)W_{\Delta_4,+}(z,-p_4)\notag\\
    &~~~~+\int_0^\infty dz~W_{\Delta_1,-}(z,-p_1)W_{\Delta_2,-}(z,-p_2)\mathcal{G}_{\Delta_3}(z,p_3)W_{\Delta_4,-}(z,-p_4)\notag\\&~~~~+\int_{0}^\infty dz~W_{\Delta_1,-}(z,-p_1)W_{\Delta_2,-}(z,-p_2)W_{\Delta_3,-}(z,-p_3)\mathcal{G}_{\Delta_4}(z,p_4)\bigg).
\end{align}
Using the explicit forms of the propagators from table \ref{tab:deltagenmomspacescalarpropagators} results in,
\small
\begin{align}\label{O4pointonestep}
    &\langle\langle 0|O_{\Delta_1}(p_1)O_{\Delta_2}(p_2)O_{\Delta_3}(p_3)O_{\Delta_4}(p_4)|0\rangle\rangle_{\order{\lambda}}=\frac{\pi \lambda 2^{4-\nu_1-\nu_2-\nu_3-\nu_4}}{\prod_{i=1}^{4}\Gamma(\nu_i+\frac{3}{2})}\theta(-p_1^2)\theta(-p_1^0)\theta(-p_4^2)\theta(p_4^0)\mathcal{A}_{\Delta_1,\Delta_2,\Delta_3,\Delta_4},
\end{align}
where\footnote{Comparing this result with its three point counterpart \eqref{threepointgenscalar} shows that it is a natural generalization and gives an interesting pattern as to how this result could potentially generalize to higher point contact interactions.},
\begin{align}\label{scalarcontact4point1}
&\mathcal{A}_{\Delta_1,\Delta_2,\Delta_3,\Delta_4}=|p_1|^{\nu_1}|p_2|^{\nu_2}|p_3|^{\nu_3}|p_4|^{\nu_4}\Bigg(\theta(p_2^2)\theta(-p_3^2)\theta(p_3^0)\int_{0}^{\infty} dz~z^2 J_{\nu_1}(z|p_1|)K_{\nu_2}(z|p_2|)J_{\nu_3}(z|p_3|)J_{\nu_4}(z|p_4)\notag\\
&+\theta(p_3^2)\theta(-p_2^2)\theta(-p_2^0)\int_{0}^{\infty}dz~z^2 J_{\nu_1}(z|p_1|)J_{\nu_2}(z|p_2|)K_{\nu_3}(z|p_3)J_{\nu_4}(z|p_4|)\notag\\
&-\frac{\pi}{2}\theta(-p_2^2)\theta(-p_3^2)\theta(-p_2^0)\theta(-p_3^0)\int_{0}^{\infty}dz~z^2J_{\nu_1}(z|p_1|)J_{\nu_2}(z|p_2|)\big(J_{\nu_3}(z|p_3|)Y_{\nu_4}(z|p_4|)+Y_{\nu_3}(z|p_3|)J_{\nu_4}(z|p_4|)\notag\\
&-\frac{\pi}{2}\theta(-p_2^2)\theta(-p_3^2)\theta(-p_2^0)\theta(p_3^0)\int_{0}^{\infty}dz~z^2J_{\nu_1}(z|p_1|)J_{\nu_4}(z|p_4|)\big(J_{\nu_2}(z|p_2|)Y_{\nu_3}(z|p_3|)+Y_{\nu_2}(z|p_2|)J_{\nu_3}(z|p_3|)\notag\\
&-\frac{\pi}{2}\theta(-p_2^2)\theta(-p_3^2)\theta(p_2^0)\theta(p_3^0)\int_{0}^{\infty}dz~z^2J_{\nu_3}(z|p_3|)J_{\nu_4}(z|p_4|)\big(J_{\nu_1}(z|p_1|)Y_{\nu_2}(z|p_2|)+Y_{\nu_1}(z|p_1|)J_{\nu_2}(z|p_2|)\Bigg).
\end{align}
\normalsize
Note that in \eqref{O4pointonestep}, $p_{1}^{\mu}$ and $p_4^\mu$ are time-like and have negative and positive energies respectively respecting the spectral condition \eqref{spectralcondition} whereas we see from \eqref{scalarcontact4point1} that $p_2^\mu$ and $p_3^\mu$ can be either space-like or time-like\footnote{Interestingly, there is no term with support when both $p_2^\mu$ and $p_3^\mu$ are space-like. Although this is clear mathematically, it would be interesting to see if there is a physical reason for this fact.\label{footnote:contactmiddlespacelike}}.
 Let us evaluate this expression for the simple case of $\Delta_1=\Delta_2=\Delta_3=\Delta_4=1$ with $p_2^2>0$. The result is,
\footnotesize
\begin{align}\label{phi4delta1contact}
    &\mathcal{A}_{1,1,1,1}\theta(p_2^2)=\frac{1}{|p_1||p_2||p_3||p_4|}\bigg(\frac{|p_2|}{|p_2|^2+(|p_1|+|p_3|-|p_4|)^2}+\frac{|p_2|}{|p_2|^2+(|p_1|-|p_3|+|p_4|)^2}\notag\\&+\frac{|p_2|}{|p_2|^2+(-|p_1|+|p_3|+|p_4|)}+\frac{|p_2|}{|p_2|^2+(|p_1|+|p_3|+|p_4|)^2}\bigg)\notag\\&=\frac{1}{2p_1p_2p_3p_4}\bigg(\frac{1}{E}+\frac{1}{E-2p_1}-\frac{1}{E-2p_2}+\frac{1}{E-2p_3}+\frac{1}{E-2p_4}+\frac{1}{E-2p_1-2p_3}+\frac{1}{E-2p_1-2p_4}+\frac{1}{E-2p_3-2p_4}\bigg),
\end{align}
\normalsize
where we have re-expressed the results in terms of the $p_i$ following the Wightman $i\epsilon$ prescription \eqref{Wightmaniepsilon}. We have also defined the four point total energy $E=p_1+p_2+p_3+p_4$.
For another case, let us consider the same scaling dimensions but take all the momenta to be time-like with $p_2$ and $p_3$ also having negative energy. 
\small
\begin{align}\label{phi4delta1contactalltimelike}
    \mathcal{A}_{1,1,1,1}\theta(-p_2^2)\theta(-p_3^2)\theta(-p_2^0)\theta(-p_3^0)=-\frac{i\pi}{2p_1 p_2 p_3 p_4}\bigg(\frac{1}{E-2p_2-2p_4}+\frac{1}{E-2p_1-2p_4}+\frac{1}{E-2p_4}-\frac{1}{E-2p_3}\bigg).
\end{align}
\normalsize
One can similarly obtain the other cases by evaluating the integrals in \eqref{scalarcontact4point1}.
\subsubsection*{Via Analytic Continuation}
To obtain the entire expression \eqref{scalarcontact4point1} via Wick rotation from Euclidean space would entail extending the careful analysis of \cite{Bautista:2019qxj} to four points and we do not attempt it here although it should be straightforward, but tedious to check. What we do is focus on the analytic continuation required to obtain the Wightman function in the two kinematic cases we discussed above. We also look at example when all operators have $\Delta=1$ for simplicity though one can perform a similar analysis for generic $\Delta$.

In the first case, We want $p_1,p_3$ and $p_4$ to be time-like and $p_2$ spacelike. The Euclidean correlator found using standard Witten diagramatics is,
\begin{align}
    \langle\langle O_{\Delta}(p_1)O_{\Delta}(p_2)O_{\Delta}(p_3)O_{\Delta}(p_4)\rangle\rangle&=\frac{1}{p_1p_2p_3p_4}\int_0^\infty dz~e^{-(p_1+p_2+p_3+p_4)z}
    =\frac{1}{p_1 p_2 p_3 p_4(p_1+p_2+p_3+p_4)}.
\end{align}
In this case, taking its discontinuity with respect to the three time-like momenta results in,
\begin{align}
    &\text{Disc}_{p_1^2}\text{Disc}_{p_3^2}\text{Disc}_{p_4}^2\big( \frac{1}{p_1 p_2 p_3 p_4(p_1+p_2+p_3+p_4)}\bigg)\notag\\
    &=\frac{1}{2p_1p_2p_3p_4}\bigg(\frac{1}{E}+\frac{1}{E-2p_1}-\frac{1}{E-2p_2}+\frac{1}{E-2p_3}+\frac{1}{E-2p_4}\notag\\
    &+\frac{1}{E-2p_1-2p_3}+\frac{1}{E-2p_1-2p_4}+\frac{1}{E-2p_3-2p_4}\bigg),
\end{align}
which is exactly 
\eqref{phi4delta1contact} thus confirming our result. 

Next, let us Wick rotate to the configuration where all momenta are time-like with $p_2$ and $p_3$ past pointing. This is already a more complicated Wick rotation. We need to perform a combination of discontinuities to obtain the Wightman function:
\begin{align}
    \text{Disc}_{p_1^2}\text{Disc}_{p_2^2}\bigg(\text{Disc}_{p_3^2}-\text{Disc}_{p_4^2}\bigg)\bigg(\frac{1}{p_1p_2p_3p_4(p_1+p_2+p_3+p_4)}\bigg).
\end{align}
This results in the correct answer \eqref{phi4delta1contactalltimelike}. Thus, we see that there is no obvious pattern on the required procedure to obtain Wightman functions from Euclidean correlators. This highlights the utility of the equation of motion method with the result \eqref{scalarcontact4point1} directly giving the full answer avoiding these non obvious Wick rotations. 
\subsubsection{$\phi^3$ theory exchange}
We proceed to the case of exchange interactions now. We consider a scalar field $\phi$ with scaling dimension $\Delta$ exchanging a scalar $\chi$ with scaling dimension $\Delta'$. The results are easily generalizable (albeit with a lot more algebra and more coupling constants to deal with) to the case with non-identical scalars.
\subsubsection*{Via the Equation of Motion}
The action is given by,
\begin{align}\label{scalar4ptphi3action}
    &S=-\frac{1}{2}\int dz d^3 x\big( (\partial_z \phi)^2+(\partial_\mu \phi)^2+\frac{(m^2+2)}{z^2}\phi^2\big)-\frac{1}{2}\int dz d^3 x \big((\partial_z \chi)^2+(\partial_\mu \chi)^2+\frac{(m_{\chi}^2+2)}{z^2}\chi^2\big)\notag\\&-g \int \frac{dz}{z} d^3 x~\phi^2\chi.
\end{align}
The relevant equations of motion read,
\begin{align}\label{scalar4pointphi3EOM}
    &(\partial_z^2+\Box-\frac{(m^2+2)}{z^2})\phi(z,x)=\frac{2g}{z} \phi(z,x)\chi(z,x),\notag\\
    &(\partial_z^2+\Box-\frac{(m_{\chi}^2+2)}{z^2})\chi(z,x)=\frac{g}{z}\phi^2.
\end{align}
We solve the above equations perturbatively by expanding the fields as follows:
\begin{align}
    &\phi(z,x)=\sum_{n=0}^{\infty}g^n \phi^{(n)}(z,x),\notag\\
    &\chi(z,x)=\sum_{n=0}^{\infty}g^n \chi^{{(n)}}(z,x).
\end{align}
Substituting these expansions into the EOMs, we solve it perturbatively. For the correlator of interest to us, we require the expression at $\order{g^2}$. The corrections that we shall require are,
\begin{align}\label{phi3fieldcorrection}
&\phi^{(1)}(z_1,p_1)=2i\int_{0}^{\infty}\frac{dz}{z} \int \frac{d^3 l}{(2\pi)^3}~ \mathcal{G}_{\Delta}(z,z_1,p_1)\phi^{(0)}(z,l)\chi^{(0)}(z,p_1-l),\notag\\
&\chi^{(1)}(z_5,p)=i\int_{0}^{\infty}\frac{dz}{z}\int\frac{ d^3 l}{(2\pi)^3}~\mathcal{G}_{\Delta'}(z_5,z,p)\phi^{(0)}(z,l)\phi^{(0)}(z,p-l),
\end{align}
and,
\begin{align}
    &\phi^{(2)}(z_1,p_1)=2i\int_{0}^{\infty}dz \int \frac{d^3 l}{(2\pi)^3}~ \mathcal{G}_{\Delta}(z,z_1,p_1)\phi^{(0)}(z,l)\chi^{(1)}(z,p_1-l)\notag\\
    &=-2\int_{0}^{\infty}\frac{dz}{z}\int \frac{d^3 l}{(2\pi)^3} \int_{0}^{\infty}\frac{dz'}{z'}\int \frac{d^3 l'}{(2\pi)^3}\mathcal{G}_{\Delta}(z,z_1,p_1)\mathcal{G}_{\Delta'}(z,z',p_1-l)\phi^{(0)}(z,l)\phi^{(0)}(z',l')\phi^{(0)}(z',p_1-l-l').
\end{align}
The four point function of interest is then given by a sum of ten terms:
\begin{align}
    &\langle 0|O_{\Delta}(p_1)O_{\Delta}(p_2)O_{\Delta}(p_3)O_{\Delta}(p_4)|0\rangle_{\order{g^2}}\notag\\&=g^2\big(\prod_{i=1}^{4}\lim_{z_i\to 0}z_i^{1-\Delta}\big)\bigg(\langle 0|\phi^{(0)}(z_1,p_1)\phi^{(1)}(z_2,p_2)\phi^{(1)}(z_3,p_3)\phi^{(0)}(z_4,p_4)\ket{0}\notag\\
    &+\langle 0|\phi^{(1)}(z_1,p_1)\phi^{(0)}(z_2,p_2)\phi^{(1)}(z_3,p_3)\phi^{(0)}(z_4,p_4)\ket{0}+\langle 0|\phi^{(1)}(z_1,p_1)\phi^{(0)}(z_2,p_2)\phi^{(0)}(z_3,p_3)\phi^{(1)}(z_4,p_4)\ket{0}\notag\\
    &+\langle 0|\phi^{(0)}(z_1,p_1)\phi^{(1)}(z_2,p_2)\phi^{(1)}(z_3,p_3)\phi^{(0)}(z_4,p_4)\ket{0}+\langle 0|\phi^{(0)}(z_1,p_1)\phi^{(1)}(z_2,p_2)\phi^{(0)}(z_3,p_3)\phi^{(1)}(z_4,p_4)\ket{0}\notag\\
    &+\langle 0|\phi^{(0)}(z_1,p_1)\phi^{(0)}(z_2,p_2)\phi^{(1)}(z_3,p_3)\phi^{(1)}(z_4,p_4)\ket{0}+\langle 0|\phi^{(2)}(z_1,p_1)\phi^{(0)}(z_2,p_2)\phi^{(0)}(z_3,p_3)\phi^{(0)}(z_4,p_4)\ket{0}\notag\\
    &+\langle 0|\phi^{(0)}(z_1,p_1)\phi^{(2)}(z_2,p_2)\phi^{(0)}(z_3,p_3)\phi^{(0)}(z_4,p_4)\ket{0}+\langle 0|\phi^{(0)}(z_1,p_1)\phi^{(0)}(z_2,p_2)\phi^{(2)}(z_3,p_3)\phi^{(0)}(z_4,p_4)\ket{0}\notag\\
    &+\langle 0|\phi^{(0)}(z_1,p_1)\phi^{(0)}(z_2,p_2)\phi^{(0)}(z_3,p_3)\phi^{(2)}(z_4,p_4)\ket{0}\bigg).
\end{align}
Evaluating these correlators using the field corrections results in the following long expression for the boundary correlator:
\scriptsize
\begin{align}\label{scalarphi3exchangefull}
    &-4 g^2\theta(-p_1^2)\theta(-p_1^0)\theta(-p_4^2)\theta(p_4^0)\theta(-s^2)\theta(-s^0)\Bigg(\int_{0}^{\infty}\frac{dz}{z}\int_{0}^{\infty}\frac{dz'}{z'}\notag\\&\Bigg[~~\textcolor{blue}{\mathcal{G}_{\Delta}(z,p_1)\mathcal{G}_{\Delta}(z',p_2)\bigg(W_{\Delta',+}(z,z',u)W_{\Delta,+}(z,-p_3)W_{\Delta,+}(z',-p_4)+W_{\Delta',+}(z,z',t)W_{\Delta,+}(z',-p_3)W_{\Delta, +}(z,-p_4)\bigg)}\notag\\&+\textcolor{blue}{\mathcal{G}_{\Delta}(z,p_1)\mathcal{G}_{\Delta}(z',p_3)\bigg(W_{\Delta',+}(z,z',s)W_{\Delta,+}(z,-p_2)W_{\Delta,+}(z',-p_4)+W_{\Delta',+}(z,z',t)W_{\Delta,-}(z',-p_2)W_{\Delta,+}(z,-p_4)\bigg)}\notag\\
    &+\textcolor{blue}{\mathcal{G}_{\Delta}(z,p_1)\mathcal{G}_{\Delta}(z',p_4)\bigg(W_{\Delta',+}(z,z',s)W_{\Delta,+}(z,-p_2)W_{\Delta, -}(z',-p_3)+W_{\Delta',+}(z,z',u)W_{\Delta,+}(z,-p_3)W_{\Delta,-}(z',-p_2)\bigg)}\notag\\
    &+\textcolor{blue}{\mathcal{G}_{\Delta}(z,p_2)\mathcal{G}_{\Delta}(z',p_3)\bigg(W_{\Delta',+}(z,z',s)W_{\Delta,-}(z,-p_1)W_{\Delta,+}(z',-p_4)+W_{\Delta',+}(z,z',-u)W_{\Delta,+}(z,-p_4)W_{\Delta,-}(z',-p_1)\bigg)}\notag\\
    &+\textcolor{blue}{\mathcal{G}_{\Delta}(z,p_2)\mathcal{G}_{\Delta}(z',p_4)\bigg(W_{\Delta',+}(z,z',s)W_{\Delta,-}(z,-p_1)W_{\Delta,-}(z',-p_3)+W_{\Delta'+}(z,z',-t)W_{\Delta,+}(z,-p_3)W_{\Delta,-}(z',-p_1)\bigg)}\notag\\
    &+\textcolor{blue}{\mathcal{G}_{\Delta}(z,p_3)\mathcal{G}_{\Delta}(z',p_4)\bigg(W_{\Delta',+}(z,z',u)W_{\Delta,-}(z,-p_1)W_{\Delta,-}(z',-p_2)+W_{\Delta',+}(z,z',-t)W_{\Delta,-}(z,-p_2)W_{\Delta,-}(z',-p_1)\bigg)}\notag\\
    &+\textcolor{violet}{\mathcal{G}_{\Delta}(z,p_1)\bigg(\mathcal{G}_{\Delta'}(z,z',s)W_{\Delta,+}(z,-p_2)W_{\Delta,+}(z',-p_3)W_{\Delta, +}(z',-p_4)+\mathcal{G}_{\Delta'}(z,z',u)W_{\Delta,+}(z,-p_3)W_{\Delta,-}(z',-p_2)W_{\Delta,-}(z',-p_4)}\notag\\
    &+\textcolor{violet}{\mathcal{G}_{\Delta'}(z,z',t))W_{\Delta,+}(z,-p_4)W_{\Delta,+}(z',-p_2)W_{\Delta,+}(z',-p_3)\bigg)+\mathcal{G}_{\Delta}(z,p_2)\bigg(\mathcal{G}_{\Delta'}(z,z',s)W_{\Delta,-}(z,-p_1)W_{\Delta,+}(z',-p_3)W_{\Delta,+}(z',-p_4)}\notag\\
    &+\textcolor{violet}{\mathcal{G}_{\Delta'}(z,z',-t)W_{\Delta,+}(z,-p_3)W_{\Delta,-}(z',-p_1)W_{\Delta,+}(z',-p_4)+\mathcal{G}_{\Delta'}(z,z',-u)W_{\Delta,+}(z,-p_4)W_{\Delta,-}(z',-p_1)W_{\Delta,+}(z',-p_3)\bigg)}\notag\\
    &+\textcolor{violet}{\mathcal{G}_{\Delta}(z,p_3)\bigg(G_{\Delta'}(z,z',u)W_{\Delta,-}(z,-p_1)W_{\Delta,-}(z',-p_2)W_{\Delta,+}(z',-p_4)+\mathcal{G}_{\Delta'}(z,z',-t)W_{\Delta,-}(z,-p_2)W_{\Delta,-}(z',-p_1)W_{\Delta,+}(z',-p_4)}\notag\\
    &+\textcolor{violet}{\mathcal{G}_{\Delta'}(z,z',-s)W_{\Delta,+}(z,-p_4)W_{\Delta,-}(z',-p_1)W_{\Delta,-}(z',-p_2)\bigg)+\mathcal{G}_{\Delta}(z,p_4)\bigg(\mathcal{G}_{\Delta'}(z,z',t)W_{\Delta,-}(z,-p_1)W_{\Delta,-}(z',-p_2)W_{\Delta,-}(z',-p_3)}\notag\\
    &+\textcolor{violet}{\mathcal{G}_{\Delta'}(z,z',-u)W_{\Delta, -}(z,-p_2)W_{\Delta,-}(z',-p_1)W_{\Delta,-}(z',-p_3)+\mathcal{G}_{\Delta'}(z,z',-s)W_{\Delta,-}(z,-p_3)W_{\Delta,-}(z',-p_1)W_{\Delta,-}(z',-p_2)\bigg)}\Bigg]\Bigg),
\end{align}
\normalsize
where $s^\mu=p_1^\mu+p_2^\mu, t^\mu=p_1^\mu+p_4^\mu$ and $u^\mu=p_1^\mu+p_3^\mu$. Their magnitudes are denoted $|s|,|t|$ and $|u|$. The first six lines of the above expression (blue colour) represent the contributions due to the first order corrected field occurring at two locations in the correlator. The remaining six lines (violet colour) of the expression are the contribution due to the second order correction of a single field in the correlator. 

\subsection*{Via analytic continuation}
One can try to ask whether \eqref{scalarphi3exchangefull} can be obtained via analytic continuation from Euclidean space. However, based on the behemoth expression \eqref{scalarphi3exchangefull}, an analytic continuation to obtain it from the corresponding Euclidean correlator is clearly going to be extremely difficult. Generalizing the methods of \cite{Bautista:2019qxj} to four point exchange diagrams is certainly possible but would be quite challenging in practice.  This shows the importance of intrinsically computing Wightman functions rather than resorting to analytic continuation from Euclidean space.

\subsection{Four point functions in special kinematics}\label{subsec:fourpointspecial}
The general expression for the scalar exchange correlator \eqref{scalarphi3exchangefull} is quite complicated.  However, for particular kinematics where we take the middle operators to have space-like momenta viz $p_2^2>0,p_3^2>0$, the expression dramatically simplifies which causes all but one of the above terms in \eqref{scalarphi3exchangefull} to vanish This 
is what we shall refer to as special kinematics\footnote{The reason for all but one term in \eqref{scalarphi3exchangefull} to drop out in these kinematics is that the only terms that contribute with $p_2^2>0,p_3^2>0$ are when the $\mathcal{G}$ propagator occurs with momentum $p_2$ and $p_3$ as we see from table \ref{tab:deltagenmomspacescalarpropagators}. When they occur with a Wightman propagator, they do not contribute since it has support only for time-like momenta as we one can see from table \ref{tab:deltagenmomspacescalarpropagators} yet again.\label{footnote:specialkinematics}}. One can check that this statement is not just true for the scalar correlator but also for ones involving particles with spin. These kinematics make four point Wightman functions extremely tractable and simplify them greatly. Further, we shall find that performing analytic continuation from the Euclidean correlator to the Wightman functions with these kinematics is a simple extension of the procedure at three points \eqref{threepointdiscexample}. We simply take discontinuities with respect to the momenta we want time-like.

We now proceed to discuss a variety of examples including the previously discussed scalar ones in these kinematics.
\subsubsection{Scalar contact and exchange}
First of all, as we see from \eqref{scalarcontact4point1}, when we work in the special kinematics with $p_2^2>0,p_3^2>0$, the contact contribution vanishes as we also discussed in footnote \ref{footnote:contactmiddlespacelike}.
\begin{align}
    \langle\langle 0|O_{\Delta_1}(p_1)O_{\Delta_2}(p_2)O_{\Delta_3}(p_3)O_{\Delta_4}(p_4)|0\rangle\rangle_{\order{\lambda}}\theta(p_2^2)\theta(p_3^2)=0.
\end{align}

Let us move on to the exchange contribution example \eqref{scalarphi3exchangefull}. We obtain in the special kinematics,
\footnotesize
\begin{align}
    &\langle\langle 0|O_{\Delta}(p_1)O_{\Delta}(p_2)O_{\Delta}(p_3)O_{\Delta}(p_4)|0\rangle\rangle_{\order{g^2}}\theta(p_2^2)\theta(p_3^2)\notag\\&=\theta(p_2^2)\theta(p_3)^2\big(\prod_{i=1}^{4}\lim_{z_i\to 0}z_i^{1-\Delta_i}\big)\langle\langle 0|\phi^{(0)}(z_1,p_1)\phi^{(1)}(z_2,p_2)\phi^{(1)}(z_3,p_3)\phi^{(0)}(z_4,p_4)=\theta(-p_1^0)\theta(-p_4^2)\theta(p_4^0)\theta(p_2^2)\theta(p_3^2)\mathcal{A}_{\Delta,\Delta,\Delta,\Delta},\notag\\
    &\mathcal{A}_{\Delta,\Delta,\Delta,\Delta}=-4g^2\Bigg[\int_0^\infty\frac{dz}{z}\int_0^\infty \frac{dz'}{z'}\mathcal{G}_{\Delta}(z,p_2)\mathcal{G}_{\Delta}(z',p_3)\bigg(W_{\Delta',+}(z,z',s)W_{\Delta,-}(z,-p_1)W_{\Delta,+}(z',-p_4)\notag\\&~~~~~~~~~~~~~~~~~~~~~~~~~~~~~~~~~~~~~~~~~~~~~~+W_{\Delta',+}(z,z',-u)W_{\Delta,+}(z,-p_4)W_{\Delta,-}(z',-p_1)\bigg)\Bigg]
\end{align}
\normalsize
Although there are two contributions in the above equation, only the first one is non-zero due to kinematic constraints. For the second term to be non-zero, we require the conditions,
\begin{align}
p_1^2<0,p_3^2>0,p_{1}^{0}<0~\text{with}~u^0>0,-u^2<0.
\end{align} which come from the Wightman propagator with momentum $-u^\mu=-p_1^{\mu}-p_3^{\mu}$. This is kinematically inconsistent as the sum of a backward pointing time-like vector and a space-like vector cannot result in a forward pointing time-like vector. Thus our result is only one term viz,
\begin{align}
    \mathcal{A}_{\Delta,\Delta,\Delta,\Delta}=-4g^2\int_0^\infty\frac{dz}{z}\int_0^\infty \frac{dz'}{z'}\mathcal{G}_{\Delta}(z,p_2)\mathcal{G}_{\Delta}(z',p_3)W_{\Delta',+}(z,z',s)W_{\Delta,-}(z,-p_1)W_{\Delta,+}(z',-p_4).
\end{align}
Further, using the expressions provided in table \ref{tab:deltagenmomspacescalarpropagators} we see that the bulk to bulk Wightman propagator factorizes into a product of bulk to boundary propagators resulting in,
\begin{align}
    W_{\Delta,+}(z,z',p)=\frac{4^{\Delta-2}\Gamma(\Delta)\Gamma(\Delta-\frac{1}{2})}{\sqrt{\pi}}W_{\Delta',+}(z,p)\frac{1}{|p|^{2\Delta-3}}W_{\Delta',+}(z',p).
\end{align}
Thus, the above integral factorizes\footnote{We used the property $W_{\Delta,+}(z,p)=W_{\Delta,-}(z,-p)$ as can be ascertained from table \ref{tab:deltagenmomspacescalarpropagators}.}:
\footnotesize
\begin{align}\label{phi3fourpointgendelta}
    \mathcal{A}_{\Delta\Delta\Delta\Delta}&=c(\Delta')\int_{0}^{\infty}\frac{dz}{z}W_{\Delta,-}(z,-p_1)\mathcal{G}_{\Delta}(z,p_2)W_{\Delta',+}(z,s)\frac{1}{|s|^{2\Delta'-3}}\int_0^\infty\frac{dz'}{z'}W_{\Delta',-}(z,-s)\mathcal{G}_{\Delta}(z',p_3)W_{\Delta,+}(z',-p_4),
\end{align}
\normalsize
with $c(\Delta')=-\frac{4^{\Delta'-1}\Gamma(\Delta')\Gamma(\Delta'-\frac{1}{2})}{\sqrt{\pi}}g^2$. This is a great simplification as there are no nested bulk integrals in contrast to traditional Witten diagrams involving the Feynman bulk to bulk propagator. This is a general feature of these kinematics and holds for spinning correlators too. Evaluating the integrals in terms of Appell $F_4$ functions using \eqref{AppellF4Integral} and replacing $g$ with the OPE coefficient \eqref{gtobautistaanswer}, we get,
\footnotesize
\begin{align}\label{GenDeltafourpointWightman}
    \mathcal{A}_{\Delta\Delta\Delta\Delta}&=\frac{c_{123}^2 4^{7-2\Delta}\pi^{\frac{19}{2}}\Gamma(\Delta')}{\Gamma(\Delta-\frac{1}{2})^2\Gamma(\Delta-\frac{\Delta'}{2})^2\Gamma(\Delta'-\frac{1}{2})\Gamma(\frac{\Delta'}{2})^2}\theta(-s^2)\theta(-s^0)|p_1|^{2\Delta-3}|p_2|^{-\Delta'}|s|^{2\Delta'-3}|p_3|^{-\Delta'}|p_4|^{2\Delta-3}\notag\\
    &\times F_4(\frac{\Delta'}{2},\Delta+\frac{\Delta'}{2}-\frac{3}{2};\Delta-\frac{1}{2},\Delta'-\frac{1}{2};-\frac{|p_1|^2}{|p_2|^2},-\frac{|s|^2}{|p_2|^2})\frac{1}{|s|^{2\Delta'-3}}F_4(\frac{\Delta'}{2},\Delta+\frac{\Delta'}{2}-\frac{3}{2};\Delta-\frac{1}{2},\Delta'-\frac{1}{2};-\frac{|p_4|^2}{|p_3|^2},-\frac{|s|^2}{|p_3|^2}).
\end{align}
\normalsize
For example, consider all the scaling dimensions to be $1$. In this case the result is,
\begin{align}\label{Delta1fourpointWightman}
    \mathcal{A}_{1111}=1024 \pi^6 c_{123}^2  \frac{1}{|p_1||p_2||p_3||p_4|}\frac{1}{|s|}\theta(-s^2)\theta(-s^0).
\end{align}
One important point to note is that the general result \eqref{GenDeltafourpointWightman} is perfectly finite for any $\Delta,\Delta'$ that satisfy the unitarity bound. This is in contrast to the Euclidean case where one needs to perform renormalization in many cases \cite{Bzowski:2015pba}\footnote{These occur when $\frac{d}{2}\pm(\Delta_1-\frac{d}{2})\pm(\Delta_2-\frac{d}{2})\pm(\Delta_3-\frac{d}{2})=-2k,k\in\mathbb{Z}_{\ge 0}$.} due to the presence of UV divergences for the CFT correlator which are not present for Wightman functions. This can even be seen at the level of two points, check out appendix $K$ of \cite{Bala:2025gmz} for instance.
\subsubsection*{Via Analytic Continuation}
Let us discuss how to obtain this correlator via analytic continuation. Clearly, as we discussed earlier, there is no obvious pattern to obtain the general result \eqref{scalarphi3exchangefull} except of course by guessing the pattern via trial and error and taking a sum of products of discontinuities with respect to the squares of the momenta taking a leaf out of the three point case \eqref{threepointdiscexample}! For the special kinematics, however, the procedure is simple as we discussed at the beginning of this section. Let us see this now. Start with the Euclidean correlator,
\footnotesize
\begin{align}
    &\langle\langle O_{\Delta}(p_1)O_{\Delta}(p_2)O_{\Delta}(p_3)O_{\Delta}(p_4)\rangle\rangle\notag\\&=(p_1p_2p_3p_4)^{\nu}\int_0^\infty dz~ \sqrt{z}\int_0^\infty dz' \sqrt{z'}~K_{\nu}(p_1 z)K_{\nu}(p_2 z)\bigg(K_{\nu}(s z)I_{\nu}(s z')\theta(z-z')+(z\leftrightarrow z')\bigg)K_{\nu}(p_3 z)K_{\nu}(p_4 z)\notag\\
    &+\text{t-channel}+\text{u-channel}
\end{align}
\normalsize
We want to reach the Wightman function in the kinematics where the middle operators have spacelike momenta as well as the exchanged momentum $s$ being time-like. For this, we take a discontinuity with respect to $p_1^2,p_4^2$ and $s^2$.
\begin{align}
    \text{Disc}_{p_1^2}\text{Disc}_{p_4^2}\text{Disc}_{s^2}\langle\langle O_{\Delta}(p_1)O_{\Delta}(p_2)O_{\Delta}(p_3)O_{\Delta}(p_4)\rangle\rangle.
\end{align}
We then Wick rotate following \eqref{Wightmaniepsilon}. The t and u channels drop out in this process (or are not kinematically consistent) yielding the correct result \eqref{phi3fourpointgendelta}.

\subsubsection{Scalar-Photon Bhabha Scattering}
Let us move on to an example of scalars exchanging a photon. The action we work with is the scalar QED action \eqref{scalarqedaction1} with the equations of motion given in \eqref{scalarqedEOM}. We require the $\order{e^2}$ correction for this correlator. 
\subsubsection*{Via the Equation of Motion}
We obtain,
\begin{align}
    &\langle\langle 0|O_{\Delta}(p_1)O_{\Delta}^*(p_2)O_{\Delta}(p_3)O_{\Delta}^*(p_4)|0\rangle\rangle_{\order{e^2}}\theta(p_2^2)\theta(p_3^2)\notag\\&=-e^2\prod_{i=1}^{4}(\lim_{z_i\to 0}z_i^{1-\Delta_i})\theta(p_2^2)\theta(p_3^2)\langle\langle 0|\phi^{(0)}(z_1,p_1)\phi^{*(1)}(z_2,p_2)\phi^{(1)}(z_3,p_3)\phi^{*(0)}(z_4,p_4)|0\rangle\rangle\notag\\
    &=-e^2\theta(p_2^2)\theta(p_3^2)\int_{0}^{\infty}dz\int_{0}^{\infty}dz'\mathcal{G}_{\Delta}(z,p_2)\mathcal{G}_{\Delta}(z,p_3)\notag\\&\times\Bigg(\int \frac{d^3 l}{(2\pi)^3}\int\frac{d^3 l'}{(2\pi)^3}l_\mu l'_{\nu}~W_{\Delta,-}(z,-p_1)\pi^{\mu \nu}(s)W_{\Delta=2,+}(z,z',s)W_{\Delta,+}(z',-p_4)\delta^3(l+p_1)\delta^3(l'+p_4)\Bigg)\notag\\&=-e^2\theta(p_2^2)\theta(p_3^2)\theta(-p_1^2)\theta(-p_4^2)\theta(-p_1^0)\theta(p_4^0)\mathcal{A}_{\Delta\Delta\Delta\Delta}^{\text{Bhabha}},
\end{align}
with,
\footnotesize
\begin{align}\label{scalarbhabha4ptgen}
    &\mathcal{A}_{\Delta\Delta\Delta\Delta}^{\text{Bhabha}}=\pi^{\mu\nu}(s)p_{1\mu}p_{4\nu}\theta(-s^0)\theta(-s^2)\int_{0}^{\infty}dz \int_{0}^{\infty}dz'W_{\Delta,-}(z,-p_1)\mathcal{G}_{\Delta}(z,p_2)W_{\Delta=2,+}(z,z',s)\mathcal{G}(z',p_3)W_{\Delta,+}(z',-p_4)\notag\\
    &=\frac{\pi^{\mu\nu}(s)p_{1\mu}p_{4\nu}\theta(-s^0)\theta(-s^2)}{2|s|}\int_0^{\infty}dzW_{\Delta,-}(z,-p_1)\mathcal{G}_{\Delta}(z,p_2)W_{\Delta=2,+}(z,s)\int_{0}^{\infty}dz' W_{\Delta=2,-}(z',-s)\mathcal{G}_{\Delta}(z',p_3)W_{\Delta,+}(z',-p_4).
\end{align}
\normalsize
Therefore the result is simply an $s-$ channel contribution. Evaluating the integrals result in the extremely simple answer viz,
\begin{align}
    \mathcal{A}_{\Delta\Delta\Delta\Delta}^{\text{Bhabha}}&=\frac{\pi^{\mu \nu}(s)p_{1\mu}p_{4\nu}\theta(-s^0)\theta(-s^2)}{|s|}\bigg(-\frac{2^{5-4\nu}\pi}{\Gamma(1+\nu)^2\Gamma(\frac{3}{2}+\nu)^2}\bigg)\frac{|p_1|^{2\nu}|p_4|^{2\nu}|s|^2}{|p_2|^3|p_3|^3}\notag\\
    &\times F_4(\frac{3}{2},\frac{3}{2}+\nu;1+\nu,\frac{3}{2};-\frac{|p_1|^2}{|p_2|^2},-\frac{|s|^2}{|p_2|^2})F_4(\frac{3}{2},\frac{3}{2}+\nu;1+\nu,\frac{3}{2};-\frac{|p_4|^2}{|p_3|^2},-\frac{|s|^2}{|p_3|^2}),
\end{align}
where $\nu=\Delta-\frac{3}{2}$.
\subsubsection*{Via Analytic Continuation}
Let us start with the Euclidean correlator. Performing the standard Witten diagram computation results in,
\begin{align}
    &\langle\langle O_\Delta(p_1)O_{\Delta}^*(p_2)O_{\Delta}(p_3)O_{\Delta}^*(p_4)\rangle\rangle=-e^2 \pi^{\mu\nu}(s)p_{1\mu}p_{4\nu}\notag\\&\times\int_0^{\infty}dz\int_0^\infty dz'G_{F,\Delta}(z,p_1)G_{F,\Delta}(z,p_2)G_{F,\Delta=2}(z,z',s)G_{F,\Delta}(z',p_3)G_{F,\Delta}(z',p_4)+\text{t+u channels},
\end{align}
where since these are Euclidean momenta, only the spacelike contributions to the Feynman propagators in table \ref{tab:deltagenmomspacescalarpropagators} contribute to this expression. Rather than evaluating this expression and then performing the analytic continuation, we choose to do it at the integrand level. Simply taking a discontinuity with respect to three momenta we want time-like viz $p_1^2,p_4^2$ and $s^2$ shows that the associated Euclidean propagators turn into the corresponding Wightman propagator in table \ref{tab:deltagenmomspacescalarpropagators}, which yields the result \eqref{scalarbhabha4ptgen}.

We now proceed to discuss three more examples: Compton scattering, Yang-Mills gluon four point function and Einstein gravity four point function. The details of the calculations are identical to the ones discussed albeit with more tensor structures involved. Thus, we skip the steps and present the final results when the middle two operators have spacelike momenta.

\subsubsection{Scalar-Photon Compton scattering}
Next, we consider the classic Compton scattering process involving two photons and two scalars. The relevant action and equations of motion are those in scalar QED which are respectively \eqref{scalarqedaction1} and \eqref{scalarqedEOM}. Solving the equation of motion at $\order{e^2}$ and computing the Wightman function with $p_2^2>0,p_3^2>0$ results in,
\footnotesize
\begin{align}
    &\epsilon_1^\mu \epsilon_4^\nu \langle\langle 0|J_\mu(p_1)O_{\Delta}(p_2)O^*_{\Delta}(p_3)J_\nu(p_4)|0\rangle\rangle\theta(p_2^2)\theta(p_3^2)=\epsilon_1^\mu \epsilon_4^\nu\lim_{z_i\to 0}\frac{z_2^{1-\Delta}z_3^{1-\Delta}}{z_1 z_4}\langle\langle 0|A_\mu^{(0)}(z_1,p_1)\phi^{(1)}(z_2,p_2)\phi^{*(1)}(z_3,p_3)A_\nu^{(0)}(z_4,p_4)|0\rangle\rangle\notag\\&=e^2\theta(p_2^2)\theta(p_3^2)(\epsilon_1\cdot p_2)(\epsilon_4\cdot p_3)\int_{0}^{\infty}dz\int_{0}^\infty dz'~W_{\Delta=2,-}(z,-p_1)\mathcal{G}_{\Delta}(z,p_2)W_{\Delta,+}(z,z',s)\mathcal{G}_{\Delta}(z',p_3)W_{\Delta=2,+}(z',-p_4)\notag\\
    &=e^2\theta(-p_1^2)\theta(p_2^2)\theta(p_3^2)\theta(-p_4^2)\theta(-p_1^0)\theta(p_4^0)(\epsilon_1\cdot p_4)(\epsilon_4\cdot p_1)\mathcal{A}_{\gamma\phi_{\Delta}\phi_{\Delta}^*\gamma}.
\end{align}
\normalsize
The function $\mathcal{A}_{\gamma\phi_{\Delta}\phi_{\Delta}^*\gamma}$ can be evaluated in terms of a product of Appell $F_4$ function resulting in,
\begin{align}
    \mathcal{A}_{\gamma\phi_{\Delta}\phi_{\Delta}^*\gamma}&=\frac{-4^{4-\Delta}\sqrt{\pi}|p_1||p_4|}{|p_2|^3 |p_3|^3\Gamma(\Delta)\Gamma(\Delta-\frac{1}{2})}F_4(\frac{3}{2},\Delta;\frac{3}{2},\Delta-\frac{1}{2};-\frac{|p_1|^2}{|p_2|^2},-\frac{|s|^2}{|p_2|^2})\notag\\&\times\frac{1}{|s|^{3-2\Delta}}F_4(\frac{3}{2},\Delta;\frac{3}{2},\Delta-\frac{1}{2};-\frac{|p_4|^2}{|p_3|^2},-\frac{|s|^2}{|p_3|^2}),
\end{align}
For example, consider conformally coupled scalars that have $\Delta=2$. In that case, the result is simply,
\footnotesize
\begin{align}
&\mathcal{A}_{\gamma\phi_{\Delta=2}\phi^*_{\Delta=2}\gamma}\notag\\&=32i\frac{p_1 p_2 sp_3 p_4}{(p_1+p_2+s)(p_1+p_2-s)(p_1-p_2+s)(-p_1+p_2+s)(p_3+p_4+s)(p_3+p_4-s)(p_3-p_4+s)(-p_3+p_4+s)}.
\end{align}
\normalsize

\subsubsection{Yang-Mills theory}
We now proceed to an example involving gluons. The Yang-Mills action was given in \eqref{YMaction} with the equation of motion provided in \eqref{YMEOM}. Following the usual method, we find the four point function,
\begin{align}
    &\langle\langle 0|J^{A_1}(p_1,\epsilon_1)J^{A_2}(p_2,\epsilon_2)J^{A_3}(p_3,\epsilon_3)J^{A_4}(p_4,\epsilon_4)|0\rangle\rangle\theta(p_2^2)\theta(p_3^2)\notag\\
    &=\big(\prod_{i=1}^{4}z_i^{-1}\big)\langle\langle 0|A^{A_1(0)}(z_1,p_1,\epsilon_1)A^{A_2(1)}(z_2,p_2,\epsilon_2)A^{A_3(1)}(z_3,p_3,\epsilon_3)A^{A_4(0)}(z_4, p_4,\epsilon_4)|0\rangle\rangle\theta(p_2^2)\theta(p_3^2)\notag\\&=\theta(-p_1^2)\theta(p_2^2)\theta(p_3^2)\theta(-p_4^2)\theta(-p_1^0)\theta(p_4^0)\theta(-s^2)\theta(-s^0)\mathcal{A}_{4,YM},
\end{align}
with,
\footnotesize
\begin{align}\label{YMfourpoint}
    &\mathcal{A}_{4,YM}=\frac{g^2}{|s|} f^{A_1 A_2 A}f^{A_3 A_4}_AV_{3,YM}^\mu(\epsilon_1,\epsilon_2)\pi_{\mu\nu}(s)V_{3,YM}^{\nu}(\epsilon_3,\epsilon_4)\int_{0}^{\infty}dz\sin(|p_1|z)e^{-|p_2|z}\sin(|s|z)\int_{0}^{\infty}dz' \sin(|s|z')e^{-|p_3|z}\sin(|p_4|z')\notag\\
    &=\frac{4g^2 f^{A_1 A_2 A}f_A^{A_3 A_4}V_{3,YM}^{\mu}(\epsilon_1,\epsilon_2)\pi_{\mu\nu}(s)V_{3,YM}^{\nu}(\epsilon_3,\epsilon_4)|p_1||p_2||s||p_3||p_4|}{((|p_1|^2+|p_2|^2)^2+|s|^4+2|s|^2(|p_2|^2-|p_1|^2))((|p_3|^2+|p_4|^2)^2+|s|^4+2|s|^2(|p_4|^2-|p_3|^2))}
\end{align}
\normalsize
Note that the Yang-Mills contact diagram drops out in these kinematics for the same reasons discussed in footnote \ref{footnote:contactmiddlespacelike}.
\subsubsection{Einstein Gravity}
The computation in Einstein gravity for the graviton four point function yields the result,
\begin{align}
    &\langle\langle 0|T(p_1,\epsilon_1)T(p_2,\epsilon_2)T(p_3,\epsilon_3)T(p_4,\epsilon_4)|0\rangle\rangle\theta(p_2^2)\theta(p_3^2)\notag\\
    &=\theta(-p_1^2)\theta(p_2^2)\theta(p_3^2)\theta(-p_4^2)\theta(-p_1^0)\theta(p_4^0)\theta(-s^2)\theta(-s^0)\mathcal{A}_{4,GR},
\end{align}
where,
\begin{align}\label{A4GR}
    \mathcal{A}_{4,GR}=\frac{4 \kappa^2 V_{3,GR}^{\mu\nu}(\epsilon_1,\epsilon_2)\Pi_{\mu\nu\rho\sigma}(s)V_{3,GR}^{\rho\sigma}(\epsilon_3,\epsilon_4)|p_1|^3|p_2|^3|s|^3|p_3|^3|p_4|^3}{\bigg(((|p_1|^2+|p_2|^2)^2+|s|^4+2|s|^2(|p_2|^2-|p_1|^2))((|p_3|^2+|p_4|^2)^2+|s|^4+2|s|^2(|p_4|^2-|p_3|^2))\bigg)^2}
\end{align}
One can easily check that the discontinuity with respect to $p_1^2,p_4^2,s^2$ of the Euclidean AdS graviton correlator results in this expression. Note that yet again, the quartic contact does not contribute in these kinematics. Comparing \eqref{A4GR} to its Yang-Mills counterpart \eqref{YMfourpoint} suggests a double copy relation. We will have much more to say on this in section \ref{sec:DoubleCopy}.
\subsection{Factorization and Wightman Conformal Partial Waves}\label{sec:FactorizationandCPW}
In the previous subsection, we have observed that every four point function we have considered factorizes into a product of three point functions when we take the middle two operators to have spacelike momenta.  For example, the Yang-Mills theory four point function \eqref{YMfourpoint} can be written as,
\begin{align}\label{YMCPW}
    &\langle\langle 0|J^{A_1}(p_1,\epsilon_1)J^{A_2}(p_2,\epsilon_2)J^{A_3}(p_3,\epsilon_3)J^{A_4}(p_4,\epsilon_4)|0\rangle\rangle_{YM}\theta(p_2^2)\theta(p_3^2)\notag\\
    &=\langle\langle 0|J^{A_1}(p_1,\epsilon_1)J^{A_2}(p_2,\epsilon_2)J^{\mu A}(-s)|0\rangle\rangle\frac{\pi_{\mu\nu}(s)}{|s|}\langle\langle 0|J^{\nu}_A(s)J^{A_3}(p_3,\epsilon_3)J^{A_4}(p_4,\epsilon_4)|0\rangle\rangle\notag\\
    &=\mathcal{W}^{(s)}_{(JJ|J|JJ)}(p_1,p_2|s|p_3,p_4),
\end{align}
where $\mathcal{W}^{(s)}_{(JJ|J|JJ)}$ is a conformal partial wave (CPW). Indeed, all our factorized expressions can be re-interpreted in terms of these quantities. In momentum space, CPWs are simply given by a product of three point functions \cite{Simmons-Duffin:2012juh,Gillioz:2020wgw,Jain:2024bza},
\begin{align}\label{CPW}
&\mathcal{W}^{(s)}_{(O_{1} O_{2}|J_s|O_{3}O_{4})}(p_1,p_2|s|p_3,p_4)\notag\\&=\frac{\Pi_{\mu_1\cdots \mu_s,\nu_1\cdots \nu_2}(s)}{|s|^{2s-1}}\langle\langle 0|\mathcal{O}_1(p_1)\mathcal{O}_2(p_2)J^{\mu_1\cdots \mu_s}(-p_1-p_2)|0\rangle\rangle\langle\langle 0|J^{\nu_1\cdots \nu_s}(-p_3-p_4)\mathcal{O}_3(p_3)\mathcal{O}_4(p_4)|0\rangle\rangle,\notag\\
&\mathcal{W}^{(s)}_{(O_{1} O_{2}|O_{\Delta}|O_{3}O_{4})}(p_1,p_2|s|p_3,p_4)\notag\\&=\frac{1}{|s|^{2\Delta-3}}\langle\langle 0|\mathcal{O}_1(p_1)\mathcal{O}_2(p_2)O_{\Delta}(-p_1-p_2)|0\rangle\rangle\langle\langle 0|O_{\Delta}(-p_3-p_4)\mathcal{O}_3(p_3)\mathcal{O}_4(p_4)|0\rangle\rangle.
\end{align}
The superscript $(s)$ denotes that this is an s-channel exchange. $\Pi_{\mu_1\cdots\mu_s\nu_1\cdots\nu_s}$ is the transverse traceless projector which is the transverse projector for spin-1 and the transverse traceless projector for spin-2. Our factorized Wightman functions written just in terms of these s-channel conformal partial waves are provided in appendix \ref{app:CPWexamples}.

Let us now compare in more detail to the construction of \cite{Meltzer:2021bmb}. The author shows for a time-ordered correlator with all external momentum space-like,
\begin{align}\label{timeordered1}
    &\text{Disc}_{s^2}\langle 0|T\{\mathcal{O}_1(p_1)\mathcal{O}_1(p_2)\mathcal{O}_3(p_3)\mathcal{O}_4(p_4)\}|0\rangle_s\theta(p_1^2)\theta(p_2^2)\theta(p_3^2)\theta(p_4^2)\notag\\&=\langle T\{\mathcal{O}_1(p_1)\mathcal{O}_2(p_2)\}\Bar{T}\{\mathcal{O}_3(p_3)\mathcal{O}_4(p_4)\}\rangle+(1,2)\leftrightarrow (3,4)=g^{(s)}_{(\mathcal{O}_1\mathcal{O}_2|\mathcal{O}_{\text{exchange}}|\mathcal{O}_3\mathcal{O}_4)}(p_1,p_2|s|p_3,p_4),
\end{align}
where the kinematics enforcing that all momenta are space-like is also implicit in the right hand side of this equation. Of course, $s$ is assumed to be time-like in this construction as well and we further make the choice that $s^0<0$\footnote{See appendix B of \cite{Meltzer:2021bmb} for the details. The author considers the $t-$ channel which can easily be generalized to the $s-$ channel case.}. The object $g^{(s)}_{(\mathcal{O}_1\mathcal{O}_2|\mathcal{O}_{\text{exchange}}|\mathcal{O}_3\mathcal{O}_4)}(p_1,p_2|s|p_3,p_4)$ is the conformal partial wave associated to the causal double commutator.

To obtain the Wightman conformal partial wave \eqref{CPW}, we need to take two more discontinuities with respect to $p_1^2,p_4^2$ in addition to \eqref{timeordered1}. This is because the spectral conditions \eqref{spectralcondition} demand in a Wightman function (with the ordering $1>2>3>4$) that $p_1^2<0,p_4^2<0$ as well as the fact that $p_1^0<0,p_4^0>0$. We also make no assumptions about $s^\mu$ as since the spectral conditions already render it time-like with $s^0<0$. Taking a discontinuity of \eqref{timeordered1} with respect to $p_1^2$ and $p_4^2$ assuming they have negative and positive energy respectively results in,
\begin{align}
    &\text{Disc}_{p_1^2}\text{Disc}_{p_4^2}\text{Disc}_{s^2}\langle 0|T\{O_{\Delta}(p_1)O_{\Delta}(p_2)O_{\Delta}(p_3)O_{\Delta}(p_4)\}|0\rangle_s \bigg(\prod_{i=2}^{3}\theta(p_i^2)\bigg)\theta(-s^2)\theta(-p_1^2)\theta(-p_4^2)\theta(-p_1^0)\theta(p_4^0)\notag\\&=\text{Disc}_{p_1^2}\text{Disc}_{p_4^2}g^{(s)}_{(\mathcal{O}_1\mathcal{O}_2|\mathcal{O}_{\text{exchange}}|\mathcal{O}_3\mathcal{O}_4)}(p_1,p_2|s|p_3,p_4)\notag\\&=\mathcal{W}_{(O_{\Delta}O_{\Delta}|O_{\Delta}|O_{\Delta}O_{\Delta})}^{(s)}(p_1,p_2|s|p_3,p_4).
\end{align}
which is precisely our factorized expression equal to the Wightman conformal partial wave\footnote{Note that the Feynman propagators and EOM inverter propagators are equal for space-like momenta, see table \ref{tab:deltagenmomspacescalarpropagators}.}. To sum up the discussion, Wightman functions in the special kinematics naturally arise by taking three cuts of the associated time-ordered correlator: One with respect to the exchanged momenta square and the remaining two with respect to the momenta squares of the first and fourth operators. This result is naturally the Wightman conformal partial wave\footnote{It would be interesting to understand if there is a generalization of the geodesic Witten diagram construction of conformal partial waves \cite{Hijano:2015zsa} for these real-time Wightman conformal partial wave. Perhaps a construction using time-like geodesics such as in \cite{Bohra:2025mhb} would be a potential starting point. We thank Allic Sivaramakrishnan for this comment.}. 
\subsection{Higher point correlators}\label{subsec:fivepoint}
The formalism outline so far can easily be extended to the computation of higher point Wightman functions. Here, we show this for the five point function of identical scalars with interacting with a cubic potential. The action and equations of motion are the same as in \eqref{scalar4ptphi3action} and \eqref{scalar4pointphi3EOM}. We shall also focus on special kinematics with the middle three operators having space-like momenta for concreteness. This receives contribution iff there is a field correction for all three middle operators. This is because we need a EOM inverter propagator with their momenta since only it has support for space-like momenta as we see from table \ref{tab:deltagenmomspacescalarpropagators}. This is exactly the same reasoning we gave for four point functions earlier in subsection \ref{subsec:fourpoint}. Thus we have,
\begin{align}
    &\langle\langle 0|O_{\Delta}(p_1)O_{\Delta}(p_2)O_{\Delta}(p_3)O_{\Delta}(p_4)O_{\Delta}(p_5)|0\rangle\rangle_{\order{g^3}}\theta(p_2^2)\theta(p_3^2)\theta(p_4^2)\notag\\
    &=\prod_{i=1}^{5}\big(\lim_{z_i\to 0}z_i^{1-\Delta_i}\big)\langle\langle 0|\phi^{(0)}(z_1,p_1)\phi^{(1)}(z_2,p_2)\phi^{(1)}(z_3,p_3)\phi^{(1)}(z_4,p_4)\phi^{(0)}(z_5,p_5)|0\rangle\rangle
\end{align}
Plugging in the field corrections \eqref{phi3fieldcorrection} and simplifying the resulting expression yields,
\footnotesize
\begin{align}
    &\langle\langle 0|O_{\Delta}(p_1)O_{\Delta}(p_2)O_{\Delta}(p_3)O_{\Delta}(p_4)O_{\Delta}(p_5)|0\rangle\rangle_{\order{g^3}}\theta(p_2^2)\theta(p_3^2)\theta(p_4^2)\notag\\
    &=(2ig)^3\theta(-p_1^2)\theta(-p_1^0)\theta(p_2^2)\theta(p_3^2)\theta(p_4^2)\theta(-p_5^2)\theta(p_5^0)\mathcal{A}_{5\Delta\Delta\Delta\Delta\Delta},\notag\\
    &\mathcal{A}_{5\Delta\Delta\Delta\Delta\Delta}=(\prod_{i=2}^{4}\frac{dz_i}{z_i})\mathcal{G}_{\Delta}(z_2,p_2)\mathcal{G}_{\Delta}(z_3,p_3)\mathcal{G}_{\Delta}(z_4,p_4)W_{\Delta,-}(z_2,-p_1)W_{\Delta,+}(z_2,z_3,s_{12})W_{\Delta,+}(z_3,z_4,s_{123})W_{+}(z_4,-p_5)\notag\\
    &=\Bigg(\frac{i 2^{\frac{29}{2}-5\Delta}\Gamma(\Delta-\frac{1}{2})^2\sqrt{\pi}}{\Gamma(\Delta)^7}\big(\prod_{i=1}^{5}|p_i|^{\nu}\big)\int_{0}^{\infty}dz_2 \sqrt{z_2}J_{\nu}(|p_1|z_2)K_{\nu}(|p_2|z_2)J_{\nu}(|s_{12}|z_2)\int_{0}^{\infty}dz_3 \sqrt{z_3}J_{\nu}(|s_{12}|z_3)K_{\nu}(|p_3|z_3)J_{\nu}(|s_{123}|z_3)\notag\\
    &\times\int_{0}^{\infty}dz_4 \sqrt{z_4}J_{\nu}(|s_{123}|z_4)K_{\nu}(|p_4|z_4)J_{\nu}(|p_5|z_4)\Bigg)\theta(-s_{12}^2)\theta(-s_{12}^0)\theta(-s_{123}^2)\theta(-s_{123}^0).
\end{align}
\normalsize
We have defined $s_{12}^\mu=p_1^\mu+p_2^\mu$ and $s_{123}^{\mu}=p_1^\mu+p_2^\mu+p_3^\mu$ as well as $\nu=\Delta-\frac{3}{2}$ in the above equation. Note that in these special kinematics, the five point functions also factorize.

As an example, let us evaluate the above integrals for $\Delta=1$ scalars. The result is,
\begin{align}\label{fivepointscalarDelta1}
    \mathcal{A}_{5,11111}\propto \frac{1}{|p_1||p_2||p_3||p_4||p_5|}\frac{1}{|s_{12}||s_{123}|}.
\end{align}
We leave a more detailed analysis of higher point functions to the future. The purpose of this section was to illustrate that our formalism provides a simple and systematic way to compute them.

\section{From Wightman functions to Euclidean correlators: How to avoid nested bulk integrals}\label{sec:WightmanToEuclid}
In this section, we demonstrate how one can recover a Euclidean correlator from the four point Wightman functions in the special kinematics. The reader might be worried since it seems like we lose a lot of information restricting to these special kinematics. However, as we shall in this section, we can actually recover the Euclidean correlator up to contact diagram contributions. 
We shall show on the other hand is\footnote{The subscript $s$ denotes that this is an s-channel contribution.},
\footnotesize
\begin{align}
    &\langle 0|T\{\mathcal{O}_1(p_1)\mathcal{O}_2(p_2)\mathcal{O}_3(p_3)\mathcal{O}_4(p_4)\}|0\rangle_s=F(p_1,p_2,p_3,p_4,s)\notag\\&\sim \int_0^\infty\frac{\omega_4d\omega_4}{\omega_4^2-p_4^2}\int_0^\infty\frac{\omega_1d\omega_1}{\omega_1^2-p_1^2}\int_0^\infty \frac{\omega_s d\omega_s}{\omega^2-s^2}(\langle 0|\mathcal{O}_1(p_1)\mathcal{O}(p_2)\mathcal{O}(p_3)\mathcal{O}(p_4)|0\rangle'/.\{p_1\to \omega_1,p_4\to \omega_4,s\to\omega_s\}),
\end{align}
\normalsize
where the prime indicates that our Wightman function is in the special kinematics $p_2^2>0,p_3^2>0$. The contour choice is also decided by the Wightman $i\epsilon$ prescription \eqref{Wightmaniepsilon}. Similarly, one can add the contributions of the other channels by permutations\footnote{This is similar to the statement in \cite{Meltzer:2021bmb} for recovering time-ordered correlators from their discontinuities up to contact diagrams. 
\begin{align}
    \langle 0|T\{\mathcal{O}_1(p_1)\mathcal{O}_2(p_2)\mathcal{O}_3(p_3)\mathcal{O}_4(p_4)\}|0\rangle_s=F(p_1,p_2,p_3,p_4,s)\sim \int_0^\infty \frac{\omega_s d\omega_s}{\omega^2-s^2}\text{Disc}_{\omega_s^2}F(p_1,p_2,p_4,p_4,\omega_s).
\end{align}
This RHS of this equation also has the nice property that it factorizes just like the Wightman functions in the special kinematics. However, for going to twistor space, we require the Wightman function for reasons argued in the introduction such as the fact that they satisfy current conservation Ward-Takahashi identities without contact terms and hence enjoy a simple representation in twistor space.}. 

We consider Yang-Mills theory as a illustrative, yet non trivial example to show this procedure. Also, note that after performing the dispersive integrals, we implicitly Wick rotate following the Wightman to obtain the Euclidean correlator from its time-ordered counterpart. First, to set the stage, let us do so at the level of three point functions.
\subsection{Three point functions}
We have seen earlier \eqref{threepointdiscexample} that in order to obtain a three point Wightman function with the middle operator having space-like momenta, from a Euclidean correlator, we need to take discontinuities with respect to the time-like momenta and Wick rotate following the Wightman $i\epsilon$ prescription \eqref{Wightmaniepsilon}. In this subsection, we want to essentially reverse this process. The Yang-Mills three point function \eqref{JJJWightman} repeated here for convenience is,
\begin{align}
    \frac{ig}{2}\theta(p_2^2)\theta(-p_1^2)\theta(-p_3^2)\theta(-p_{1}^{0})\theta(p_{3}^{0})f^{ABC}V_{3}^{YM}\bigg(\frac{1}{E}-\frac{1}{E-2p_1}-\frac{1}{E-2p_2}-\frac{1}{E-2p_3}\bigg),
\end{align}
where $V_{3}^{YM}$ is the Yang-Mills three point vertex \eqref{YMvertex3pt}.

Stripping off the theta functions, we get,
\begin{align}
    \langle\langle 0|J^{A}(p_1,\epsilon_1)J^{B}(p_2,\epsilon_2)J^{C}(p_3,\epsilon_3)|0\rangle\rangle'=i\frac{g}{2}f^{ABC}V_{3}^{YM}f_W(p_1,p_2,p_3),
\end{align}
with,
\begin{align}
    f_W(p_1,p_2,p_3)=\bigg(\frac{1}{E}-\frac{1}{E-2p_1}-\frac{1}{E-2p_2}-\frac{1}{E-2p_3}\bigg).
\end{align}
To obtain the Euclidean form-factor we need to ``invert" the discontinuity procedure. This can be achieved by performing dispersive integrals over the two time-like momenta magnitudes as follows\footnote{The three point dispersive integrals have previously been performed in \cite{Baumann:2024ttn,Bala:2025gmz}. The difference from our analysis here is that they performed dispersive integrals with respect to all three momenta. What we see here is that it is sufficient to do it with respect to just any two of them.}:
\begin{align}
    f_E(p_1,p_2,p_3)=\int_{0}^{\infty} \frac{d\omega_1^2}{\omega_1^2-p_1^2}\int_{0}^{\infty} \frac{d\omega_3^2}{\omega_3^2-p_3^2}f_W(\omega_1,p_2,\omega_3).
\end{align}
As for the contour choice to avoid the poles at $\omega_i=\pm p_i,i=1,3$ there is no ambiguity thanks to the Wightman $i\epsilon$ prescription \eqref{Wightmaniepsilon} using the fact that $p_1^0<0,p_3^0>0$ for this Wightman function. Evaluating the integrals results in,
\begin{align}
    f_E(p_1,p_2,p_3)=\frac{1}{p_1+p_2+p_3},
\end{align}
which is the correct Euclidean form-factor. The same method holds for all the other spinning three point functions which one can readily check with the results of subsection \ref{subsec:Threepoint} and comparing with the known Euclidean results. For three point correlators involving generic scalar operators such as \eqref{threepointgenscalar}, \eqref{JOOgendelta}, this procedure is much more complicated due to the fact that they are not simple rational functions of the $p_i^2$ but rather have a more intricate branch cut structure (that of the Appell $F_4$ functions in the referenced expressions). Thus, the dispersive integrals have to be performed taking this into account and will be more involved .

\subsection{Four point functions}
Consider the the Yang-Mills four point function in the special kinematics \eqref{YMfourpoint}. Stripping off the theta functions and pre-factors we obtain from \eqref{YMfourpoint},
\small
\begin{align}
    &\langle\langle 0|J^{A_1}(p_1,\epsilon_1)J^{A_2}(p_2,\epsilon_2)J^{A_3}(p_3,\epsilon_3)J^{A_4}(p_4,\epsilon_4)|0\rangle\rangle'\notag\\
    &=f^{A_1A_2A}f_A^{A_3A_4}\frac{p_1p_2 s}{(p_1^4+(p_2^2-s^2)^2-2p_1^2(p_2^2+s^2))}V_3^{YM,\mu}(\epsilon_1,\epsilon_2)\pi_{\mu\nu}(s)V_{3}^{YM,\nu}(\epsilon_3,\epsilon_4)\frac{p_3p_4 s}{(p_3^4+(p_4^2-s^2)^2-2p_3^2(p_4^2+s^2))}\notag\\
    &=f^{A_1A_2A}f_A^{A_3A_4}V_{3}^{YM,\mu}(\epsilon_1,\epsilon_2)V_{3}^{YM,\nu}f^W(p_1,p_2,s,p_3,p_4)\big(\eta_{\mu\nu}-\frac{s_\mu s_\nu}{s^2}\big).
\end{align}
\normalsize
To obtain the Euclidean correlator, let us take a leaf out of the three point case and perform dispersive integrals with respect to the squares of the three time-like momenta viz $p_1,s$ and $p_4$. We consider,
\small
\begin{align}
    &f^{A_1A_2A}f_A^{A_3A_4}V_{3}^{YM,\mu}(\epsilon_1,\epsilon_2)V_{3}^{YM,\nu}\int_0^\infty \frac{d\omega_1^2}{\omega_1^2-p_1^2}\int_0^\infty \frac{d\omega_4^2}{\omega_4^2-p_4^2}\int_0^\infty \frac{d\omega_s^2}{\omega_s^2-s^2}f^W(\omega_1,p_2,\omega_s,p_3,\omega_4)(\eta_{\mu\nu}-\frac{p_{12\mu}p_{12\nu}}{\omega_s^2}).
\end{align}
\normalsize
To perform the $\omega_1$ integral, we enclose the poles at $\omega_1=p_1,\omega_1=p_2-\omega_s$ and $\omega_1=p_2+\omega_s$ which is the generalization of the prescription used for three point functions in \cite{Baumann:2024ttn}. Then, to perform the $\omega_4$ integral we enclose the poles at $\omega_4=p_4,\omega_4=p_3-\omega_s$ and $\omega_4=p_3+\omega_s$. Finally, the $\omega_s$ integral is performed by enclosing the poles at $\omega_s=p_1+p_2,\omega_s=p_3+p_4$ and $\omega_s=s$. The result of this computation is,
\footnotesize
\begin{align}
    V_{3}^{YM,\mu}(\epsilon_1,\epsilon_2)V_{3}^{YM,\nu}(\epsilon_3,\epsilon_4)\frac{1}{(p_1+p_2+p_3+p_4)(p_1+p_2+s)(p_3+p_4+s)}\bigg(\eta_{\mu\nu}+\frac{s_\mu s_\nu}{(p_1+p_2)(p_3+p_4)s}(p_1+p_2+p_3+p_4+s)\bigg),
\end{align}
\normalsize
which is precisely the correct answer for the s-channel contribution to the Euclidean correlator, see for instance \cite{Albayrak:2018tam}.

To obtain the $t$ and $u$ channels, we simply re-instate colour factors, permute labels and add them up to obtain the sum of all the exchange diagrams. This gives us,
\begin{align}
    &\langle\langle J^{A_1}(p_1,\epsilon_1)J^{A_2}(p_2,\epsilon_2)J^{A_3}(p_3,\epsilon_3)J^{A_4}(p_4,\epsilon_4)\rangle\rangle_{\text{exchange}}\notag\\&=\text{Disp}_{p_1^2}\text{Disp}_{p_4^2}\text{Disp}_{s^2}\langle\langle 0| J^{A_1}(p_1,\epsilon_1)J^{A_2}(p_2,\epsilon_2)J^{A_3}(p_3,\epsilon_3)J^{A_4}(p_4,\epsilon_4)|0\rangle\rangle'\notag\\
    &+\text{Disp}_{p_1^2}\text{Disp}_{p_4^2}\text{Disp}_{u^2}\langle\langle 0| J^{A_1}(p_1,\epsilon_1)J^{A_3}(p_3,\epsilon_3)J^{A_2}(p_2,\epsilon_2)J^{A_4}(p_4,\epsilon_4)|0\rangle\rangle'\notag\\
    &+\text{Disp}_{p_1^2}\text{Disp}_{p_3^2}\text{Disp}_{t^2}\langle\langle 0| J^{A_1}(p_1,\epsilon_1)J^{A_4}(p_4,\epsilon_4)J^{A_2}(p_2,\epsilon_2)J^{A_3}(p_3,\epsilon_3)|0\rangle\rangle'.
\end{align}
Disp is shorthand for the dispersive integrals we performed above. Every term on the RHS is a stripped Wightman function in special kinematics where the middle two operators have space-like momenta in each of the terms. This procedure yields the correct exchange contribution to the four point function as can be checked using our expressions and comparing with the known results. 

This method however, misses the contact diagram contribution.  However, this is not too bad since contact diagrams are in general very simple and can be added to the final result by hand by demanding that the current conservation Ward-Takahashi identtiy for the Euclidean correlator has to be obeyed. Thus we get,
\footnotesize
\begin{align}
    &\langle\langle J^{A_1}(p_1,\epsilon_1)J^{A_2}(p_2,\epsilon_2)J^{A_3}(p_3,\epsilon_3)J^{A_4}(p_4,\epsilon_4)\rangle\rangle\notag\\&=\langle\langle J^{A_1}(p_1,\epsilon_1)J^{A_2}(p_2,\epsilon_2)J^{A_3}(p_3,\epsilon_3)J^{A_4}(p_4,\epsilon_4)\rangle\rangle_{\text{exchange}}+\langle\langle J^{A_1}(p_1,\epsilon_1)J^{A_2}(p_2,\epsilon_2)J^{A_3}(p_3,\epsilon_3)J^{A_4}(p_4,\epsilon_4)\rangle\rangle_{\text{contact}},
\end{align}
\normalsize
where the contact diagram expression can be found in the literature.

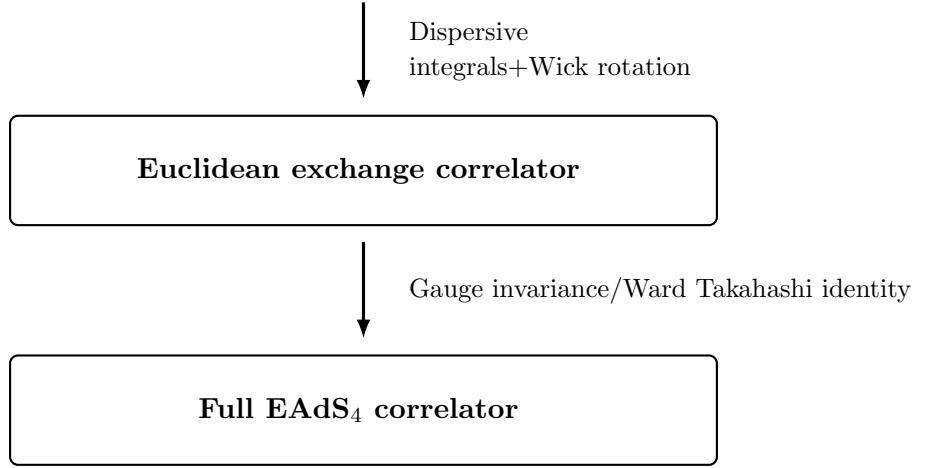
\begin{figure}[h!]
\centering

\begin{tikzpicture}[
    node distance=1.7cm,
    box/.style={
        rectangle,
        rounded corners=3pt,
        draw=black,
        thick,
        align=center,
        minimum width=9.3cm,
        minimum height=1.45cm,
        inner sep=6pt
    },
    arrow/.style={-{Latex[length=2.5mm]}, very thick, shorten >=6pt, shorten <=6pt},
    lab/.style={midway, right=0.45cm, align=left, font=\small}
]

\node[box] (b2) {
    \textbf{Momentum space Wightman function (Factorized)
    $=$Wightman conformal partial wave}
};

\node[box, below=of b2] (b3) {
    \textbf{Euclidean exchange correlator}
};

\node[box, below=of b3] (b4) {
    \textbf{Full EAdS$_4$ correlator}
};


\draw[arrow] (b2) -- (b3)
    node[lab] {Dispersive\\ integrals$+$Wick rotation};

\draw[arrow] (b3) -- (b4)
    node[lab] {Gauge invariance/Ward Takahashi identity};

\end{tikzpicture}

\caption{Recovering the full correlator from just the special kinematics Wightman function.  Note that there is not a single nested bulk integral in sight in this procedure! Once we have the EAdS$_4$ correlator, one can in principle obtain any Lorentzian correlator in any desired kinematics.}
\label{fig:WightmanToEuclid}
\end{figure}

We summarize this procedure in figure \ref{fig:WightmanToEuclid}.

Let us note that we have obtained the result for the Yang-Mills Euclidean four point function without ever having to perform a nested bulk integral that one would do in traditional Witten diagram approaches. Just starting from the Wightman function in these special kinematics allows us to construct the Euclidean correlator, which has knowledge of all kinematics\footnote{Of course, it is a different matter of performing the analytic continuation from the Euclidean correlator to a Wightman function in arbitrary kinematics in practice. Look at \eqref{scalarphi3exchangefull} for example. The point is, the Euclidean correlator in principle knows about all kinematic configurations of Wightman functions. It would be very interesting to carefully analyze why this is the case and find out how general it is. At the very least, it works for all the spinning four point examples we have considered in this paper (including those with $\Delta=1,2$ scalars).}.

\section{Wightman functions in Twistor space}\label{sec:Twistors}
As indicated in the introduction, the  motivation for us to study holographic Wightman functions is due to the fact that they enjoy a simple representation in the twistor space construction for 3d CFT correlators. In this section, after briefly reviewing the construction of twistor variables from momentum space, we first show that the two and three point Wightman functions we have explicitly calculated in the earlier section are precisely the results obtained from twistor space via a half-Fourier transform and a suitable ``analytic" continuation. We then take first steps in extending the 3d CFT twistor framework to four point functions.

\subsection{From momentum space to twistors}\label{subsec:SHtoTwistor}
 Let us first review the construction of spinor helicity variables from momentum space, see \cite{S:2025pmh} for a more detailed discussion. If $p_\mu$ is a space-like vector, we can trade it for a pair of real $SL(2,\mathbb{R})$ spinors $\lambda$ and $\Bar{\lambda}$ as follows:
\begin{align}\label{SHvariables}
    p_\mu\to p^{ab}= p_\mu (\sigma^\mu)^{ab}=\frac{(\lambda^a\Bar{\lambda}^b+\lambda^b\Bar{\lambda}^a)}{2},
\end{align}
where $(\sigma^\mu)$ are the Lorentzian Pauli matrices. This description has the little group redundancy,
$\lambda\to r \lambda,\Bar{\lambda}\to\frac{\Bar{\lambda}}{r},p_{ab}\to p_{ab},~r\in\mathbb{R}/\{0\}$. Lorentz invariant quantities in this description are obtained by contracting spinors using the $SL(2,\mathbb{R})$ Levi-Civita symbol:
\begin{align}
    \langle i j\rangle=\lambda_{ia}\lambda_j^b,\langle \Bar{i}\Bar{j}\rangle=\Bar{\lambda}_{ia}\Bar{\lambda}_j^b,\langle i \Bar{j}\rangle=\lambda_{ia}\Bar{\lambda}_j^a,
\end{align}
where $i,j$ label spinors associated to momenta $p_{i\mu},p_{j\mu}$.

Twistor space is related to spinor helicity variables by a \textit{half-Fourier transform} first introduced by Witten \cite{Witten:2003nn}. We go from the projective coordinates $(\lambda,\Bar{\lambda})$ to projective coordinates $(\lambda,\Bar{\mu})$. This is a twistor, which is a fundamental representation of the double cover of the 3d conformal group Sp$(4;\mathbb{R})$. 
\begin{align}
    Z^A=(\lambda^a,\Bar{\mu}_{a'}),
\end{align}
Given a conserved current $J_s^{\mu_1\cdots \mu_s}(p_\mu)$ dual to a massless boson, we can trade it for two quantities viz positive and negative helicity components as follows\footnote{The remaining components of the current constructed using \eqref{polarizations} like in \eqref{Jspm} are all proportional to its divergence which is zero for a conserved current.}:
\begin{align}\label{Jspm}
    J_s^{\pm}(\lambda,\Bar{\lambda})=\epsilon_{\mu_1}^{\pm}\cdots \epsilon_{\mu_s}^{\pm}J_s^{\mu_1\cdots\mu_s}(p_\mu),
\end{align}
with,
\begin{align}\label{polarizations}
    \epsilon_{\mu}^-=\frac{-1}{\lambda\cdot\Bar{\lambda}}(\sigma_\mu)_a^b \lambda^a\lambda_b~,\epsilon_{\mu}^+=\frac{-1}{\lambda\cdot\Bar{\lambda}}(\sigma_\mu)_a^b \Bar{\lambda}^a\Bar{\lambda}_b.
\end{align}
It is also not too difficult to show that \cite{S:2025pmh},
\begin{align}\label{JsintermsofJspm}
    J_{s\mu_1\cdots \mu_s}=\bigg(-\frac{1}{2}\bigg)^s\bigg(\epsilon_{\mu_1}^{-}\cdots\epsilon_{\mu_s}^{-}J_s^{+}+\epsilon_{\mu_1}^{+}\cdots\epsilon_{\mu_s}^{+}J_s^-\bigg).
\end{align}
  The twistor space counterparts of these currents are obtained via a \textit{half Fourier transform}:
\begin{align}\label{halfFourier}
    \hat{J}_s^{\pm}(Z)=\hat{J}_s^{\pm}(\lambda,\Bar{\mu})=\int \frac{d^2\Bar{\lambda}}{(2\pi)^2}e^{i\Bar{\lambda}\cdot\Bar{\mu}}\frac{J_s^{\pm}(\lambda,\Bar{\lambda})}{p^{s-1}},
\end{align}
The conformal generators in these variables combine to form a Sp$(4;\mathbb{R})$ covariant quantity,
\begin{align}\label{TAB}
    T^{AB}=Z^{(A}\frac{\partial}{\partial Z_{B)}}.
\end{align}
Note that we have lowered one of the indices in the above expression which we can naturally do using the Sp$(4;\mathbb{R})$ invariant symplectic form $\Omega$ as follows:
\begin{align}
    Z^A=\Omega^{AB}Z_B,Z_A=\Omega_{BA}Z^B, \Omega=\begin{pmatrix}
        0&&1_{2\times 2}\\
        -1_{2\times 2}&& 0
    \end{pmatrix}.
\end{align}
Thus, one can expect simple results for current correlators due to the simplicity of \eqref{TAB}. In particular the, twistor space currents are dimensionless\footnote{The position space operator has $\Delta=s+1$ which the Fourier transform and rescaling by $\frac{1}{p^{s-1}}$ bring down to $-1$ with the $d^2\Bar{\lambda}$ in the half-Fourier transform \eqref{halfFourier} contributing scaling dimension $+1$ thus yielding a dimensionless twistor space current. Note in particular that this is also true for $\Delta=1$ scalars which can be obtained by setting $s=0$. On the other hand, the twistor space description of other operators are  complicated and involve non-local terms \cite{Bala:2025qxr}.\label{footnote:twistorspaceops}}.  
The twistor space correlators obey the conformal Ward identities,
\begin{align}\label{TABWardId}
    \sum_{i=1}^{N}\langle 0|\cdots [T^{AB},\hat{J}_{s_i}^{\pm}(Z_i)]\cdots 0|\rangle=0.
\end{align}
The solutions to this equation come in two flavors. Symplectic dot products of twistors and the $\delta^4$ distribution
\footnotesize
\begin{align}\label{sp4invariants}
    \text{Sp}(4;\mathbb{R})~\text{Invariants}:\bigg \{Z_i\cdot Z_j=-\Omega_{AB}Z_i^{A}Z_j^B=\lambda_i\cdot\Bar{\mu}_j-\lambda_j\cdot\Bar{\mu}_i\bigg\},\bigg\{\delta^4(\sum_{k=1}^{N}c_k Z_k)=\delta^2(\sum_{k=1}^{N}c_k \lambda_k)\delta^2(\sum_{k=1}^{N}c_k\Bar{\mu}_k), c_k\in \mathbb{R}\bigg\}.
\end{align}
\normalsize
Our results below shall include both classes of solutions. 
We will also work with scalar operators which in twistor space are given by,
\begin{align}\label{scalarhalffourier}
    \hat{O}_{\Delta}(Z)=\int \frac{d^2\Bar{\lambda}}{(2\pi)^2}e^{i\Bar{\lambda}\cdot\Bar{\mu}}p~O_\Delta(\lambda,\Bar{\lambda}).
\end{align}
For correlators involving $\Delta=1$ scalars, the invariants will indeed be of the form \eqref{sp4invariants}. 
Before we proceed to the examples, let us write down the inverse of the half-Fourier transforms \eqref{halfFourier}, \eqref{scalarhalffourier} which will be useful later.
\begin{align}\label{inverseHalfFourier}
    &\hat{J}_s^{\pm}(\lambda,\Bar{\lambda})=\frac{J_s^{\pm}(\lambda,\Bar{\lambda})}{p^{s-1}}=\int d^2\Bar{\mu}e^{-i\Bar{\lambda}\cdot \Bar{\mu}}\hat{J}_s^{\pm}(Z),\notag\\
    &\hat{O}_\Delta(p)=p O_\Delta(p)=\int d^2\Bar{\mu}e^{-i\Bar{\lambda}\cdot\Bar{\mu}}\hat{O}_\Delta(Z).
\end{align}
\subsection{What are we computing?}

The objects of interest to us are twistor space ``Wightman functions" of these operators. Strictly speaking, we have defined twistor space via a half-Fourier transform with momenta that are space-like \eqref{SHvariables}. However, even Wightman two point functions have support only when the momenta are time-like which is required by the spectral conditions. As we shall see, what we compute in twistor space are the Wightman functions with the spectral theta functions stripped off and the momenta analytically continued to be spacelike, thus clarifying the relation between twistor correlators and Wightman functions: The former are close spacelike ``cousins" of the latter. Let us also note that we re-instate the momentum conserving delta functions in this section as it is required when transforming to twistor space.
\subsection{Two point functions}\label{subsec:twistor2point}
The general spin-s current two point function in twistor space takes the form\footnote{One can also work in an ambidextrous basis, choosing to represent positive helicity currents in twistor space with coordinates $Z^A$  and negative helicity currents in the dual twistor space with coordinates $W_A$. One can go form one description to the other using \begin{align}\label{twistorfouriertransform}
    \hat{J}_s^{h}(W)=\int\frac{d^4 Z}{(2\pi)^2}e^{iW\cdot Z}\hat{J}_s^{h}(Z).
\end{align}
},
\begin{align}\label{twopointgenans}
    \langle 0|\hat{J}_s^{h_1}(Z_1)\hat{J}_s^{h_2}(Z_2)|0\rangle=\delta_{h_1,h_2}\frac{i^{2 h_1+2}}{(Z_1\cdot Z_2)^{2h_1+2}}.
\end{align}
Its half-Fourier transform to spinor helicity variables using \eqref{halfFourier} yields,
\begin{align}\label{twopointtwistortoSH}
    \langle 0|\hat{J}_s^{+}(p_1)\hat{J}_s^{+}(p_2)|0\rangle'\propto \frac{\langle\Bar{1}\Bar{2}\rangle^{2s}}{\langle 1\Bar{1}\rangle^{2s-1}}\text{Sign}(\langle 1\Bar{1}\rangle)\delta^3(p_1+p_2),\langle 0|\hat{J}_s^{-}(p_1)\hat{J}_s^{-}(p_2)|0\rangle'\propto \frac{\langle 12 \rangle^{2s}}{\langle 1\Bar{1}\rangle^{2s-1}}\text{Sign}(\langle 1\Bar{1}\rangle)\delta^3(p_1+p_2).
\end{align}
Let us relate this result to our two point Wightman functions starting with the spin-1 example \eqref{JJtwopoint}. We first contract it with general transverse null polarization vectors $\epsilon_i$.
\begin{align}
        \langle 0|J(\epsilon_1,p_1)J(\epsilon_2,p_2)|0\rangle&=2|p_1|\epsilon_{1\mu}\epsilon_{2\nu}(\eta^{\mu\nu}-\frac{p^\mu p^\nu}{p^2})\theta(-p_1^2)\theta(-p_1^0)(2\pi)^3\delta^3(p_1+p_2)\notag\\
        &=2|p_1|(\epsilon_1\cdot \epsilon_2)\theta(-p_1^2)\theta(-p_1^0)(2\pi)^3\delta^3(p_1+p_2).
\end{align}
We then define a stripped Wightman function denoted with a prime:
\begin{align}\label{JJstrippedWightman}
    &\langle 0|J(\epsilon_1,p_1)J(\epsilon_2,p_2)|0\rangle=\langle 0|J(\epsilon_1,p_1)J(\epsilon_2,p_2)|0\rangle' \theta(-p_1^2)\theta(-p_1^0),\notag\\
    &\langle 0|J(\epsilon_1,p_1)J(\epsilon_2,p_2)|0\rangle'=2|p_1|(\epsilon_1\cdot \epsilon_2)(2\pi)^3\delta^3(p_1+p_2).
\end{align}
Although the full Wightman function has support only for time-like momenta, we can analytically continue the stripped Wightman function \eqref{JJstrippedWightman} to spacelike momenta. Making use of the spinor helicity construction \eqref{SHvariables} and choosing the polarization vectors to have positive and negative helicity \eqref{polarizations} respectively yields,
\begin{align}\label{JJstrippedtwopoint}
    &\langle 0|J^{+}(p_1)J^{+}(p_2)|0\rangle'=2\text{Sign}(\langle 1\Bar{1}\rangle)\frac{\langle \Bar{1}\Bar{2}\rangle^2}{\langle 1\Bar{1}\rangle}(2\pi)^3\delta^3(p_1+p_2),\notag\\
    &  \langle 0|J^-(p_1)J^-(p_2)|0\rangle'=2\text{Sign}(\langle 1\Bar{1}\rangle)\frac{\langle 12\rangle^2}{\langle 1\Bar{1}\rangle}(2\pi)^3\delta^3(p_1+p_2)
\end{align}
 Note that the stripped correlators have $(1\leftrightarrow 2)$ permutation symmetry since the spectral theta functions that distinguish the operator ordering have been stripped off. Thus, we see that the twistor space two point function found in earlier works is actually the counterpart of the stripped Wightman function \eqref{JJstrippedWightman}. This also explains the origin of the sign factor in the above equations first found in \cite{Baumann:2024ttn}. 
Similarly, for the stress tensor two point function, one can follow the same procedure showing that its stripped Wightman function corresponds to \eqref{twopointgenans} with $h_i=\pm 2$.

For the general scalar case \eqref{ODeltaODelta2point}, the twistor space answer is more complicated as it involves what is known as the \textit{infinity twistor} of $\mathbb{R}^{2,1}$ \cite{Bala:2025qxr}. The half-Fourier transform for generic scalar operators from momentum space to twistor space is given in \eqref{scalarhalffourier}. The twistor space result corresponding to the scalar two point function was found in \cite{Bala:2025qxr},
\begin{align}
    \langle 0|\hat{O}_{\Delta}(Z_1)\hat{O}_{\Delta}(Z_2)|0\rangle=\frac{|\langle Z_1 I Z_2\rangle|^{2(\Delta-1)}}{(Z_1\cdot Z_2)^{2\Delta}},
\end{align}
where,
\begin{align}
    \langle Z_1 I Z_2\rangle=Z_1^{A}I_{AB}Z_2^B=\langle 1 2\rangle~\text{with}~I_{AB}=\begin{pmatrix}
        \epsilon_{ab}&0\\
        0&0.
    \end{pmatrix}
\end{align}
Going to momentum space using \eqref{inverseHalfFourier} yields,
\begin{align}
    \langle 0|\hat{O}_{\Delta}(p_1)\hat{O}_{\Delta}(p_2)|0\rangle'\propto (2\pi)^3\delta^3(p_1+p_2)p_1^{2\Delta-3}\text{Sign}(p_1),
\end{align}
which is precisely the scalar two point Wightman function \eqref{ODeltaODelta2point} with the spectral theta functions stripped and the result analytically continued to spacelike momenta.
\subsection{Three point functions}\label{subsec:twistor3point}
We now move on to the case of three point functions. We focus on Wightman functions with $p_2^2>0$. Analogous to the two point case, we define the stripped three point Wightman functions,
\begin{align}\label{threepointstripped}
    \langle 0|\mathcal{O}_1(p_1)\mathcal{O}_2(p_2)\mathcal{O}_3(p_3)|0\rangle\theta(p_2^2)=\theta(-p_1^2)\theta(-p_1^0)\theta(p_2^2)\theta(-p_3^2)\theta(p_3^0)\langle 0|\mathcal{O}_1(p_1)\mathcal{O}_2(p_2)\mathcal{O}_3(p_3)|0\rangle',
\end{align}
where $\mathcal{O}_i$ are arbitrary (spinning) operators. When converting to twistor space, we analytically continue the momenta in the stripped correlator to be spacelike. When some of these operators are identical, we shall see that the stripped correlator also possesses permutation symmetry. 

Below, we discuss the twistor counterparts of the various three point Wightman functions we have computed earlier. Let us also mention that the scalars we shall focus on henceforth are conformally coupled ones with Neumann boundary conditions imposed. This implies that the dual conformal operator has $\Delta=1$ which as we discussed in footnote \ref{footnote:twistorspaceops}, enjoys a simpler description in twistor space compared to its generic $\Delta$ counterparts. Also, we ignore overall numerical constants as well as the bulk couplings in the expressions for notational simplicity but it is a simple matter to reinstate them if desired.  

For general spins $s_1,s_2$ and $s_3$ (including the $\Delta=1$ scalar by setting spin to zero), the general solution at the conformal Ward identities is \cite{Bala:2025qxr}\footnote{One can also choose to work in a mixture of twistor and dual twistor variables like in \cite{Baumann:2024ttn,Bala:2025gmz}. These are related to the solutions in \eqref{twistorspace3pointgensols1s2s3} by a twistor Fourier transform \eqref{twistorfouriertransform}.},
\begin{align}\label{twistorspace3pointgensols1s2s3}
    &\langle 0|\hat{J}_{s_1}^{h_1}(Z_1)\hat{J}_{s_2}^{h_2}(Z_2)\hat{J}_{s_3}^{h_3}(Z_3)|0\rangle\notag\\&=i^{\alpha+\beta+\gamma}\bigg(c_{h_1h_2h_3}\delta^{[\alpha]}(Z_1\cdot Z_2)\delta^{[\beta]}(Z_2\cdot Z_3)\delta^{[\gamma]}(Z_3\cdot Z_1)\notag\\
    &+k_{h_1h_2h_3}\delta^{[\alpha+\beta+\gamma]}(Z_1\cdot Z_2)\int dc_1 dc_2 c_1^{-\beta}c_2^{-\gamma}\delta^4(c_1 Z_1+c_2 Z_2-Z_3)\bigg),
\end{align}
where the coefficients $c_{h_1h_2h_3},k_{h_1h_2h_3}$ and $\alpha,\beta,\gamma$ depend on the helicity. Using the fact that the conserved currents have helicity $h_i=\pm s_i$, we have,
\begin{align}
    (h_1 h_2 h_3)~\text{helicity}:~\alpha=h_1 +h_2 -h_3 ,\beta=h_2 +h_3 -h_1 ,\gamma=h_3 +h_1 -h_2 .
\end{align}
Our notation for the delta function derivatives is as follows:
\begin{align}
    \delta^{[n]}(x)=\int_{-\infty}^\infty dc (-i c)^n e^{-i c x}.
\end{align}
We also make sense of this formula by suitably regularizing the cases where $n$ is a negative integer following earlier works. For example, the anti-derivative of $\delta^{[0]}(x)=\delta(x)$ is $\delta^{[-1]}(x)=\frac{\text{Sgn}(x)}{2}$,$\delta^{[-2]}(x)=\frac{|x|}{2}$ and so on.

The inverse half-Fourier transform to spinor helicity variables \eqref{inverseHalfFourier} of the general solution \eqref{twistorspace3pointgensols1s2s3} results in,
\begin{align}\label{threepointtwistortoSHgensol}
    &\langle 0|\hat{J}_{s_1}^{h_1}(p_1)\hat{J}_{s_2}^{h_2}(p_2)\hat{J}_{s_3}^{h_3}(p_3)|0\rangle'\notag\\&=\frac{(-1)^{\alpha+\beta+\gamma}}{4}(2\pi)^3\delta^3(p_1+p_2+p_3)\langle \Bar{1}\Bar{2}\rangle^{\alpha}\langle\Bar{2}\Bar{3}\rangle^\beta\langle\Bar{3}\Bar{1}\rangle^\gamma\bigg(\frac{c_{h_1h_2h_3}}{E^{\alpha+\beta+\gamma}}+\frac{k_{h_1h_2h_3}}{(E-2p_1)^\alpha (E-2p_2)^\gamma (E-2p_3)^\beta}\bigg),
\end{align}
where $E=p_1+p_2+p_3$ as usual. We will now see how \eqref{threepointtwistortoSHgensol} reproduces all the three point results we have computed so far.
\subsubsection{Conformally coupled scalars}
Let us begin with the simplest three point function: That of a $\Delta=1$ scalar operator.
The twistor space expression corresponding to this correlator is found by setting $k_{s_1s_2s_3}=1$ and $c_{s_1s_2s_3}=0$ in our general solution \eqref{twistorspace3pointgensols1s2s3}, see \cite{Bala:2025qxr}.
\begin{align}\label{O1O1O1twistor}
    \langle 0|\hat{O}_1(Z_1)\hat{O}_1(Z_2)\hat{O}_1(Z_3)|0\rangle=\delta(Z_1\cdot Z_2)\int_{-\infty}^{\infty}dc_1\int_{-\infty}^{\infty}dc_2~ \delta^4(c_1 Z_1+c_2 Z_2-Z_3).
\end{align}
The inverse half Fourier transform found using \eqref{threepointtwistortoSHgensol} yields,
\begin{align}
  &\langle 0|\hat{O}_1(p_1)\hat{O}_1(p_2)\hat{O}_1(p_3)|0\rangle'= \frac{1}{4}(2\pi)^3\delta^3(p_1+p_2+p_3)\notag\\
  &\implies  \langle 0|O_1(p_1)O_1(p_2)O_1(p_3)|0\rangle'=\frac{(2\pi)^3\delta^3(p_1+p_2+p_3)}{p_1 p_2 p_3}
\end{align}
This is indeed the correct stripped Wightman function which can be obtained by evaluating the integral with $\theta(p_2^2)$ in \eqref{threepointgenscalar}, setting all the scaling dimensions to $1$ and stripping the theta functions as in \eqref{threepointstripped}.

\subsubsection{Photon-Scalar-Scalar}
Similar to the scalar example, one can choose the values of the coefficients in the general solution \eqref{threepointtwistortoSHgensol} to match with the stripped Wightman functions corresponding to this scalar QED example. 
\footnotesize
\begin{align}\label{JOOtwistor}
    &\langle 0|\hat{J}^{+}(Z_1)\hat{O}_1(Z_2)\hat{O}_1(Z_3)|0\rangle=ic_{100}\bigg(\delta^{[1]}(Z_1\cdot Z_2)\text{Sgn}(Z_2\cdot Z_3)\delta^{[1]}(Z_3\cdot Z_1)+\delta^{[1]}(Z_1\cdot Z_2)\int \frac{dc_1}{c_1}\frac{dc_2}{c_2}\delta^4(c_1 Z_1+c_2Z_2-Z_3)\bigg),\notag\\
    &\langle 0|\hat{J}^{-}(Z_1)\hat{O}_1(Z_2)\hat{O}_1(Z_3)|0\rangle=ic_{-100}\bigg(\text{Sgn}(Z_1\cdot Z_2)\delta^{[1]}(Z_2\cdot Z_3)\text{Sgn}(Z_3\cdot Z_1)+\text{Sgn}(Z_1\cdot Z_2)\int dc_1 dc_2 c_1 c_2\delta^4(c_1 Z_1+c_2Z_2-Z_3)\bigg),
\end{align}
\normalsize
with their spinor helicity counterparts found using \eqref{threepointtwistortoSHgensol}.
Contracting our Wightman function result \eqref{JOOgendelta},\eqref{JO1O1andJO2O2} with the null transverse polarization vector $\epsilon_{1\mu}$, stripping off the spectral functions as in \eqref{threepointstripped}and converting to spinor helicity variables shows that it matches with \eqref{JOOtwistor}.

\subsubsection{Yang-Mills theory}
Let us move on to a case where all three operators have spin. In this case, we have eight helicity configurations to deal with. In appendix \ref{app:YmandGRallhelicities}, we provide all these expressions both in spinor helicity and twistor space.

For example, consider the $(+++)$ helicity configutation. As the reader can easily verify using \eqref{inverseHalfFourier}, the twistor space expression corresponding to our Wightman function \eqref{JJJWightman} is none other than \eqref{twistorspace3pointgensols1s2s3} with $s_1=s_2=s_3=1,c_{111}=0,k_{111}=1$ viz\footnote{It is easy to show using \eqref{twistorfouriertransform} that this in dual twistor space equals,
\begin{align}
    \langle 0|\hat{J}^{+A}(W_1)\hat{J}^{+B}(W_2)\hat{J}^{+C}(W_3)|0\rangle=f^{ABC} \text{Sgn}(W_1\cdot W_2)\text{Sgn}(W_2\cdot W_3)\text{Sgn}(W_3\cdot W_1),
\end{align}
matching with the result of \cite{Baumann:2024ttn}.}.
\begin{align}\label{Jppptwistor}
    \langle 0|\hat{J}^{+A}(Z_1)\hat{J}^{+B}(Z_2)\hat{J}^{+C}(Z_3)|0\rangle=f^{ABC}\text{Sgn}(Z_1\cdot Z_2)\int \frac{dc_1 dc_2}{c_1 c_2}\delta^4(c_1 Z_1+c_2 Z_2-Z_3).
\end{align}

\subsubsection{More examples also including gravity}
Similar analysis can be carried out for the remaining examples. Here, we quote the final results in twistor space which give rise to the stripped momentum space correlators.

\subsubsection*{Minimal coupling between scalars and gravitons}
For $\langle 0|T^{+}O_1O_1|0\rangle$, we obtain the positive helicity stripped Wightman function using our result \eqref{TO1O1Wightman} and the definition \eqref{threepointstripped}. The twistor space counterpart to it is obtained by setting $\alpha=2,\beta=-2,\gamma=2,k_{200}=1,c_{200}=1$, which results in,
\begin{align}
    \langle 0|\hat{T}^{+}(Z_1)\hat{O}_1(Z_2)\hat{O}_1(Z_3)|0\rangle'&=-\delta^{[2]}(Z_1\cdot Z_2)\delta^{[-2]}(Z_2\cdot Z_3)\delta^{[2]}(Z_3\cdot Z_1)\notag\\
    &+\delta^{[2]}(Z_1\cdot Z_2)\int dc_1dc_2 c_1^2 c_2^{-2}\delta^4(c_1 Z_1+c_2 Z_2-Z_3),
\end{align}
as can be verified using \eqref{inverseHalfFourier} and the result \eqref{threepointtwistortoSHgensol}. It is easy to obtain the negative helicity case similarly.

\subsubsection*{Einstein Gravity}
Finally, we move on to the example of Einstein gravity in the $(+++)$ helicity which can be found using \eqref{TTTWightman}, writing down the stripped correlator \eqref{threepointstripped} and choosing all helicities to be positive using \eqref{Jspm}. The spinor helicity and twistor results are provided in appendix \ref{app:YmandGRallhelicities}. For example we find\footnote{Using the twistor Fourier transform \eqref{twistorfouriertransform} we see that the dual twistor space expression  equals,
\begin{align}
    \langle 0|\hat{T}^{+A}(W_1)\hat{T}^{+B}(W_2)\hat{T}^{+C}(W_3)|0\rangle=|(W_1\cdot W_2)||W_2\cdot W_3||W_3\cdot W_1|,
\end{align}
which matches with the result of \cite{Baumann:2024ttn}.},
\begin{align}
    \langle 0|\hat{T}^{+}(Z_1)\hat{T}^{+}(Z_2)\hat{T}^{+}(Z_3)|0\rangle=\int \frac{dc_{12}dc_{23}dc_{31}}{c_{12}^2c_{23}^2c_{31}^2}\delta^4(c_{12}Z_3+c_{23}Z_1+c_{31}Z_2)e^{\frac{iZ_1\cdot Z_2}{c_{12}}}.
\end{align}
This concludes our discussion of three point functions in twistor space. What we have established is the fact that the two and three point twistor space correlators considered in literature so far actually correspond to stripped Wightman functions.

\subsection{Four point functions}\label{subsec:twistor4point}
Given the fact that we have obtained factorized formulae for Wightman functions when the middle two operators have spacelike momenta, it is natural to leverage our knowledge of the three point functions in twistor space to obtain the results at four points in this kinematic regime. In this section, we focus exactly on this problem. To begin with, we define the stripped four point Wightman function,
\begin{align}\label{fourpointstrippedWightman}
    &\langle 0|\mathcal{O}_1(p_1)\mathcal{O}_2(p_2)\mathcal{O}_3(p_3)\mathcal{O}_4(p_4)|0\rangle\theta(p_2^2)\theta(p_3^2)\notag\\&=\theta(-p_1^2)\theta(-p_1^0)\theta(p_2^2)\theta(-s^2)\theta(-s^0)\theta(p_3^2)\theta(-p_4^2)\theta(p_4^0)\langle 0|\mathcal{O}_1(p_1)\mathcal{O}_2(p_2)\mathcal{O}_3(p_3)\mathcal{O}_4(p_4)|0\rangle'.
\end{align}
As we showed in subsection \ref{sec:FactorizationandCPW}, this coincides with a Wightman conformal partial wave which through figure \ref{fig:WightmanToEuclid} has enough information to recover the full correlator. Thus, it is an extremely important object to construct in twistor space. 
\subsubsection{Conformally coupled scalars}
We begin with the simplest example of $\Delta=1$ scalars exchanging $\Delta=1$ scalars. In this case we have the extremely simple expression for the rescaled correlator,
\begin{align}
    \langle 0|\hat{O}_1(p_1)\hat{O}_1(p_2)\hat{O}_1(p_3)\hat{O}_1(p_4)|0\rangle'=\frac{1}{|s|}(2\pi)^3\delta^3(p_1+p_2+p_3+p_4).
\end{align}
Introducing an auxiliary integral we can write it as,
\begin{align}
    \int \frac{d^3 p}{(2\pi)^3|p|}\langle 0|\hat{O}_1(p_1)\hat{O}_1(p_2)\hat{O}_1(p)|0\rangle'\langle 0|\hat{O}_1(-p)\hat{O}_1(p_3)\hat{O}_1(p_4)|0\rangle'.
\end{align}
In this form, conversion to twistor space is straightforward!. For the integrated over momentum $p$, we re-express the measure in spinor helicity using the formula \cite{Bala:2025qxr},
\begin{align}\label{d3pinlambdalambdabar}
    \int d^3 p=\frac{1}{2\text{Vol}(GL(1,\mathbb{R}))}\int d^2 \lambda d^2\Bar{\lambda} |p|.
\end{align}
As for the operators with this momenta we use,
\begin{align}\label{O1pandO1minusp}
    &\hat{O}_1(p)=\hat{O}_1(\lambda,\Bar{\lambda})\notag\\
    &\hat{O}_1(-p)=\hat{O}_1(\lambda,-\Bar{\lambda}).
\end{align}
Essentially we describe a momenta using $(\lambda,\Bar{\lambda})$ and minus of the momenta by flipping the sign of $\Bar{\lambda}$. For the external operators we directly perform the half-Fourier transform \eqref{scalarhalffourier}. For the exchanged one, we simply write it in terms of its twistor space counterpart using \eqref{inverseHalfFourier}. Putting all these steps together we have,
\begin{align}\label{O14pointintermidiatestep}
     &\langle 0|\hat{O}_1(Z_1)\hat{O}_1(Z_2)\hat{O}_1(Z_3)\hat{O}_1(Z_4)|0\rangle\notag\\&=\frac{1}{2\text{Vol}(GL(1,\mathbb{R}))}\int\frac{d^2\lambda d^2\Bar{\lambda}d^2\Bar{\mu}d^2\Bar{\nu}}{(2\pi)^3}\langle  0|\hat{O}_1(Z_1)\hat{O}_2(Z_2)\hat{O}_1(\lambda,\Bar{\mu})|0\rangle \langle 0|\hat{O}_1(\lambda,\Bar{\nu})\hat{O}_1(Z_3)\hat{O}_1(Z_4)|0\rangle e^{i\Bar{\lambda}\cdot(\Bar{\nu}-\Bar{\mu})}\notag\\
     &=\frac{1}{2\text{Vol}(GL(1,\mathbb{R}))}\int\frac{d^2\lambda d^2\Bar{\mu}d^2\Bar{\nu}}{(2\pi)}\delta^2(\Bar{\nu}^a-\Bar{\mu}^a)\langle  0|\hat{O}_1(Z_1)\hat{O}_2(Z_2)\hat{O}_1(\lambda,\Bar{\mu})|0\rangle \langle \langle 0|\hat{O}_1(\lambda,\Bar{\nu})\hat{O}_1(Z_3)\hat{O}_1(Z_4)|0\rangle\notag\\
     &=\frac{1}{4\pi\text{Vol}(GL(1,\mathbb{R}))}\int d^2\lambda d^2\Bar{\mu}~ \langle 0|\hat{O}_1(Z_1)\hat{O}_2(Z_2)\hat{O}_1(\lambda,\Bar{\mu})|0\rangle\langle  0|\hat{O}_1(\lambda,\Bar{\mu})\hat{O}_1(Z_3)\hat{O}_1(Z_4)|0\rangle\notag\\
     &=\frac{1}{4\pi\text{Vol}(GL(1,\mathbb{R}))}\int d^4 Z~ \langle 0|\hat{O}_1(Z_1)\hat{O}_2(Z_2)\hat{O}_1(Z)|0\rangle\langle  0|\hat{O}_1(Z)\hat{O}_1(Z_3)\hat{O}_1(Z_4)|0\rangle,
\end{align}
where to go from the first to the second line, we performed the $d^2\Bar{\lambda}$ integral and in going to the third from the second, we used the resulting delta function for the $d^2\Bar{\nu}$ integral setting $\Bar{\nu}^a=\Bar{\mu}^a$. Finally, we defined the twistor space $\mathbb{RP}^3$ measure,
\begin{align}
    \frac{d^4 Z}{\text{Vol}(GL(1,\mathbb{R}))}=\frac{d^2\lambda d^2\Bar{\lambda}}{\text{Vol}(GL(1,\mathbb{R}))}~,Z^A=(\lambda^a,\Bar{\mu}_{a'}).
\end{align}
Note that this four point function is simply a product of three point functions with the common twistor integrated over all of $\mathbb{RP}^3$.
Let us now use the expressions for these three point functions \eqref{O1O1O1twistor} in the above expression. We obtain the beautiful expression\footnote{As we discussed, this correlator is the twistor space counterpart of the Wightman function in the special kinematics. It would be interesting to explore the twistor correlators corresponding to more general kinematic configurations. It would also be interesting to explore if there exist twistor space expressions corresponding to time-ordered or Euclidean correlators. One possible way would be to come up with a prescription for analytic continuation in twistor space or in Schwinger parameter space to obtain these other correlators. We thank Guilherme Pimentel for this comment.},
\begin{align}\label{O14pointtwistor}
     &\langle 0|\hat{O}_1(Z_1)\hat{O}_1(Z_2)\hat{O}_1(Z_3)\hat{O}_1(Z_4)|0\rangle\notag\\&=\frac{c_{123}^2}{8\pi\text{Vol}(GL(1,\mathbb{R}))}\delta(Z_1\cdot Z_2)\delta(Z_3\cdot Z_4)\int dc_1 dc_2 dc_3 dc_4\int d^4 Z \delta^4(c_1 Z_1+c_2 Z_2-Z)\delta^4(c_3 Z_3+c_4 Z_4+Z)\notag\\
     &=\frac{c_{123}^2}{8\pi \text{Vol}(GL(1,\mathbb{R}))}\delta(Z_1\cdot Z_2)\delta(Z_3\cdot Z_4)\int dc_1 dc_2 dc_3 dc_4 \delta^4(c_1 Z_1+c_2 Z_2+c_3 Z_3+c_4 Z_4)\notag\\
     &=\frac{c_{123}^2}{8\pi|Z_1\cdot Z_3||Z_2\cdot Z_4|}\delta(Z_1\cdot Z_2)\delta(Z_3\cdot Z_4)\delta(v-1),
\end{align}
where the conformal cross ratio $v$ is defined as,
\begin{align}\label{crossratiov}
    v=\frac{Z_1\cdot Z_4 Z_2\cdot Z_3}{Z_1\cdot Z_3 Z_2\cdot Z_4}.
\end{align}
Following our discussion in subsection \ref{sec:FactorizationandCPW}, we can identify the final expression as a twistor space Wightman conformal partial wave corresponding to the $\Delta=1$ scalar exchange:
\begin{align}
    \mathcal{W}^{(s)}_{(O_1O_1|O_1|O_1O_1)}(Z_1,Z_2,Z_3,Z_4)=\frac{1}{8\pi|Z_1\cdot Z_3||Z_2\cdot Z_4|}\delta(Z_1\cdot Z_2)\delta(Z_3\cdot Z_4)\delta(v-1).
\end{align}
Note the remarkable simplicity of this result. The expression is localized at the value of the cross ratio $v=1$.
\subsubsection{Yang Mills theory}
Let us start with the stripped momentum space correlator viz,
\begin{align}
    &\langle 0|J^{A_1}(p_1,\epsilon_1)J^{A_2}(p_2,\epsilon_2)J^{A_3}(p_3,\epsilon_3)J^{A_4}(p_4,\epsilon_4)|0\rangle'=(2\pi)^3\delta^3(p_1+p_2+p_3+p_4)\notag\\&\times\langle\langle 0|J^{A_1}(p_1,\epsilon_1)J^{A_2}(p_2,\epsilon_2)J^{A\mu}(-s)|0\rangle\rangle'\frac{\pi_{\mu\nu}(s)}{|s|}\langle\langle 0|J_{A}^{\nu}(-p_3-p_4)J^{A_3}(p_3,\epsilon_3)J^{A_4}(p_4,\epsilon_4)|0\rangle'\notag\\
    &=\int \frac{d^3 p}{(2\pi)^3|p|}\int \langle 0|J^{A_1}(p_1,\epsilon_1)J^{A_2}(p_2,\epsilon_2)J^{A\mu}(p)|0\rangle'\pi_{\mu\nu}(p)\langle 0|J_{A}^{\nu}(-p)J^{A_3}(p_3,\epsilon_3)J^{A_4}(p_4,\epsilon_4)|0\rangle'.
\end{align}
The most convinient way to deal with this expression is the helicity basis. For the exchanged current, we use the formula,
\begin{align}\label{JJprojector}
    J^{A\mu}(p)|0\rangle\pi_{\mu\nu}(p)\langle 0|J^{\nu}_A(-p)=\bigg(J^{A+}(p)|0\rangle\langle 0|J_A^{-}(-p)+J^{A-}(p)|0\rangle\langle 0| J_A^{+}(-p)\bigg).
\end{align}
For the external operators we consider the MHV configuration with $\epsilon_1,\epsilon_3$ being positive helicity polarizations and $\epsilon_2,\epsilon_4$ being negative helicity ones, using their form in \eqref{polarizations}. This converts the above expression into,
\begin{align}\label{JJJJintermidiatestep1}
    &\langle 0|J^{A_1+}(p_1)J^{A_2-}(p_2)J^{A_3+}(p_3)J^{A_4-}(p_4)|0\rangle'\notag\\&=\frac{1}{2(2\pi)^3}\int d^2\lambda d^2\Bar{\lambda}\langle 0|J^{A_1+}(p_1)J^{A_2-}(p_2)J^{A+}(p)|0\rangle'\langle 0|J_A^{-}(-p)J^{A_3+}(p_3)J^{A_4-}(p_4)|0\rangle'+(1\leftrightarrow 3,2\leftrightarrow 4).
\end{align}
where we used \eqref{d3pinlambdalambdabar} for writing the measure in spinor helicity variables. Also note that we are using the fact that these stripped Wightman functions have permutation symmetry so we can perform a $(1\leftrightarrow 3,2\leftrightarrow 4)$ exchange to obtain the contribution due to the second term in \eqref{JJprojector}. Performing a half-Fourier transform \eqref{inverseHalfFourier} for the external operators and expressing the exchanged operator following the scalar case viz \eqref{O1pandO1minusp}, \eqref{O14pointintermidiatestep}, we obtain,
\begin{align}
    &\langle 0|J^{A_1 +}(Z_1)J^{A_2 -}(Z_2)J^{A_3 +}(Z_3)J^{A_4 -}(Z_4)|0\rangle\notag\\&=\frac{1}{4\pi\text{Vol}(GL(1,\mathbb{R}))}\int d^4 Z\langle 0|J^{A_1+}(Z_1)J^{A_2 -}(Z_2)J^{A+}(Z)|0\rangle\langle J_A^{-}(Z)J^{A_3 +}(Z_3)J^{A_4 -}(Z_4)|0\rangle+(1\leftrightarrow 3,2\leftrightarrow 4).
\end{align}
Plugging in the explicit forms of these three point functions results in\footnote{One can obtain the dual twistor space expressions using \eqref{twistorfouriertransform} if desired.},
\small
\begin{align}\label{JJJJMHVTwistor1}
     &\frac{\langle 0|J^{A_1 +}(Z_1)J^{A_2 -}(Z_2)J^{A_3 +}(Z_3)J^{A_4 -}(Z_4)|0\rangle}{f^{A_1 A_2 A}f_A^{A_3 A_4}}\notag\\&=\frac{8 i^2\text{Sgn}(Z_1\cdot Z_2)\text{Sgn}(Z_3\cdot Z_4)}{4\pi\text{Vol}(GL(1,\mathbb{R}))}\int dc_3\int dc_4\frac{c_3^3}{c_4}\int d^4 Z \text{Sgn}(Z_2\cdot Z)\delta^{[3]}(Z\cdot Z_1)\delta^4(Z-c_3 Z_3-c_4 Z_4)+(1\leftrightarrow 3,2\leftrightarrow 4)\notag\\
     &=\frac{2 i^2\text{Sgn}(Z_1\cdot Z_2)\text{Sgn}(Z_3\cdot Z_4)}{\pi\text{Vol}(GL(1,\mathbb{R}))}\int dc_3 \int dc_4 \frac{c_3^3}{c_4}\text{Sgn}(c_3 Z_2\cdot Z_3+c_4 Z_2\cdot Z_4)\delta^{[3]}(c_3 Z_3\cdot Z_1+c_4 Z_4\cdot Z_1)+(1\leftrightarrow 3,2\leftrightarrow 4)\notag\\
     &=\frac{4i^4\text{Sgn}(Z_1\cdot Z_2)\text{Sgn}(Z_3\cdot Z_4)}{\pi\text{Vol}(GL(1,\mathbb{R}))}\int dc_1 dc_2 dc_3 dc_4 \frac{c_1^3 c_3^3}{c_2 c_4}e^{-i c_2 c_3 Z_2\cdot Z_3-i c_2 c_4 Z_2\cdot Z_4}e^{i c_1 c_3 Z_1\cdot Z_3+i c_1 c_4 Z_1\cdot Z_4}+(1\leftrightarrow 3,2\leftrightarrow 4)\notag\\
     &=\frac{4i^4\text{Sgn}(Z_1\cdot Z_2)\text{Sgn}(Z_3\cdot Z_4)}{\pi\text{Vol}(GL(1,\mathbb{R}))}\int dc_1 dc_2 dc_3 dc_4 \frac{c_1^3 c_3^3}{c_2 c_4}e^{ic_1c_4Z_1\cdot Z_4-i c_2 c_3 Z_2\cdot Z_3}\cos\big(c_1 c_3 Z_1\cdot Z_3- c_2 c_4 Z_2\cdot Z_4\big)\notag\\
     &=\frac{16}{\pi}\text{Sgn}(Z_1\cdot Z_2)\text{Sgn}(Z_2\cdot Z_3)\text{Sgn}(Z_3\cdot Z_4)\text{Sgn}(Z_4\cdot Z_1)\frac{1}{|Z_1\cdot Z_3|^4}\frac{d^3}{dv^3}(v^3\text{Sign}(1-\frac{1}{v})),
\end{align}
\normalsize
which is a simple and pleasing result. $v$ is the conformal cross ratio \eqref{crossratiov}. Although it is not obvious from the above, we note that the $(1\leftrightarrow 3,2\leftrightarrow 4)$ exchange term simply contributed an extra factor of $2$. The fact that the two terms are equal is not obvious in spinor helicity variables due to the presence of degeneracies and Schouten identities\footnote{These arise due to the fact that spinor helicity variables deal with two component spinors out of which only two can be linearly independent. The same could be said for twistor space where the variables are $\lambda_{a}$ and $\Bar{\mu}_a$ for each operator out of which only two of them are linearly independent. However, the results for conserved currents and $\Delta=1$ scalars reorganize themselves in terms of $Z^A$ as can be seen by the structure of the conformal generators \eqref{TAB}. In these variables, there are no degeneracies at four points since $Z^A$ is itself a four component object.}. Given the fact that the two terms are identical in twistor space, we can now go back to spinor helicity variables to verify it. We
have checked numerically that,
\begin{align}
    &\langle 0|J^{A_1+}(p_1)J^{A_2-}(p_2)J^{A+}(-p_1-p_2)|0\rangle'\frac{1}{|s|}\langle 0|J_A^{-}(p_1+p_2)J^{A_3+}(p_3)J^{A_4-}(p_4)|0\rangle'\notag\\
    &=\langle 0|J^{A_1+}(p_1)J^{A_2-}(p_2)J_A^{-}(-p_1-p_2)|0\rangle'\frac{1}{|s|}\langle 0|J^{A+}(p_1+p_2)J^{A_3+}(p_3)J^{A_4-}(p_4)|0\rangle'.
\end{align}
Therefore, the answer in spinor helicity variables \eqref{JJJJintermidiatestep1} is simply,
\begin{align}\label{YMMHVSH1}
     &\langle 0|J^{A_1+}(p_1)J^{A_2-}(p_2)J^{A_3+}(p_3)J^{A_4-}(p_4)|0\rangle\notag\\&=\frac{1}{(2\pi)^3}\int d^2\lambda d^2\Bar{\lambda}\langle 0|J^{A_1+}(p_1)J^{A_2-}(p_2)J^{A+}(p)|0\rangle\langle 0|J_A^{-}(-p)J^{A_3+}(p_3)J^{A_4-}(p_4)|0\rangle\notag\\
     &=\frac{4}{|s|}\langle 0|J^{A_1+}(p_1)J^{A_2-}(p_2)J^{A_3+}(-p_1-p_2)|0\rangle'\langle 0|J_A^{-}(-p_3-p_4)J^{A_3+}(p_3)J^{A_4-}(p_4)|0\rangle'
\end{align}
Twistor space expressions on the other hand, are much more easier to manipulate and handle at four points.

Finally, we note that the  pre-factor in \eqref{JJJJMHVTwistor1} is 
\begin{align}
    \text{Sgn}(Z_1\cdot Z_2)\text{Sgn}(Z_2\cdot Z_3)\text{Sgn}(Z_3\cdot Z_4)\text{Sgn}(Z_4\cdot Z_1),
\end{align} 
which is identical in form to the flat space gluon MHV amplitude \cite{Arkani-Hamed:2013jha}! 
Similar to the scalar case, we identify this twistor space expression \eqref{JJJJMHVTwistor1} with the Wightman CPW corresponding to the exchange of a spin-1 current:
\small
\begin{align}
    \mathcal{W}^{(s)}_{(JJ|J|JJ)}(Z_1,Z_2,Z_3,Z_4)=\frac{16}{\pi}\text{Sgn}(Z_1\cdot Z_2)\text{Sgn}(Z_2\cdot Z_3)\text{Sgn}(Z_3\cdot Z_4)\text{Sgn}(Z_4\cdot Z_1)\frac{1}{|Z_1\cdot Z_3|^4}\frac{d^3}{dv^3}(v^3\text{Sign}(1-\frac{1}{v})).
\end{align}
\normalsize
We proceed to the case of  Einstein gravity next.
\subsubsection{Einstein Gravity}
Moving on to Einstein gravity, the stripped four point function in the MHV configuration is given by,
\begin{align}\label{TTTTGRexp1MHV}
&\langle 0|T^{+}(p_1)T^{-}(p_2)T^{+}(p_3)T^{-}(p_4)|0\rangle'\notag\\&=\int \frac{d^3 p}{(2\pi)^3|p|}\langle 0|T^{+}(p_1)T^{-}(p_2)T^{\mu\nu}(p)|0\rangle'\Pi_{\mu\nu\rho\sigma}(p)\langle 0|T^{\rho\sigma}(-p)T^{+}(p_3)T^{-}(p_4)|0\rangle'.
\end{align}
We then analytically continuing all momenta to be space-like and use the helicity basis for the exchanged operator \eqref{JsintermsofJspm}. Performing a half-Fourier transform for the external operators \eqref{halfFourier} and expressing the exchanged ones using the inverse half-Fourier transform \eqref{inverseHalfFourier} and performing the same sequence of steps as we did for the previous examples results in,
\begin{align}
    &\langle 0|T^{+}(Z_1)T^{-}(Z_2)T^{+}(Z_3)T^{-}(Z_4)|0\rangle\notag\\&=\frac{1}{4\pi\text{Vol}(GL(1,\mathbb{R}))}\int d^4 Z\langle 0|T^{+}(Z_1)T^{-}(Z_2)T^{+}(Z)|0\rangle\langle T^{-}(Z)T^{+}(Z_3)T^{-}(Z_4)|0\rangle+(1\leftrightarrow 3,2\leftrightarrow 4).
\end{align}
Using the expressions for the three point functions and performing the $d^4 Z$ integral results in,
\begin{align}\label{TTTTMHVTwistor1}
    &\langle 0|T^{+}(Z_1)T^{-}(Z_2)T^{+}(Z_3)T^{-}(Z_4)|0\rangle\notag\\&=\frac{4}{\pi\text{Vol}(GL(1,\mathbb{R}))}|Z_1\cdot Z_2||Z_3\cdot Z_4|\int dc_1dc_2dc_3dc_4 \frac{c_1^6 c_3^6}{c_2^2 c_4^2}e^{i c_1 c_4 Z_1\cdot Z_4-i c_2 c_3 Z_2\cdot Z_3}\cos\big(c_1 c_3 Z_1\cdot Z_3-c_2 c_4 Z_2\cdot Z_4\big)\notag\\
    &=32|Z_1\cdot Z_2||Z_2\cdot Z_3||Z_3\cdot Z_4||Z_4\cdot Z_1|\frac{1}{|Z_1\cdot Z_3|^8}\frac{1}{v}\frac{d^6}{dv^6}\bigg(v^7|1-\frac{1}{v}|\bigg).
\end{align}
Very interestingly, the pre-factor $|Z_1\cdot Z_2||Z_2\cdot Z_3||Z_3\cdot Z_4||Z_4\cdot Z_1|$ is exactly of the same form as the four graviton scattering amplitude in flat space \cite{Arkani-Hamed:2013jha} exactly like the gluon case.  Similar to the previous examples, we can interpret this result as the Wightman CPW corresponding to the exchange of the stress tensor,
\begin{align}
    \mathcal{W}^{(s)}_{(TT|T|TT)}(Z_1,Z_2,Z_3,Z_4)=32|Z_1\cdot Z_2||Z_2\cdot Z_3||Z_3\cdot Z_4||Z_4\cdot Z_1|\frac{1}{|Z_1\cdot Z_3|^8}\frac{1}{v}\frac{d^6}{dv^6}\bigg(v^7|1-\frac{1}{v}|\bigg).
\end{align}

Also note the similarity of \eqref{TTTTMHVTwistor1} to the gluon MHV result \eqref{JJJJMHVTwistor1}. In fact, that $(1\leftrightarrow 3),(2\leftrightarrow 4)$ term just contributes an extra factor of $2$ just like the gluon case. This implies that even in spinor helicity variables where there appear to be two independent terms \eqref{TTTTGRexp1MHV} as we see using \eqref{JsintermsofJspm}, they are actually equal! This is not obvious to show directly in spinor helicity due to the degeneracies but the twistor space analysis makes it clear! Thus we have in spinor helicity,
\small
\begin{align}\label{GRMHVSH1}
    \langle\langle 0|T^{+}(p_1)T^{-}(p_2)T^{+}(p_3)T^{-}(p_4)|0\rangle\rangle'=\frac{4}{|s|^3}\langle\langle 0|T^{+}(p_1)T^{-}(p_2)T^{+}(-p_1-p_2)|0\rangle\rangle'\langle\langle 0|T^{-}(-p_3-p_4)T^{+}(p_3)T^{-}(p_4)|0\rangle\rangle',
\end{align}
\normalsize
which one can verify numerically. It is however more satisfying that twistor space on the other hand has given us simple analytic proof that they are equal.
This indicates a double copy with the gluon Wightman function which we will make this much more concrete in section \ref{sec:DoubleCopy}.

This concludes our discussion of correlators in twistor space. One can similarly convert our other results such as for Bhabha and Compton scattering to twistor space if desired. In short, what we have seen in this section is that twistor space yields extremely simple and elegant expressions for correlators, whose conformal invariance is manifest. We leave a more comprehensive analysis to the future.

\subsection{Higher point functions}\label{subsec:twistorfivepoint}
In this brief subsection, we convert our result for the $\Delta=1$ scalar five point function to twistor space. We have the factorized expression \eqref{fivepointscalarDelta1}. After rescaling it for suitablitiy to twistor space \eqref{inverseHalfFourier}, we can re-write the stripped five point Wightman function as,
\begin{align}
    &\langle 0|\hat{O}_1(p_1)\hat{O}_1(p_2)\hat{O}_1(p_3)\hat{O}_1(p_4)\hat{O}_1(p_5)|0\rangle'=\frac{1}{|s_{12|}|s_{123}|}(2\pi)^3\delta^3(p_1+p_2+p_3+p_4+p_5)\notag\\
    &=\int\frac{d^3 p}{(2\pi)^3|p|}\frac{d^3 q}{(2\pi)^3|q|}\langle 0|\hat{O}_1(p_1)\hat{O}_1(p_2)\hat{O}_1(p)|0\rangle'\langle 0|\hat{O}_1(-p)\hat{O}_1(p_3)\hat{O}_1(q)|0\rangle'\langle 0|\hat{O}_1(-q)\hat{O}_1(p_4)\hat{O}_1(p_5)|0\rangle'.
\end{align}
Performing steps analogous to the above four point examples to convert this into twistor space, we obtain the simple result,
\small
\begin{align}
    &\langle 0|\hat{O}_1(Z_1)\hat{O}_1(Z_2)\hat{O}_1(Z_3)\hat{O}_1(Z_4)\hat{O}_1(Z_5)|0\rangle\notag\\&=\frac{1}{(4\pi)^2}\delta(Z_1\cdot Z_2)\delta(Z_4\cdot Z_5)\int \frac{dc_1dc_2dc_3dc_4dc_5}{\text{Vol}(GL(1,\mathbb{R}))}\delta(c_1 Z_1\cdot Z_3+c_2 Z_2\cdot Z_3)\delta^4(c_1 Z_1+c_2 Z_2+c_3 Z_3+c_4 Z_4+c_5 Z_5)\notag\\
    &=\frac{1}{4\pi^2}\frac{\delta(Z_1\cdot Z_2)\delta(Z_4\cdot Z_5)}{|Z_1\cdot Z_3||Z_2\cdot Z_4||Z_3\cdot Z_5|}\delta(u_{14,23}-1-u_{34,25}(u_{15,23}-1)),
\end{align}
\normalsize
where the five-point cross ratios are given by,
\begin{align}
    u_{ij,kl}=\frac{(Z_i\cdot Z_j)(Z_k\cdot Z_l)}{(Z_i\cdot Z_l)(Z_k\cdot Z_j)}.
\end{align}
We leave a systematic and comprehensive analysis of higher point functions in twistor space for a future work.

\section{Double Copy}\label{sec:DoubleCopy}
Broadly speaking, the double copy in the context of scattering amplitudes states that graviton amplitudes can (roughly) be obtained by squaring gauge theory amplitudes. Since the initial works, a lot of progress has been made including double copy at loop level, double copy for classical solutions in gauge theory and gravity, double copy involving many other theories, see \cite{Adamo:2022dcm} and citations thereof for a review of the developments. Given the fact that gravity is generally much harder to study than gauge theory, it is desirable to want to further understand and utilize this phenomena.

In this section, we study the double copy in the context of Yang-Mills theory and Einstein gravity. First, we discuss three point functions and show a double copy between stripped gluon and graviton Wightman functions in momentum space, spinor helicity and twistor variables. For four point functions, we stick to twistor variables since momentum space and spinor helicity approaches become more complicated due to the presence of too many degenerate tensor structures. We derive an extremely simple double copy relation between the correlators in Yang-Mills theory and Einstein gravity obtainaing the latter as a simple square of the former. This correlator corresponds in momentum space to the stripped factorized Wightman function.

\subsection{Three point functions}
 The Einstein gravity three point scattering amplitude is the square of its colour stripped Yang-Mills counterpart.
\begin{align}
    A_{3,GR}=A_{YM}^2=\bigg((\epsilon_1\cdot \epsilon_2)(\epsilon_3\cdot p_1)+(\epsilon_2\cdot \epsilon_3)(\epsilon_1\cdot p_3)+(\epsilon_3\cdot \epsilon_1)(\epsilon_2\cdot p_3)\bigg)^2.
\end{align}
For three point AdS/CFT correlators, double copy structures have been found first in the Euclidean context \cite{Farrow:2018yni,Jain:2021qcl} and more recently in the  Lorentzian setting \cite{Baumann:2024ttn}. Let us check this property for our Wightman functions. We focus on the case where the middle operator has spacelike momenta. We begin by removing the colour factor and spectral theta functions and squaring our result for $\langle JJJ\rangle$ \eqref{JJJWightman} yielding,
\begin{align}
    (\langle\langle 0|JJJ|0\rangle\rangle'_{YM})^2\sim V_{3,YM}^2\frac{p_1^2p_2^2p_3^2}{E^2(E-2p_1)^2(E-2p_2)^2(E-2p_3)^2}.
\end{align}
Comparing with the expression for the graviton three point Wightman function \eqref{TTTWightman} shows that,
\begin{align}
     \langle\langle 0|TTT|0\rangle\rangle'_{GR}\sim p_1 p_2 p_3  (\langle\langle 0|JJJ|0\rangle\rangle'_{YM})^2,
\end{align}
As a consequence of this relation, one can also check in spinor helicity variables that the double copy holds in every helicity configuration. The expressions for these eight helicity configurations are provided in appendix \ref{app:YmandGRallhelicities}. Similarly, one can see using the twistor space expressions in appendix \ref{app:YmandGRallhelicities}, the double copy in twistor space corresponds to squaring the Schwinger parameters when the correlator is written in said parametrization. For concreteness, consider the $(+++)$ helicity configuration. We have,
\begin{align}\label{JJJ3pttwistor1}
    &\langle 0|J^{+A}(Z_1)J^{+B}(Z_2)J^{+C}(Z_3)|0\rangle'=-4 f^{ABC}\delta^{[3]}(Z_1\cdot Z_2)\int \frac{dc_1}{c_1}\int \frac{dc_2}{c_2}\delta^4(c_1 Z_1+c_2 Z_2-Z_3)\notag\\
    &=-4i f^{ABC}\int dc_1 dc_2 dc_{12} \frac{c_{12}^3}{c_1 c_2}e^{-ic_{12}Z_1\cdot Z_2}\delta^4(c_1 Z_1+c_2 Z_2-Z_3).
\end{align}
Similarly, for gravity we have,
\begin{align}\label{TTT3pointtwistor1}
    &\langle 0|T^{+}(Z_1)T^{+}(Z_2)T^{+}(Z_3)|0\rangle=-4\delta^{[6]}(Z_1\cdot Z_2)\int \frac{dc_1}{c_1^2}\int \frac{dc_2}{c_2^2}\delta^4(c_1 Z_1+c_2 Z_2-Z_3)\notag\\
    &=4 \int dc_1 dc_2 dc_{12}\bigg(\frac{c_{12}^3}{c_1 c_2}\bigg)^2e^{-ic_{12}Z_1\cdot Z_2}\delta^4(c_1 Z_1+c_2 Z_2-Z_3),
\end{align}
thus showing that the squaring the Schwinger parameter function $\frac{c_{12}^3}{c_1 c_2}$ in the gluon correlator \eqref{JJJ3pttwistor1} and stripping off the colour factor results in \eqref{TTT3pointtwistor1}. Similarly, one can check in every helicity configuration using the expressions provided in appendix \ref{app:YmandGRallhelicities} that this is a general feature.

\subsection{Four point functions}
Consider the expressions for our MHV four point functions in twistor space viz \eqref{JJJJMHVTwistor1} and \eqref{TTTTMHVTwistor1}. Let us re-write these expressions also expressing the pre-factors in Schwinger parametrization. For the gluon case we have,
\small
\begin{align}\label{JJJJYMtwistor1exp}
    &\frac{\langle 0|J^{A_1 +}(Z_1)J^{A_2 -}(Z_2)J^{A_3 +}(Z_3)J^{A_4 -}(Z_4)|0\rangle}{f^{A_1 A_2 A}f_A^{A_3 A_4}}\notag\\
    &=\int \frac{-4dc_1dc_2dc_3dc_4 dc_{12}dc_{34}}{\pi\text{Vol}(GL(1,\mathbb{R}))}\frac{c_1^3 c_3^3}{c_{12}c_{34}c_2 c_4}e^{ic_1c_4Z_1\cdot Z_4-i c_2 c_3 Z_2\cdot Z_3}e^{-ic_{12}Z_1\cdot Z_2-ic_{34}Z_3\cdot Z_4}\cos\big(c_1 c_3 Z_1\cdot Z_3- c_2 c_4 Z_2\cdot Z_4\big),
\end{align}
\normalsize
We define the Schwinger parameter Yang-Mills correlator viz,
\begin{align}
    \mathcal{M}_{4,YM}(c_1,c_2,c_3,c_4,c_{12},c_{34})=\frac{c_1^3 c_3^3}{c_{12}c_{34}c_2 c_4}.
\end{align}
Let us now do the same for the graviton case. We have,
\small
\begin{align}\label{TTTTYMtwistor1exp}
     &\langle 0|T^{+}(Z_1)T^{-}(Z_2)T^{+}(Z_3)T^{-}(Z_4)|0\rangle\notag\\&=\int\frac{4dc_1 dc_2dc_3dc_4dc_{12}dc_{34}}{\pi\text{Vol}(GL(1,\mathbb{R}))}\bigg(\frac{c_1^3c_3^3}{c_{12}c_{34}c_2 c_4}\bigg)^2e^{ic_1c_4Z_1\cdot Z_4-i c_2 c_3 Z_2\cdot Z_3}e^{-ic_{12}Z_1\cdot Z_2-ic_{34}Z_3\cdot Z_4}\cos\big(c_1 c_3 Z_1\cdot Z_3- c_2 c_4 Z_2\cdot Z_4\big).
\end{align}
\normalsize
The graviton Schwinger parameter correlator is thus,
\begin{align}\label{maindoublecopy}
    \mathcal{M}_{4,GR}(c_1,c_2,c_3,c_4,c_{12},c_{34})=\bigg(\frac{c_1^3 c_3^3}{c_{12}c_{34}c_2 c_4}\bigg)^2=\mathcal{M}_{4,YM}^2,
\end{align}
which is simply the square of its Yang-Mills counterpart! Using the helicity flipping operator of \cite{Baumann:2024ttn}, one can induce the double copy in every other helicity configuration. For example, consider the $(+++-)$ helicity. Our answers for the $(+-+-)$ helicity are expressed in the variables $(Z_1,Z_2,Z_3,Z_4)$. To $(+++-)$, we simply replace $Z_2$ by $W_2$ which converts $J^{A_2 -}(Z_2)$ to $J^{A_2 +}(W_2)$. The expressions for gluons \eqref{JJJJYMtwistor1exp} and gravitons \eqref{TTTTYMtwistor1exp} becomes,
\small
\begin{align}
    &\frac{\langle 0|J^{A_1 +}(Z_1)J^{A_2 +}(W_2)J^{A_3 +}(Z_3)J^{A_4 -}(Z_4)|0\rangle}{f^{A_1 A_2 A}f_A^{A_3 A_4}}\notag\\
    &=\int \frac{-4dc_1dc_2dc_3dc_4 dc_{12}dc_{34}}{\pi\text{Vol}(GL(1,\mathbb{R}))}\frac{c_1^3 c_3^3}{c_{12}c_{34}c_2 c_4}e^{ic_1c_4Z_1\cdot Z_4-i c_2 c_3 W_2\cdot Z_3}e^{-ic_{12}Z_1\cdot W_2-ic_{34}Z_3\cdot Z_4}\cos\big(c_1 c_3 Z_1\cdot Z_3- c_2 c_4 W_2\cdot Z_4\big),\notag\\
      &\langle 0|T^{+}(Z_1)T^{+}(W_2)T^{+}(Z_3)T^{-}(Z_4)|0\rangle\notag\\&=\int\frac{4dc_1 dc_2dc_3dc_4dc_{12}dc_{34}}{\pi\text{Vol}(GL(1,\mathbb{R}))}\bigg(\frac{c_1^3c_3^3}{c_{12}c_{34}c_2 c_4}\bigg)^2e^{ic_1c_4Z_1\cdot Z_4-i c_2 c_3 W_2\cdot Z_3}e^{-ic_{12}Z_1\cdot W_2-ic_{34}Z_3\cdot Z_4}\cos\big(c_1 c_3 Z_1\cdot Z_3- c_2 c_4 W_2\cdot Z_4\big),
\end{align}
\normalsize
and similarly for other helicity configurations. Thus, the double copy is present in general.
This analysis really shows us how twistor space can shed light on structures that are obscured in momentum space and spinor helicity variables. Given the fact that our twistor analysis unveiled that the spinor helicity expressions correspond to \eqref{YMMHVSH1} and \eqref{GRMHVSH1}, we obtain the following induced double copy in spinor helicity variables:
\begin{align}
    \langle 0|T^{+}(p_1)T^{-}(p_2)T^{+}(p_3)T^{-}(p_4)|0\rangle'=|s| p_1 p_2 p_3 p_4 (\langle 0|J^{A_1+}(p_1)J^{A_2-}(p_2)J^{A_3+}(p_3)J^{A_4-}(p_4)|0\rangle')^2,
\end{align}
highlighting the utility of twistor space to derive such formulae\footnote{It would be interesting to compare our results with other double copy constructions, such as for on-shell correlators \cite{Cheung:2022pdk}, or using a differential representation for AdS boundary correlators such as in \cite{Herderschee:2022ntr}.}.
\subsection{From the twistor double copy to the full graviton correlator}
We have see that twistor space provides a pleasing and simple double copy relation. Given this fact, we can easily obtain a graviton four point Wightman function from its gluon counterpart using \eqref{maindoublecopy}. These are of course, expressions that correspond to the factorized results in momentum space \eqref{YMCPW} and \eqref{GRCPW} where the middle two operators have space-like momenta. However, we have shown that there is a straightforward and simple way to obtain the Euclidean AdS correlators from this factorized expression in section \ref{sec:WightmanToEuclid}. We emphasize yet again that there is not a single nested bulk integral in sight in this process and the process is straightforward.
\section{Summary and Discussion}\label{sec:Discussion}
In this paper, we have set the stage for the analysis of higher point Wightman functions in twistor space. We began our discussion by setting up the computation of holographic Wightman functions in the context of AdS$_4$. We systematically computed many examples of two, three and four point Wightman functions involving general scalars, photons, gluons and gravitons and also five point scalar correlators.
Interestingly, we found that the four point functions factorize into a product of three point functions in special kinematics when the momenta of the middle operators are taken to be space-like. This expression is a conformal partial wave associated to the operator dual to the particle exchanged in the bulk. Converting the resulting expressions into spinor helicity variables and performing a half-Fourier transform to twistor space yielded results for two and three point functions consistent with previous works. We then proceeded to the case of four point functions in twistor space associated to the factorized expressions in momentum space. Utilizing the helicity basis, we found extremely simple and compact results. In particular, we discovered a simple double copy relation between gluon and graviton correlators with the latter simply being a square of the former in Schwinger parameter space. Along the way, we discussed the analytic continuation from our factorized Lorentzian Wightman functions to Euclidean AdS showing that even the simple factorized expression, contains enough information to recover the Euclidean correlator up to contact diagram contributions.

There are a number of interesting future works that this paper opens the door towards:
\subsection*{Twistor conformal bootstrap}
One of our goals is to set up the conformal bootstrap program in twistor space. Given our holographic results, it seems promising to pursue this program. General four point functions receive contributions from intermediate states involving scalars and non-conserved spinning operators and thus it is important to compute three point functions involving at least one insertion of these operators. The twistor framework was extended in \cite{Bala:2025qxr} to accommodate generic operators. However, it remains an open problem to bootstrap their three point functions. Once that is done, the stage will be set for the bootstrap.  

\subsection*{Generalization to higher dimensions}
The machinery to compute Wightman functions in this paper cam easily extended to any spacetime dimension. The twistor space framework on the other hand, is quite dimension dependent \cite{Adamo:2017qyl}. An AdS$_5$ twistor formulation was presented in \cite{Adamo:2016rtr}. It would be interesting to develop the twistor space machinery for 4d CFT and connect it to the formalism of \cite{Adamo:2016rtr}.

\subsection*{Black hole backgrounds}
Given the simplicity of our algebraic approach using the equation of motion to compute real-time Wightman functions, a natural question is to extend the formalism of this paper to AdS black hole and more general backgrounds. From the CFT perspective, these correspond to real time correlators and as such are an important and interesting direction to pursue.

\subsection*{General kinematics and other correlators}
In this paper, we have mostly focused on four point spinning correlators in the special kinematics where the middle two operators have space-like momenta. Although these kinematics are quite powerful and even allow us to construct the Euclidean correlator which has in principle information about all possible Wightman functions and their kinematics, it is still a different matter to explicitly be able to reach these kinematics via analytic continuation. That is also of course an important problem to pursue but so is the computation of spinning Wightman functions in general kinematics much like what we did for the scalar case \eqref{phi3fourpointgendelta}.
\subsection*{Higher point functions}
The machinery developed in this paper can be generalized to compute higher point correlators as we also exemplified with the computation of a scalar five point function. Given the simplicity of the twistor answers, a possible approach is to intrinsically setup our formalism in twistor space from the get go which might lead to an even more systematic approach.

\subsection*{Recursion relations}
Recursion relations such as BCFW \cite{Britto:2005fq}, and BG \cite{Berends:1987me} for scattering amplitudes paved the way for many developments in field theory. BCFW recursion has also been developed in twistor space where they take their most natural and simplified form \cite{Arkani-Hamed:2009hub,Mason:2009sa}. In AdS, recursion relations have also been developed for time-ordered or Euclidean boundary correlators \cite{Raju:2010by,Raju:2011mp,Raju:2012zr,Raju:2012zs}. However, they are yet to reach the simplicity of their counterparts in scattering amplitudes, where at tree-level, one can simply stitch together three point amplitudes to obtain higher point ones. What we would like to emphasize yet again is that Wightman functions are simpler quantities than other correlators as we have seen in this paper. They satisfy factorization formulae and also do not possess contact terms in current conservation Ward-Takahashi identities. Thus, developing recursion relations for Wightman functions might be an interesting direction to pursue.

\subsection*{Extension to loops}
It would be interesting to extend our formalism for computing Wightman functions to one-loop and beyond, including understanding the analytic continuation at loop level that relates Wightman functions to Euclidean counterparts.

\acknowledgments
We would like to thank Guilherme Pimentel, Allic Sivaramakrishnan, Kostas Skenderis for useful comments on the draft. We also thank Mohd Ali, Aswini Bala, Nipun Bhave, Deep Mazumdar, Saurabh Pant for useful discussions. D K.S. would also like to thank Suneeta Vardarajan for an excellent lecture series on AdS/CFT which proved valuable during the preparation of this work. AA acknowledges the support from a Senior Research Fellowship, granted by the Human Resource Development Group, Council of Scientific and Industrial Research, Government of India. We would especially like to acknowledge our debt to the people of India for their steady support of research in
basic sciences.

\appendix

\section{Notation}\label{app:Notation}
Our arena is the Poincare patch of the AdS$_4$ spacetime with metric,
\begin{align}
    ds^2=\frac{dz^2+\eta_{\mu\nu}dx^\mu dx^\nu}{z^2},
\end{align}
where $\eta_{\mu\nu}=\text{diag}(-1,1,1)$ is the three dimensional Minkowski metric. Greek indices $\mu,\nu,\rho,\cdots$ run over the boundary coordinates $t,x,y$. The $z$ coordinate runs from $0$ to $\infty$. We denote the time component of a boundary vector $p^\mu$ as $p^0$. A vector $p^\mu$ is time-like if it satisfies,
\begin{align}
   (\text{time-like})~ p^\mu p^\nu \eta_{\mu\nu}=p^2<0,
\end{align}
whereas if it is space-like satisfies,,
\begin{align}
   (\text{space-like})~ p^\mu p^\nu \eta_{\mu\nu}=p^2>0.
\end{align}
Dot products between 3-vectors $p_1,p_2$ are denoted by,
\begin{align}
    p_1\cdot p_2=p_{1}^{\mu}p_2^{\nu}\eta_{\mu\nu}.
\end{align}
We also work with $SL(2,\mathbb{R})$ spinor helicity variables. Small latin alphabets $a,b,c,\cdots$ denote spinor indices. A vector index can be traded for a pair of spinors using the Lorentzian Pauli matrices.
\begin{align}
    p^\mu\to p_{a}^{b}=\frac{(\sigma^\mu)_{a}^{b}\lambda^a\Bar{\lambda}_b}{2},
\end{align}
with the slightly unconventional choice,
\begin{align}
    (\sigma^x)_a^b=\begin{pmatrix}
        0&1\\
        1&0
    \end{pmatrix},(\sigma^y)_a^b=\begin{pmatrix}
        1&0\\
        0&-1
    \end{pmatrix},(\sigma^t)_a^b=\begin{pmatrix}
        0&-1\\
        1&0
    \end{pmatrix}.
\end{align}
As for twistor variables we denote their Sp$(4;\mathbb{R})$ fundamental indices with capitalized latin alphabets $A,B,C,\cdots$. 

We deal with many types of correlators in this paper distinguished by extra brackets and primes. For the convenience of the reader we list them all here,
\begin{align}
&\langle \mathcal{O}_1(p_1)\cdots \mathcal{O}_4(p_4)\rangle\text{(Euclidean correlator)}\notag\\&\langle\langle \mathcal{O}_1(p_1)\cdots \mathcal{O}_4(p_4)\rangle\rangle\text{(Euclidean correlator with}~(2\pi)^3\delta^3(p_1+\cdots+p_4)~\text{stripped off})\notag\\&\langle 0|\mathcal{O}_1(p_1)\cdots\mathcal{O}_4(p_4)|0\rangle\text{( Wightman function)}\notag\\
    &\langle\langle 0|\mathcal{O}_1(p_1)\cdots\mathcal{O}_4(p_4)|0\rangle\rangle\text{(Wightman function with}~(2\pi)^3\delta^3(p_1+\cdots+p_4)~\text{stripped off)}\notag\\
    &\langle 0|\mathcal{O}_1(p_1)\cdots\mathcal{O}_4(p_4)|0\rangle'\text{(Wightman function with spectral theta functions stripped off and}~p_2^2>0,p_3^2>0)\notag\\
    &\langle 0|\mathcal{O}_1(Z_1)\cdots \mathcal{O}_4(Z_4)|0\rangle~(\text{Twistor space function corresponding to primed Wightman function}).
\end{align}
Essentially, double brackets $\langle\langle\cdots \rangle\rangle$ indicate the momentum conserving delta function has been stripped off. A primed quantity $\langle 0|\cdots|0\rangle'$ represents a Wightman function with the spectral theta functions stripped off in the kinematic regime $p_2^2>0,p_3^2>0$. Finally, all the twistor space correlators we consider correspond to the half-Fourier transform \eqref{halfFourier} of these primed Wightman functions.

A useful Schwinger parametrization that we consider is,
\begin{align}
    \delta^{[n]}(x)=\int_{-\infty}^{\infty}\frac{dc}{2\pi}(-i c^n) e^{-i c x}.
\end{align}
Wherever Schwinger parameter integrals appear, they run from $-\infty$ to $+\infty$. 

\section{Conformally coupled scalars}\label{app:CCscalarsReview}
In this appendix, we discuss the quadratic action and construction of Wightman and Feynman propagators for conformally coupled scalar fields, carefully discussing the $i\epsilon$ prescriptions. We consider both Dirichlet and Neumann boundary conditions as the latter is used in our twistor space analysis. 

The free action for a massless non-minimally coupled scalar field is given by,
\begin{align}\label{ActionForScalarnonminimal1}
    S_{KG}=\int \frac{dz d^3 x}{z^4}\sqrt{-g}\bigg(-\frac{1}{2}g^{AB}\partial_{A}\Phi \partial_{B}\Phi-\frac{\xi}{2} R \Phi^2\bigg),
\end{align}
where $R=-3(3+1)=-12$ in units where the $AdS_4$ radius is set to unity. The AdS metric is $g^{AB}$ For the special value $\xi=\frac{1}{6}$, it is easy to show that the quadratic part of the above action can be mapped to one in half of flat space under the following Weyl transformations:
\begin{align}\label{scalarWeyltransform1}
    \Phi(z,x)=z \phi(z,x).
\end{align}
The resulting action for $\phi$ is simply,
\begin{align}
    S_{KG}=\int dz d^3 x~ \bigg(-\frac{1}{2}\eta^{AB}\partial_A \phi \partial_B \phi\bigg).
\end{align}
The free equation of motion is,
\begin{align}
    (\partial_z^2+\Box)\phi(z,x)=0.
\end{align}
The Feynman propagator is a Green's function of the Laplacian $(\partial_z^2+\Box)$ which appears above. In particular, it satisfies,
\begin{align}\label{scalarFeynmanbulktobulk2}
    (\partial_z^2+\Box)G_F(z,z',x-x')=-i\delta(z-z)\delta^3(x-x').
\end{align}
The solution to this PDE with the usual Feynman boundary conditions enforced by an $i\epsilon$ prescription comes in two forms depending on whether we impose Neumann or Dirichlet boundary conditions for the field at $z=0$. These respectively correspond to,
\begin{align}
    &G_{F,\Delta=1}(z,z',x-x')=\frac{1}{4\pi^2}\bigg(\frac{1}{(z-z')^2+(x-x')^2+i\epsilon}+\frac{1}{(z+z')^2+(x-x')^2+i\epsilon}\bigg),\notag\\
    &G_{F,\Delta=2}(z,z',x-x')=\frac{1}{4\pi^2}\bigg(\frac{1}{(z-z')^2+(x-x')^2+i\epsilon}-\frac{1}{(z+z')^2+(x-x')^2+i\epsilon}\bigg),
\end{align}
where $\epsilon>0$, is an infinitesimal quantity, $\Delta=1$ and $\Delta=2$ denote the scaling dimension of the conformal operator dual to the bulk field $\phi$. This can be understood from the fact that for $\xi=\frac{1}{6}$, the action \eqref{ActionForScalarnonminimal1} effectively is that of a scalar with $m^2=-2$ and via the AdS$_4$/CFT$_3$ relation,
\begin{align}
    m^2=\Delta(\Delta-3),
\end{align}
we see that both $\Delta=1$ and $\Delta=2$ satisfy $m^2=-2$. Given the fact that the CFT unitarity bounds demand $\Delta\ge \frac{3-2}{2}=\frac{1}{2}$ in three dimensions, we see that both these choices are consistent.
After making this choice, this is also the tree-level time-ordered two point function of the field $\phi(z,x)$.
\begin{align}\label{scalartimeordered2point1}
    \langle 0|T\{\phi_{\Delta}(z,x)\phi_{\Delta}(z',x')\}|0\rangle=G_{F,\Delta}(z,z',x-x'),~\Delta=1,2.
\end{align}
$|0\rangle$ is the Vaccuum state of the scalar field which we take to be invariant under all the isometries of AdS$_4$.
The Wightman propagators on the other hand are homogeneous solution to \eqref{scalarFeynmanbulktobulk2}. 
\begin{align}
    (\partial_z^2+\Box)W(z,z',x-x')=0,
\end{align}
which is solved by,
\begin{align}\label{scalarWightmanbulktobulk2}
    &W_{\Delta=1,\pm}(z,z',x-x')=\frac{1}{4\pi^2}\bigg(\frac{1}{(z-z')^2+(x-x')^2\pm i\epsilon(t-t')}+\frac{1}{(z+z')^2+(x-x')^2\pm i\epsilon(t-t')}\bigg),\notag\\
    &W_{\Delta=2,\pm}(z,z',x-x')=\frac{1}{4\pi^2}\bigg(\frac{1}{(z-z')^2+(x-x')^2\pm i\epsilon(t-t')}-\frac{1}{(z+z')^2+(x-x')^2\pm i\epsilon(t-t')}\bigg).
\end{align}
where $\epsilon>0$ and the $\pm$ sign determines which Wightman propagator (of which there are two) we want. They are given by,
\begin{align}
    &\langle 0|\phi_{\Delta}(z,x)\phi_{\Delta}(z',x')|0\rangle=W_{\Delta,+}(z,z',x-x')\notag\\
    &\langle 0|\phi_{\Delta}(z',x')\phi_{\Delta}(z,x)|0\rangle=W_{\Delta,-}(z,z',x-x'),
\end{align}
with $\Delta=1,2$. It is also clear by definition that the following properties hold:
\begin{align}
    W_{\Delta,-}(z,z',x-x')=W_{\Delta,+}(z',z,x'-x)=W_{\Delta,+}^{*}(z,z',x-x').
\end{align}
The Feynman propagator \eqref{scalarFeynmanbulktobulk2} can be written in terms of the Wightman propagators \eqref{scalarWightmanbulktobulk2} as follows:
\begin{align}
    G_{\Delta}(z,z',x-x')=\theta(t-t')W_{\Delta,+}(z,z',x-x')+\theta(t'-t)W_{\Delta,-}(z,z',x-x').
\end{align}
We shall also be interested in the bulk to boundary propagators that have one point at the conformal boundary at $z=0$. They can be obtained by performing a series expansion of \eqref{scalarFeynmanbulktobulk2} and \eqref{scalarWightmanbulktobulk2} about $z'=0$. The results are,
\begin{align}
    &G_{\Delta=1}(z,x-x')=\frac{1}{2\pi^2(z^2+(x-x')^2+i\epsilon)},~G_{\Delta=2}(z,x-x')=\frac{z}{\pi^2(z^2+(x-x')^2+i\epsilon)},\notag\\
    &W_{\Delta=1,\pm}(z,x-x')=\frac{1}{2\pi^2(z^2+(x-x')^2\pm i\epsilon(t-t'))},~W_{\Delta=2,\pm}(z,x-x')=\frac{z}{\pi^2(z^2+(x-x')^2\pm i\epsilon(t-t'))}.
\end{align}
For the $\Delta=2$ case, we see that the leading order term in the expansion about $z'=0$ starts or $\order{z'}$. Thus, before setting $z'=0$, we rescale the propagator by $\frac{1}{z'}$ to cancel out this factor leading to the result quoted above. This can also be understood from the AdS/CFT extrapolate dictionary that relates the \textit{rescaled} boundary value of the bulk field to its dual conformal operator. In general, we have,
\begin{align}\label{scalarboundaryvalue}
    \phi(x)=\lim_{z\to 0}z^{-\Delta+1}\phi(z,x),
\end{align}
where $\phi(x)$ will be identified with the conformal operator. The $+1$ in the exponent of $z$ arises due to the Weyl transformation \eqref{scalarWeyltransform}. Therefore, there is no rescaling for $\Delta=1$ and a $\frac{1}{z}$ rescaling for $\Delta=2$. 

Due to the translation invariance in the $x^\mu$ variables, it is useful to Fourier transform to momentum space\footnote{In the Fourier transform, one obtains a factor of $e^{-|p^0|\epsilon}$ which we set to $1$ since $\epsilon$ is infinitesimal. However, when performing an inverse Fourier transform back to position space, this factor has to be re-instated to obtain the correct position $i\epsilon$ prescription.}.  The results for $\Delta=1$ scalars are,
\footnotesize
\begin{align}
    &G_{F,\Delta=1}(z,z',p)=\frac{\theta(z-z')}{2|p|}\big(\theta(-p^2)e^{i |p|(z-z')}-i\theta(p^2)e^{-|p|(z-z')}\big)+(z\leftrightarrow z')+\frac{1}{2|p|}\big(\theta(-p^2)e^{i|p|(z_1+z_2)}-i\theta(p^2)e^{-|p|(z_1+z_2)}\big),\notag\\
    &G_{F,\Delta=1}(z,p)=\frac{1}{|p|}\big(\theta(-p^2)e^{i|p|z}-i \theta(p^2)e^{-|p|z}\big)~,W_{\Delta=1,\pm}(z,z',p)= \frac{2\cos{|p| z}\cos{|p| z'}}{|p|}\theta(\pm p_0)\theta(-p^2),\notag\\
    &W_{\Delta=1,\pm}(z,p)=\frac{2 \cos{|p| z}}{|p|}\theta(\pm p_0)\theta(-p^2),
\end{align}
\normalsize
For $\Delta=2$ we have on the other hand,
\footnotesize
\begin{align}
    &G_{F,\Delta=2}(z,z',p)=\frac{\theta(z-z')}{2|p|}\big(\theta(-p^2)e^{i |p|(z-z')}-i\theta(p^2)e^{-|p|(z-z')}\big)+(z\leftrightarrow z')-\frac{1}{2|p|}\big(\theta(-p^2)e^{i|p|(z_1+z_2)}-i\theta(p^2)e^{-|p|(z_1+z_2)}\big),\notag\\
    &G_{F,\Delta=2}(z,p)=-i\big(\theta(-p^2)e^{i|p|z}+ \theta(p^2)e^{-|p|z}\big)~,W_{\Delta=2,\pm}(z,z',p)= \frac{2\sin{|p| z}\sin{|p| z'}}{|p|}\theta(\pm p_0)\theta(-p^2),\notag\\
    &W_{\Delta=2,\pm}(z,p)=2 \sin{|p| z}\theta(\pm p_0)\theta(-p^2),
\end{align}
\normalsize
with the propagator used to invert the equation of motion found through its definition \eqref{newpropagator}.
This concludes our discussion of the propagators for the conformally coupled scalar field. 

\section{Current conservation in Wightman functions}\label{app:LongitudinalContributions}
As is known, Wightman functions of conserved currents are identically conserved rather than satisfying non-trivial Ward-Takahashi identities with contact terms \cite{Bala:2025gmz}.

In this appendix, we show this explicitly with the characteristic example of the Yang-Mills three point Wightman function, which serves as an additional check that our formalism is correct. Let us return to the Yang-Mills equation of motion \eqref{YMEOM} with the perturbative expansion \eqref{YMperturbation1}. We then constructed the three point function by contracting with the transverse polarization vectors \eqref{JJJWightman}. This of course, causes any longitudinal contributions to drop out. These longitudinal contributions arise from the EOM inverter propagator $\mathcal{G}_{\Delta=2}$ (see \eqref{gluonprops} and \eqref{photonEOMinverterprops}) used to invert the equation of motion. We now show that even if we do keep such contributions, they go to zero after computing the bulk $z$ integral. We have,
\begin{align}
    \langle 0|J^{\mu A}(p_1)J^{\nu B}(p_2)J^{\rho C}(p_3)|0\rangle_{\text{Longitudinal}}\theta(p_2^2)=\frac{i g f^{ABC}}{2}\theta(-p_1^2)\theta(-p_1^0)\theta(p_2^2)\theta(-p_3^2)\theta(p_3^0)\mathcal{A}_{3,YM,\text{Longitudinal}}^{\mu\nu\rho},
\end{align}
with,
\footnotesize
\begin{align}
    &\mathcal{A}^{\mu\nu\rho}_{3,YM,\text{Longitudinal}}\notag\\&=-i\frac{p_2^\nu p_2^\alpha}{p_2^2}\bigg(\pi^{\beta\rho}(p_3)(2p_{1\beta}\pi^\mu_\alpha(p_1)-p_{1\alpha}\pi^\mu_{\beta}(p_1))-\pi^{\mu\beta}(p_1)(2p_{3\beta}\pi_\alpha^\rho(p_3)-p_{3\alpha}\pi_\beta^\rho(p_3))\bigg)\int_{0}^{\infty} dz\sin(|p_1|z)\sin(|p_3|z)\notag\\&=-\frac{(p_1^2-p_3^2)p_2^\nu}{2p_1^2 p_2^2 p_3^2}\bigg(p_1^\mu((p_1^2+p_2^2+p_3^2)p_1^\rho+(p_1^2+p_2^2-p_3^2)p_{2\rho})+2p_1^2(p_1^\rho p_2^\mu+p_2^\mu p_2^\rho-p_3^2\eta^{\mu\rho})\bigg)\pi\delta(|p_1|-|p_3|)=0,
\end{align}
\normalsize
where we used the distributional identity,
\begin{align}
    \int_0^\infty dz \sin(|p_1|z)\sin(|p_3|z)=\pi\delta(|p_1|-|p_3|)-\pi \delta(|p_1|+|p_3|)=\pi \delta(|p_1|-|p_3|).
\end{align}
A similar analysis can be done for other Wightman functions showing that they are identically conserved and do not contain any longitudinal contributions.

\section{Four point correlators in special kinematics as CPWs}\label{app:CPWexamples}
In this appendix, we present the interpretation of our special kinematics four point functions in terms of conformal partial waves as discusssed in subsection \ref{sec:FactorizationandCPW}.

We begin with the scaling dimension $\Delta$ scalar tree level four point function \eqref{phi3fourpointgendelta} arising due to the exchange of a $\Delta'$ scalar. The result can simply be written as,
\begin{align}
    &\langle\langle 0|\mathcal{O}_{\Delta}(p_1)\mathcal{O}_{\Delta}(p_2)\mathcal{O}_{\Delta}(p_3)\mathcal{O}_{\Delta}(p_4)|0\rangle\rangle_{\text{phi-cube}}\theta(p_2^2)\theta(p_3^2)\notag\\&\propto \langle\langle 0|O_{\Delta}(p_1)O_{\Delta}(p_2)O_{\Delta'}(-s)|0\rangle\rangle\frac{1}{|s|^{2\Delta'-3}}\langle\langle 0|O_{\Delta'}(s)O_{\Delta}(p_3)O_{\Delta}(p_4)|0\rangle\rangle  \notag\\&=\mathcal{W}_{(O_\Delta O_{\Delta}|O_{\Delta'}|O_{\Delta}O_{\Delta})}^{(s)}(p_1,p_2|s|p_3,p_4),
\end{align}
that is, the Wightman function in these kinematics is simply the s-channel Wightman scalar exchange CPW! 

Similarly, for scalar Bhabha scattering we obtain,
\begin{align}
     &\langle\langle 0|\mathcal{O}_{\Delta}(p_1)\mathcal{O}^*_{\Delta}(p_2)\mathcal{O}_{\Delta}(p_3)\mathcal{O}^*_{\Delta}(p_4)|0\rangle\rangle_{\text{Bhabha}}\theta(p_2^2)\theta(p_3^2)\notag\\&\propto \langle\langle 0|O_{\Delta}(p_1)O_{\Delta}(p_2)J^{\mu}(-s)|0\rangle\rangle\frac{\pi_{\mu\nu}(s)}{|s|}\langle\langle 0|J^{\nu}(s)O_{\Delta}(p_3)O_{\Delta}(p_4)|0\rangle\rangle\notag\\&=\mathcal{W}_{(O_{\Delta} O_{\Delta}|J|O_{\Delta}O_{\Delta})}^{(s)}(p_1,p_2|s|p_3,p_4),
\end{align}
which is yet again the s-channel conformal partial wave due to the exchange of a spin-1 current dual to the bulk photon.

For Compton scattering we obtain,
\begin{align}
    &\langle\langle 0|J(p_1,\epsilon_1)O_{\Delta}(p_2)O^*_{\Delta}(p_3)J(p_4,\epsilon_4)|0\rangle\rangle_{\text{Compton}}\theta(p_2^2)\theta(p_3^2)\notag\\&=\langle\langle 0|J(p_1,\epsilon_1)O_{\Delta}(p_2)O_{\Delta}^*(-s)|0\rangle\frac{1}{|s|^{2\Delta-3}}\langle\langle 0|O_{\Delta}(s)O_{\Delta}^*(p_3)J(p_4,\epsilon_4)|0\rangle\rangle\notag\\&=\mathcal{W}^{(s)}_{(JO_{\Delta}|O_{\Delta}|O_{\Delta}^* J)}(p_1,p_2|s|p_3,p_4).
\end{align}
Finally, for Yang-Mills and Einstein-Hilbert gravity we find,
\begin{align}\label{YMCPW}
    &\langle\langle 0|J^{A_1}(p_1,\epsilon_1)J^{A_2}(p_2,\epsilon_2)J^{A_3}(p_3,\epsilon_3)J^{A_4}(p_4,\epsilon_4)|0\rangle\rangle_{YM}\theta(p_2^2)\theta(p_3^2)\notag\\
    &=\langle\langle 0|J^{A_1}(p_1,\epsilon_1)J^{A_2}(p_2,\epsilon_2)J^{\mu A}(-s)|0\rangle\rangle\frac{\pi_{\mu\nu}(s)}{|s|}\langle\langle 0|J^{\nu}_A(s)J^{A_3}(p_3,\epsilon_3)J^{A_4}(p_4,\epsilon_4)|0\rangle\rangle\notag\\
    &=\mathcal{W}^{(s)}_{(JJ|J|JJ)}(p_1,p_2|s|p_3,p_4),
\end{align}
and,
\begin{align}\label{GRCPW}
    &\langle\langle 0|T(p_1,\epsilon_1)T(p_2,\epsilon_2)T(p_3,\epsilon_3)T(p_4,\epsilon_4)|0\rangle\rangle_{EH}\theta(p_2^2)\theta(p_3^2)\notag\\
    &=\langle\langle 0|T(p_1,\epsilon_1)T(p_2,\epsilon_2)T^{\mu\nu}(-s)|0\rangle\frac{\Pi_{\mu\nu\rho\sigma}(s)}{|s|^3}\langle\langle 0|T^{\rho\sigma}(s)T(p_3,\epsilon_3)T(p_4,\epsilon_4)|0\rangle\rangle\notag\\&=\mathcal{W}^{(s)}_{(TT|T|TT)}(p_1,p_2|s|p_3,p_4).
\end{align}

\section{Details of half-Fourier transform from twistor space to spinor helicity}\label{app:HalfFourierDetails}
The aim of this appendix is to detail the calculation from the general twistor space three point result \eqref{twistorspace3pointgensols1s2s3} to spinor helicity variables \eqref{threepointtwistortoSHgensol}.
\subsection{The $\delta\delta\delta$ solution}
We begin with the first term in \eqref{twistorspace3pointgensols1s2s3}. Let us denote it,
\begin{align}
    F_1=i^{\alpha+\beta+\gamma}\delta^{[\alpha]}(Z_1\cdot Z_2)\delta^{[\beta]}(Z_2\cdot Z_3)\delta^{[\gamma]}(Z_3\cdot Z_1).
\end{align}
Expressing this quantity using Schwinger parametrization results in,
\begin{align}
    &F_1=\frac{1}{(2\pi)^3}\int dc_{12}c_{12}^\alpha\int dc_{23}c_{23}^\beta\int dc_{31}c_{31}^\gamma e^{-ic_{12}Z_1\cdot Z_2-ic_{23}Z_2\cdot Z_3-ic_{31}Z_3\cdot Z_1}\notag\\
    &=\frac{1}{(2\pi)^3}\int dc_{12}c_{12}^\alpha\int dc_{23}c_{23}^\beta\int dc_{31}c_{31}^\gamma e^{-ic_{12}(\lambda_1\cdot\Bar{\mu}_2-\lambda_2\cdot\Bar{\mu}_1)-ic_{23}(\lambda_2\cdot\Bar{\mu}_3-\lambda_3\cdot\Bar{\mu}_2)-ic_{31}(\lambda_3\cdot\Bar{\mu}_1-\lambda_1\cdot\Bar{\mu}_3)}.
\end{align}
Performing a half-Fourier transform \eqref{halfFourier} to convert to spinor helicity variables we get,
\small
\begin{align}
    &\tilde{F}_1=\frac{1}{(2\pi)^3}\int dc_{12}c_{12}^\alpha\int dc_{23}c_{23}^\beta \int dc_{31}c_{31}^\gamma\int d^2\Bar{\mu}_1\int d^2\Bar{\mu}_2\int d^2\Bar{\mu}_3\notag\\
    &\times e^{-i(\Bar{\lambda}_1-c_{12}\lambda_2+c_{31}\lambda_3)\cdot\Bar{\mu}_1}e^{-i(\Bar{\lambda}_2+c_{12}\lambda_1-c_{23}\lambda_3)\cdot \Bar{\mu}_2}e^{-i(\Bar{\lambda}_3+c_{23}\lambda_2-c_{31}\lambda_1)\cdot\Bar{\mu}_3}\notag\\
    &=(2\pi)^3\int dc_{12}c_{12}^\alpha\int dc_{23}c_{23}^\beta \int dc_{31}c_{31}^\gamma\delta^2(\Bar{\lambda}_1-c_{12}\lambda_2+c_{31}\lambda_3)\delta^2(\Bar{\lambda}_2+c_{12}\lambda_1-c_{23}\lambda_3)\delta^2(\Bar{\lambda}_3+c_{23}\lambda_2-c_{31}\lambda_1).
\end{align}
\normalsize
To perform the three Schwinger parameter integrals, we express each of the two dimensional delta functions in a particular basis. In particular, we use the Schouten identity to write down,
\begin{align}
    \Bar{\lambda}_{1a}=\frac{\langle \Bar{1}3\rangle\lambda_{2a}-\langle \Bar{1}2\rangle\lambda_{3a}}{\langle 2 3\rangle},~\Bar{\lambda}_{2a}=\frac{\langle \Bar{2}3\rangle\lambda_{1a}-\langle \Bar{2}1\rangle\lambda_{3a}}{\langle 1 3\rangle},\Bar{\lambda}_{3a}=\frac{\langle \Bar{3}1\rangle\lambda_{2a}-\langle \Bar{3}2\rangle\lambda_{1a}}{\langle 2 1\rangle}.
\end{align}
We then substitute these formulae into the above delta function and use the identity,
\begin{align}\label{delta2id}
    \delta^2(\alpha_1 v_1+\alpha_2 v_2)=\frac{1}{\langle v_1 v_2\rangle}\delta(\alpha_1)\delta(\alpha_2),
\end{align}
where $v_1,v_2$ are linearly independent basis vectors and $\alpha_1,\alpha_2$ are their coefficients. The resulting expression is,
\begin{align}
    \tilde{F}_1&=(2\pi)^3\int dc_{12}c_{12}^\alpha\int dc_{23}c_{23}^\beta \int dc_{31}c_{31}^\gamma\Bigg(\delta(c_{12}-\frac{\langle \Bar{1}3\rangle}{\langle 23\rangle})\delta(c_{31}-\frac{\langle \Bar{1}2\rangle}{\langle 2 3\rangle})\delta(c_{12}-\frac{\langle \Bar{2}3\rangle}{\langle 3 1\rangle})\notag\\
    &\times\delta(c_{23}+\frac{\langle 1 \Bar{2}\rangle}{\langle 3 1\rangle})\delta(c_{23}-\frac{\langle \Bar{3}1\rangle}{\langle 1 2\rangle})\delta(c_{31}+\frac{\langle 2 \Bar{3}\rangle}{\langle 1 2\rangle})\frac{1}{|\langle 12\rangle||\langle 2 3\rangle||\langle 3 1\rangle|}\Bigg).
\end{align}
Thus, the Schwinger parameter integrals localize yielding,
\small
\begin{align}\label{F1tilde}
    &\tilde{F}_1=\frac{(2\pi)^3}{|\langle 12\rangle||\langle 2 3\rangle||\langle 3 1\rangle|}\bigg(\frac{\langle \Bar{1}3\rangle}{\langle 2 3\rangle}\bigg)^\alpha\bigg(\frac{-\langle 1\Bar{2}\rangle}{\langle 3 1\rangle}\bigg)^\beta\bigg(\frac{-\langle 2\Bar{3}\rangle}{\langle 1 2\rangle}\bigg)^\gamma\delta\bigg(\frac{\langle \Bar{1}3\rangle}{\langle 2 3\rangle}-\frac{\langle \Bar{2}3\rangle}{\langle 3 1\rangle}\bigg)\delta\bigg(\frac{\langle 1\Bar{2}\rangle}{\langle 31\rangle}+\frac{\langle\Bar{3}1\rangle}{\langle 12\rangle}\bigg)\delta\bigg(\frac{\langle 2\Bar{3}\rangle}{\langle 1 2\rangle}+\frac{\langle \Bar{1}2\rangle}{\langle 2 3\rangle}\bigg)\notag\\
    &=\frac{(-1)^{\alpha+\beta+\gamma}}{4}\frac{\langle \Bar{1}\Bar{2}\rangle^{\alpha}\langle \Bar{2}\Bar{3}\rangle^\beta\langle \Bar{3}\Bar{1}\rangle^\gamma}{(p_1+p_2+p_3)^{\alpha+\beta+\gamma}}(2\pi)^3\delta^3(p_1+p_2+p_3).
\end{align}
\normalsize
To obtain the final result we used the fact that,
\begin{align}\label{momentumconsinSH}
    \delta^3(p_1+p_2+p_3)=\frac{4}{|\langle 12\rangle||\langle 2 3\rangle||\langle 3 1\rangle|}\delta\bigg(\frac{\langle \Bar{1}3\rangle}{\langle 2 3\rangle}-\frac{\langle \Bar{2}3\rangle}{\langle 3 1\rangle}\bigg)\delta\bigg(\frac{\langle 1\Bar{2}\rangle}{\langle 31\rangle}+\frac{\langle\Bar{3}1\rangle}{\langle 12\rangle}\bigg)\delta\bigg(\frac{\langle 2\Bar{3}\rangle}{\langle 1 2\rangle}+\frac{\langle \Bar{1}2\rangle}{\langle 2 3\rangle}\bigg),
\end{align}
as well as several other spinor helicity identities that can all be found by contracting the momentum conservation equation $p_{1\mu}+p_{2\mu}+p_{3\mu}=0$ with suitable vectors and converting to spinor helicity using \eqref{SHvariables}. \eqref{F1tilde} is exactly the result quoted in the main text (the first term in \eqref{threepointtwistortoSHgensol}).

\subsection{The $\delta^4$ solution}
Consider the second term in \eqref{twistorspace3pointgensols1s2s3}.
\begin{align}
    F_2&=i^{\alpha+\beta+\gamma}\delta^{[\alpha+\beta+\gamma]}(Z_1\cdot Z_2)\int dc_1\int dc_2 c_1^{-\beta} c_2^{-\gamma}\delta^4(c_1 Z_1+c_2 Z_2-Z_3)\notag\\
    &=\frac{1}{2\pi}\int dc_{12} dc_1 dc_2~c_{12}^{\alpha+\beta+\gamma}c_1^{-\beta}c_2^{-\gamma} e^{-ic_{12}(\lambda_1\cdot\Bar{\mu}_2-\lambda_2\cdot\Bar{\mu}_2)}\delta^2(c_1\lambda_1+c_2\lambda_2-\lambda_3)\delta^2(c_1\Bar{\mu}_1+c_2\Bar{\mu}_2-\Bar{\mu}_3).
\end{align}
Performing the inverse half-Fourier transform \eqref{inverseHalfFourier} and performing the $d^2\Bar{\mu}_i$ integrals results in,
\begin{align}
    \tilde{F}_2=(2\pi)^3\int dc_{12}dc_1 dc_2 c_{12}^{\alpha+\beta+\gamma}c_1^{-\beta}c_2^{-\gamma}\delta^2(\Bar{\lambda}_1+c_1\Bar{\lambda}_3-c_{12}\lambda_2)\delta^2(\Bar{\lambda}_2+c_2\Bar{\lambda}_3+c_{12}\lambda_1)\delta^2(c_1\lambda_1+c_2\lambda_2-\lambda_3).
\end{align}
We then use the Schouten identities,
\begin{align}
    \lambda_{2a}=\frac{\langle 2\Bar{3}\rangle\Bar{\lambda}_{1a}-\langle 2\Bar{1}\rangle \Bar{\lambda}_{3a}}{\langle \Bar{1}\Bar{3}\rangle},\lambda_{1a}=\frac{\langle 1\Bar{3}\rangle\Bar{\lambda}_{2a}-\langle 1 \Bar{2}\rangle\Bar{\lambda}_{3a}}{\langle\Bar{2}\Bar{3}\rangle},\lambda_{3a}=\frac{\langle 3 2\rangle\lambda_{1a}-\langle 3 1\rangle \lambda_{2a}}{\langle 1 2\rangle}.
\end{align}
Substituting these identities in the two dimensional delta functions and using the formula \eqref{delta2id} results in the Schwinger integrals localizing.
\small
\begin{align}
    &\tilde{F}_2=\frac{(2\pi)^3}{|\langle 12\rangle|\langle \Bar{2}\Bar{3}\rangle||\langle\Bar{3}\Bar{1}\rangle|}\int dc_{12}dc_1 dc_2 c_{12}^{\alpha+\beta+\gamma}c_1^{-\beta}c_2^{-\gamma}\delta\bigg(1-\frac{c_{12}\langle 2 \Bar{3}\rangle}{\langle \Bar{1}\Bar{3}\rangle}\bigg)\delta\bigg(c_1+\frac{c_{12}\langle 2\Bar{1}\rangle}{\langle \Bar{1}\Bar{3}\rangle}\bigg)\delta\bigg(1+\frac{c_{12}\langle 1\Bar{3}\rangle}{\langle \Bar{2}\Bar{3}\rangle}\bigg)\notag\\
    &\times\delta\bigg(c_2-\frac{c_{12}\langle 1\Bar{2}\rangle}{\langle\Bar{2}\Bar{3}\rangle}\bigg)\delta\bigg(c_1-\frac{\langle 32\rangle}{\langle 12\rangle}\bigg)\delta\bigg(c_2+\frac{\langle 3 1\rangle}{\langle 12\rangle}\bigg)\notag\\&=\frac{(2\pi)^3}{|\langle 12\rangle|\langle \Bar{2}\Bar{3}\rangle||\langle\Bar{3}\Bar{1}\rangle|}\bigg(\frac{\langle 3 2\rangle}{\langle 12\rangle}\bigg)^{-\beta}\bigg(\frac{-\langle 31\rangle}{\langle 12\rangle}\bigg)^{-\gamma}\bigg(\frac{\langle \Bar{1}\Bar{3}\rangle}{\langle 2\Bar{3}\rangle}\bigg)^{\alpha+\beta+\gamma}\bigg|\frac{\langle \Bar{1}\Bar{3}\rangle}{\langle 2\Bar{3}\rangle}\bigg|\delta\bigg(\frac{\langle 32\rangle}{\langle 12\rangle}+\frac{2\Bar{1}\rangle}{\langle 2 \Bar{3}\rangle}\bigg)\delta\bigg(1+\frac{\langle \Bar{1}\Bar{3}\rangle\langle 1\Bar{3}\rangle}{\langle 2\Bar{3}\rangle\langle\Bar{2}\Bar{3}\rangle}\bigg)\delta\bigg(\frac{\langle 31\rangle}{\langle 1 2\rangle}+\frac{\langle \Bar{1}\Bar{3}\rangle\langle 1 \Bar{2}\rangle}{\langle 2 \Bar{3}\rangle\langle \Bar{2}\Bar{3}\rangle}\bigg).
\end{align}
\normalsize
Note that by exchanging $\lambda_3\leftrightarrow \Bar{\lambda}_3$ which keeps the left hand side of \eqref{momentumconsinSH} invariant yield the following representation for the momentum conserving delta function after some algebra:
\begin{align}
    \delta^3(p_1+p_2+p_3)=\frac{4}{|\langle 1 2\rangle\langle 2\Bar{3}\rangle\langle\Bar{2}\Bar{3}\rangle|}\delta\bigg(\frac{\langle 3 2\rangle}{\langle 12\rangle}+\frac{\langle 2 \Bar{1}\rangle}{\langle 2\Bar{3}\rangle}\bigg)\delta\bigg(1+\frac{\langle \Bar{1}\Bar{3}\rangle\langle 1 \Bar{3}\rangle}{\langle 2 \Bar{3}\rangle\langle\Bar{2}\Bar{3}\rangle}\bigg)\delta\bigg(\frac{\langle 3 1\rangle}{\langle 1 2\rangle}+\frac{\langle 1\Bar{2}\rangle\langle\Bar{1}\Bar{3}\rangle}{\langle 2 \Bar{3}\rangle\langle\Bar{2}\Bar{3}\rangle}\bigg),
\end{align}
and some spinor helicity identities we obtain the result,
\begin{align}
    \tilde{F}_2=\frac{(-1)^{\alpha+\beta+\gamma}}{4}\frac{\langle\Bar{1}\Bar{2}\rangle^\alpha\langle \Bar{2}\Bar{3}\rangle^\beta\langle\Bar{3}\Bar{1}\rangle^\gamma}{(p_1+p_2-p_3)^\alpha(p_1-p_2+p_3)^\gamma(-p_1+p_2+p_3)^\beta}(2\pi)^3\delta^3(p_1+p_2+p_3),
\end{align}
matching the result quoted in the main text (the second term in \eqref{threepointtwistortoSHgensol}).
\section{Three point functions in all helicities}\label{app:YmandGRallhelicities}
In this appendix, we present the expressions for the stripped three point Wightman functions in Yang-Mills theory and Einstein gravity in all eight helicities: First in spinor helicity variables and then in twistor space. We express all results with net negative helicity in terms of $\langle ij\rangle$ and $p_i$ and net positive helicity in terms of $\langle\Bar{i}\Bar{j}\rangle$ and $p_i$. However, there are many spinor helicity identities derived from momentum conservation and the Schouten identities\footnote{The Schouten identity is simply the statement that there are at most $d$ linearly independent vectors in a $d$ dimensional space.} that we use in the main-text to bring these results to a more convenient form depending on the application.
\subsection{Yang-Mills theory}
Consider the Yang-Mills stripped three point function in the eight helicity configurations (also suppressing the momentum conserving delta function). The chiral sector (net positive helicity) expressions are,
\begin{align}
    &\langle\langle 0|J^{+A}(p_1)J^{+B}(p_2)J^{+C}(p_3)|0\rangle\rangle'=if^{ABC}\frac{\langle \Bar{1}\Bar{2}\rangle\langle\Bar{2}\Bar{3}\rangle\langle\Bar{3}\Bar{1}\rangle}{(E-2p_1)(E-2p_2)(E-2p_3)},\notag\\
    &\langle\langle 0|J^{+A}(p_1)J^{+B}(p_2)J^{-C}(p_3)|0\rangle\rangle'=-if^{ABC}\frac{\langle\Bar{1}\Bar{2}\rangle^3}{\langle\Bar{2}\Bar{3}\rangle\langle\Bar{3}\Bar{1}\rangle E},\notag\\&\langle\langle 0|J^{+A}(p_1)J^{-B}(p_2)J^{+C}(p_3)|0\rangle\rangle'=-if^{ABC}\frac{\langle\Bar{3}\Bar{1}\rangle^3}{\langle\Bar{1}\Bar{2}\rangle\langle\Bar{2}\Bar{3}\rangle E},\notag\\&\langle\langle 0|J^{-A}(p_1)J^{+B}(p_2)J^{+C}(p_3)|0\rangle\rangle'=-if^{ABC}\frac{\langle\Bar{2}\Bar{3}\rangle^3}{\langle\Bar{1}\Bar{2}\rangle\langle\Bar{3}\Bar{1}\rangle E}.
\end{align}
Their twistor space counterparts are given by,
\begin{align}
    &\langle 0|J^{+A}(Z_1)J^{+B}(Z_2)J^{+C}(Z_3)|0\rangle'=-4 f^{ABC}\delta^{[3]}(Z_1\cdot Z_2)\int \frac{dc_1}{c_1}\int \frac{dc_2}{c_2}\delta^4(c_1 Z_1+c_2 Z_2-Z_3),\notag\\
   &\langle 0|J^{+A}(Z_1)J^{+B}(Z_2)J^{-C}(Z_3)|0\rangle'=-f^{ABC}\text{Sgn}(Z_3\cdot Z_1)\text{Sgn}(Z_2\cdot Z_3)\delta^{[3]}(Z_1\cdot Z_2),\notag\\ &\langle 0|J^{+A}(Z_1)J^{-B}(Z_2)J^{+C}(Z_3)|0\rangle'=-f^{ABC}\text{Sgn}(Z_1\cdot Z_2)\text{Sgn}(Z_2\cdot Z_3)\delta^{[3]}(Z_3\cdot Z_1),\notag\\ &\langle 0|J^{-A}(Z_1)J^{+B}(Z_2)J^{+C}(Z_3)|0\rangle'=-f^{ABC}\text{Sgn}(Z_1\cdot Z_2)\text{Sgn}(Z_3\cdot Z_1)\delta^{[3]}(Z_2\cdot Z_3).
\end{align}
The anti-chiral sector (net negative helicity) on the other hand is given by,
\begin{align}
    &\langle \langle 0|J^{-A}(p_1)J^{-B}(p_2)J^{-C}(p_3|0\rangle\rangle'=-if^{ABC}\frac{\langle 12\rangle\langle 23\rangle\langle 3 1\rangle}{(E-2p_1)(E-2p_2)(E-2p_3)},\notag\\&\langle \langle 0|J^{-A}(p_1)J^{-B}(p_2)J^{+C}(p_3|0\rangle\rangle'=if^{ABC}\frac{\langle 12\rangle^3}{\langle 2 3\rangle\langle 3 1\rangle E},\notag\\&\langle \langle 0|J^{-A}(p_1)J^{+B}(p_2)J^{-C}(p_3|0\rangle\rangle'=if^{ABC}\frac{\langle 3 1\rangle^3}{\langle 1 2\rangle\langle 23\rangle E},\notag\\&\langle \langle 0|J^{+A}(p_1)J^{-B}(p_2)J^{-C}(p_3|0\rangle\rangle'=if^{ABC}\frac{\langle 2 3\rangle^3}{\langle 1 2\rangle\langle 3 1\rangle E},\notag\\
\end{align}
with their twistor versions,
\begin{align}
    &\langle 0|J^{-A}(Z_1)J^{-B}(Z_2)J^{-C}(Z_3)|0\rangle=-\frac{1}{2}f^{ABC}\text{Sgn}(Z_1\cdot Z_2)\text{Sgn}(Z_2\cdot Z_3)\text{Sgn}(Z_3\cdot Z_1),\notag\\ &\langle 0|J^{-A}(Z_1)J^{-B}(Z_2)J^{+C}(Z_3)|0\rangle=-2f^{ABC}\text{Sgn}(Z_1\cdot Z_2)\int \frac{dc_1}{c_1}\int \frac{dc_2}{c_2}\delta^4(c_1 Z_1+c_2 Z_2-Z_3),\notag\\ &\langle 0|J^{-A}(Z_1)J^{+B}(Z_2)J^{-C}(Z_3)|0\rangle=-2f^{ABC}\text{Sgn}(Z_1\cdot Z_2)\int \frac{dc_1}{c_1}\int dc_2~c_2^3~\delta^4(c_1 Z_1+c_2 Z_2-Z_3),\notag\\ &\langle 0|J^{+A}(Z_1)J^{-B}(Z_2)J^{-C}(Z_3)|0\rangle=-2f^{ABC}\text{Sgn}(Z_1\cdot Z_2)\int dc_1 c_1^3\int \frac{dc_2}{c_2}\delta^4(c_1 Z_1+c_2 Z_2-Z_3),\notag\\
\end{align}
\subsection{Einstein Gravity}
A similar analysis can be performed for Einstein gravity. The chiral sector is given by,
\begin{align}
    &\langle\langle 0|T^{+}(p_1)T^{+}(p_2)T^{+}(p_3)|0\rangle\rangle'=\frac{\langle \Bar{1}\Bar{2}\rangle^2\langle\Bar{2}\Bar{3}\rangle^2\langle \Bar{3}\Bar{1}\rangle^2 p_1 p_2 p_3}{(E-2p_1)^2(E-2p_2)^2(E-2p_3)^2},\notag\\&\langle\langle 0|T^{+}(p_1)T^{+}(p_2)T^{-}(p_3)|0\rangle\rangle'=\frac{\langle\Bar{1}\Bar{2}\rangle^6 p_1 p_2 p_3}{\langle\Bar{2}\Bar{3}\rangle^2\langle\Bar{3}\Bar{1}\rangle^2 E^2},\notag\\&\langle\langle 0|T^{+}(p_1)T^{-}(p_2)T^{+}(p_3)|0\rangle\rangle'=\frac{\langle \Bar{3}\Bar{1}\rangle^6 p_1 p_2 p_3}{\langle\Bar{1}\Bar{2}\rangle^2\langle\Bar{2}\Bar{3}\rangle^2 E^2},\notag\\&\langle\langle 0|T^{-}(p_1)T^{+}(p_2)T^{+}(p_3)|0\rangle\rangle'=\frac{\langle\Bar{2}\Bar{3}\rangle^6 p_1 p_2 p_3}{\langle\Bar{1}\Bar{2}\rangle^2\langle\Bar{3}\Bar{1}\rangle^2 E^2},
\end{align}
with the corresponding twistor space correlators,
\begin{align}
    &\langle 0|T^{+}(Z_1)T^{+}(Z_2)T^{+}(Z_3)|0\rangle=-4\delta^{[6]}(Z_1\cdot Z_2)\int \frac{dc_1}{c_1^2}\int \frac{dc_2}{c_2^2}\delta^4(c_1 Z_1+c_2 Z_2-Z_3),\notag\\
    &\langle 0|T^{+}(Z_1)T^{+}(Z_2)T^{-}(Z_3)|0\rangle=-|Z_3\cdot Z_1||Z_2\cdot Z_3|\delta^{[6]}(Z_1\cdot Z_2),\notag\\
    &\langle 0|T^{+}(Z_1)T^{-}(Z_2)T^{+}(Z_3)|0\rangle=-|Z_1\cdot Z_2||Z_2\cdot Z_3|\delta^{[6]}(Z_3\cdot Z_1),\notag\\
    &\langle 0|T^{-}(Z_1)T^{+}(Z_2)T^{+}(Z_3)|0\rangle=-|Z_1\cdot Z_2||Z_3\cdot Z_1|\delta^{[6]}(Z_2\cdot Z_3).
\end{align}
Their counterparts in the anti-chiral sector are given by,
\begin{align}
    &\langle\langle 0|T^{-}(p_1)T^{-}(p_2)T^{-}(p_3)|0\rangle\rangle'=\frac{\langle 12 \rangle^2\langle 23\rangle^2\langle 3 1\rangle^2 p_1 p_2 p_3}{(E-2p_1)^2(E-2p_2)^2(E-2p_3)^2},\notag\\
    &\langle\langle 0|T^{-}(p_1)T^{-}(p_2)T^{+}(p_3)|0\rangle\rangle'=\frac{\langle 12\rangle^6 p_1p_2p_3}{\langle 23\rangle^2\langle 3 1\rangle^2 E^2},\notag\\&\langle\langle 0|T^{-}(p_1)T^{+}(p_2)T^{-}(p_3)|0\rangle\rangle'=\frac{\langle 3 1\rangle^6p_1p_2p_3}{\langle 12\rangle^2\langle 23\rangle^2 E^2},\notag\\&\langle\langle 0|T^{+}(p_1)T^{-}(p_2)T^{-}(p_3)|0\rangle\rangle'=\frac{\langle 2 3\rangle^6 p_1 p_2 p_3}{\langle 12 \rangle^2\langle 31\rangle^2 E^2},
\end{align}
and,
\begin{align}
    &\langle 0|T^{-}(Z_1)T^{-}(Z_2)T^{-}(Z_3)|0\rangle=-\frac{1}{2}|Z_1\cdot Z_2||Z_2\cdot Z_3||Z_3\cdot Z_1|,\notag\\
    &\langle 0|T^{+}(Z_1)T^{-}(Z_2)T^{-}(Z_3)|0\rangle=-2|Z_1\cdot Z_2|\int \frac{dc_1}{c_1^2}\int \frac{dc_2}{c_2^2}\delta^4(c_1 Z_1+c_2 Z_2-Z_3),\notag\\
    &\langle 0|T^{-}(Z_1)T^{+}(Z_2)T^{-}(Z_3)|0\rangle=-2|Z_1\cdot Z_2|\int \frac{dc_1}{c_1^2}\int dc_2 c_2^6\delta^4(c_1 Z_1+c_2 Z_2-Z_3),\notag\\
    &\langle 0|T^{+}(Z_1)T^{-}(Z_2)T^{-}(Z_3)|0\rangle=-2|Z_1\cdot Z_2|\int dc_1 c_1^6\int \frac{dc_2}{c_2^2}\delta^4(c_1 Z_1+c_2 Z_2-Z_3).
\end{align}

\bibliographystyle{JHEP}
\bibliography{biblio}

\end{document}